\documentclass[a4paper,11pt]{article}
\pdfoutput=1 

\usepackage{jheppub} 
\usepackage{etoolbox} 
\usepackage{braket}

\makeatletter
\patchcmd{\maketitle}{\@fpheader}{}{}{}
\makeatother

\usepackage[T1]{fontenc}
\usepackage{amsmath,amssymb,mathtools,slashed}
\usepackage{color,xcolor}
\definecolor{blue}{rgb}{0,0,0.5}
\usepackage[utf8]{inputenc}
\usepackage{enumerate}
\usepackage[printonlyused]{acronym}
\usepackage{footnote}
\usepackage{makecell}
\makesavenoteenv{tabular} 
\makesavenoteenv{table} 
\usepackage{hhline}
\usepackage{subcaption}
\usepackage{graphicx}
\usepackage{cancel}

\usepackage{multirow}

 \definecolor{darkgreen}{rgb}{0.1,0.1,1.0}
 \hypersetup{
    linktocpage,
     colorlinks,
     citecolor=darkgreen,
     linkcolor= darkgreen,
     urlcolor=darkgreen
 }

\usepackage{mciteplus}

\renewcommand{\arraystretch}{1.5}
\renewcommand{\Im}{\textrm{Im}}

\newcommand{\Ga}{\Gamma}

\newcommand{\la}{\lambda}

\newcommand{\sig}{\sigma}

\newcommand{\NC}{N_C}

\newcommand{\intm}{\textrm{int}^{-}}
\newcommand{\intp}{\textrm{int}^{+}}
\newcommand{\exc}{\textrm{exc}}

\newcommand{\gammaexc}[2]{ \Gamma^{#1}_{\exc}(#2)}
\newcommand{\gammaintm}[2]{ \Gamma^{#1}_{\intm}(#2)}
\newcommand{\gammaintp}[2]{ \Gamma^{#1}_{\intp}(#2)}
\newcommand{\gammaintpSL}[2]{ \Gamma^{#1,\mathrm{SL}}_{\intp}(#2)}

\newcommand{\ckm}[1]{|V_{#1}|^2}

\newcommand{\diq}{\mathcal{D}-q}

\newcommand{\Bary}{\mathcal{B}}
\newcommand{\Barycc}{{\mathcal{B}_{cc}}}

\newcommand{\Barybb}{{\mathcal{B}_{bb}}}

\newcommand{\Barybbstar}{{\mathcal{B}^*_{bb}}}

\newcommand{\Barybc}{{\mathcal{B}_{bc}}}

\newcommand{\Barybcprime}{{\mathcal{B}'_{bc}}}
\newcommand{\Barybcqprime}{{\mathcal{B}'_{bcq}}}
\newcommand{\Barybcstar}{{\mathcal{B}^*_{bc}}}
\newcommand{\Barybcqstar}{{\mathcal{B}^*_{bcq}}}

\newcommand{\Mes}{M}

\newcommand{\Xiccpp}{\Xi_{cc}^{++}}

\newcommand{\Xiccp}{\Xi_{cc}^{+}}

\newcommand{\Opsix}[2]{\langle O_{#1}^{#2}\rangle}
\newcommand{\Opsixt}[2]{\langle\tilde{O}_{#1}^{#2}\rangle}

\newcommand{\OpsevenP}[2]{\langle P_{#1}^{#2}\rangle}
\newcommand{\OpsevenPt}[2]{\langle \tilde{P}_{#1}^{#2}\rangle}

\newcommand{\Lifetime}[1]{\tau \left( #1 \right)}
\newcommand{\Spm}{S_{+-}}
\newcommand{\Spp}{S_{++}}
\newcommand{\Smm}{S_{--}}
\newcommand{\Opm}{O_{+-}}
\newcommand{\Omp}{O_{-+}}
\newcommand{\Opp}{O_{++}}
\newcommand{\Spmt}{\widetilde{S}_{+-}}
\newcommand{\Sppt}{\widetilde{S}_{++}}
\newcommand{\Smmt}{\widetilde{S}_{--}}
\newcommand{\Opmt}{\widetilde{O}_{+-}}
\newcommand{\Ompt}{\widetilde{O}_{-+}}

\newcommand{\ME}[2]{\langle #1 \rangle_{#2}}

\newcommand{\undermarker}[2]{\underbrace{#1}_{\rm #2}}

\newcommand{\wavefnsq}[2]{|\Psi_{#1}^{#2}(0)|^2}

\newcommand{\GeV}{\,\textrm{GeV}}

\definecolor{ivn}{rgb}{0.89,0.05,0.05}
\definecolor{LD}{rgb}{0.1,0.2,0.9}

\title{New Predictions for the Lifetimes of Doubly Heavy Baryons and the $B_c$ Meson}
 \author{Lovro Dulibi\' c,}
 \author{Bla\v zenka Meli\' c,}
 \author{and Ivan  Ni\v sand\v zi\'c}

\affiliation{Ru\dj er Bo\v skovi\'c Institute, Bijeni\v cka cesta 54, 10000, Zagreb, Croatia.}

\emailAdd{ldulibic@irb.hr}
\emailAdd{melic@irb.hr}
\emailAdd{ivan.nisandzic@irb.hr}

\abstract{We present updated predictions for the lifetimes of all weakly decaying doubly heavy baryons, including $bb$, $cc$, and $bc$ baryons, as well as for the $B_c$ meson. The analysis includes NNLO corrections to the leading dimension-three contribution, NLO corrections to the chromomagnetic term, and the complete set of currently known NLO corrections to the dimension-six heavy-light quark spectator contributions, including penguin terms. We also compare the results in the $\overline{\rm MS}$, kinetic, and, where applicable, $\Upsilon$ mass schemes.
For the $bc$ baryons, we present predictions for both possible ground-state diquark-spin assignments, $S_{bc}=0$ and $S_{bc}=1$. In the kinetic scheme we obtain the lifetime hierarchies 
$\tau(\Xi^0_{bb})<\tau(\Xi^-_{bb})\simeq\tau(\Omega^-_{bb})$,
$\tau(\Xi^+_{cc})<\tau(\Omega^+_{cc})<\tau(\Xi^{++}_{cc})$,
$\tau(\Xi^0_{bc})\lesssim\tau(\Omega^0_{bc})<\tau(\Xi^+_{bc})$ for $S_{bc}=0$, and
$\tau(\Xi^{\prime\, 0}_{bc})<\tau(\Omega^{\prime\, 0}_{bc})<\tau(\Xi^{\prime\, +}_{bc})$ for $S_{bc}=1$.
We also revisit the $B_c$ lifetime and discuss the impact of the newly included Darwin term.
}
\date{\today}

\begin{document}
\preprint{RBI-ThPhys-2026-06}
\maketitle
\flushbottom

\newpage
\section{Introduction}

Inclusive weak decays of heavy hadrons provide a natural setting for the application of the heavy quark expansion (HQE)~\cite{Shifman:1984wx,Chay:1990da,Bigi:1992su} and the associated Heavy Quark Effective Theory, see, for example, Refs.~\cite{Lenz:2014jha,Albrecht:2024oyn} for recent reviews. In this framework, the total decay width of a heavy hadron is evaluated using an operator product expansion in inverse powers of the heavy-quark mass $m_Q$. This expansion separates short-distance effects, encoded in perturbatively calculable Wilson coefficients, from long-distance effects, encoded in matrix elements of local operators. In the infinite-mass limit, the total decay widths reduce to the corresponding partonic decay widths of the heavy quarks contained in the hadron. Bound-state effects first appear as power-suppressed terms starting at order $1/m_Q^2$. At order $1/m_Q^3$, dimension-six terms include phase-space-enhanced four-quark operator contributions arising from spectator weak interactions between the heavy quark and the remaining valence quarks within the hadron. These contributions are responsible for the lifetime splittings between hadrons with the same heavy-quark content but different spectator flavors.

For hadrons containing a $b$ quark, the expansion parameter $1/m_b$ is small enough for the series to exhibit good convergence. Recent predictions~\cite{Lenz:2022rbq,Gratrex:2023pfn}, including both higher $1/m_b$ and $\alpha_s$ corrections, successfully reproduce the observed pattern of $b$-hadron lifetimes. For charmed hadrons, however, the $1/m_c$ expansion is more delicate. Since the charm-quark mass is only moderately larger than the characteristic nonperturbative scale of QCD, the convergence of the expansion is considerably slower. As a result, contributions of higher-dimensional operators become significant, enhancing the sensitivity of the predicted lifetimes to the corresponding nonperturbative matrix elements and thereby increasing the overall uncertainties of the predictions, as reflected in recent analyses~\cite{King:2021xqp,Gratrex:2022xpm}.

Doubly heavy hadrons provide an additional important testing ground for the HQE, as their lifetimes are gradually becoming experimentally accessible. In this work, we study the lifetimes of all weakly decaying doubly heavy baryons, together with the lifetime of the $B_c$ meson. Compared with earlier analyses~\cite{Kiselev:1999kh,Likhoded:1999yv,Kiselev:2001fw,Berezhnoy:2018bde,Cheng:2018mwu,Cheng:2019sxr,Yang:2022nps}, we incorporate a substantially updated set of short-distance contributions, including the leading decay term through NNLO, the chromomagnetic term and the dimension-six spectator contributions up to NLO and corresponding penguin contributions, and the Wilson coefficient of the dimension-six Darwin term at leading order in $\alpha_s$. In addition, we estimate the effects of subleading dimension-seven spectator operators. We provide a quantitative assessment of theoretical uncertainties, dominated by hadronic input parameters and renormalization-scale variation. To test the stability of the results with respect to the heavy-quark mass definition, we compare predictions obtained in the $\overline{\rm MS}$ and kinetic mass schemes. Furthermore, for the $bc$ baryons and the $B_c$ meson we also employ the $\Upsilon$ mass scheme. The present study also extends our previous analysis of doubly charmed baryons in Ref.~\cite{Dulibic:2023jeu} and places the lifetime predictions for all doubly heavy baryons on a common footing.

Specifically, we present new predictions for the lifetimes of the doubly bottom baryons
\begin{equation}
   \mathcal{B}_{bb}= (\Xi_{bb}^0,\, \Xi_{bb}^-,\, \Omega_{bb}^-)\,,
\end{equation}
the analogous $(bcq)$ baryons
\begin{equation}
   \mathcal{B}_{bc}= (\Xi_{bc}^{+},\, \Xi_{bc}^{0},\, \Omega_{bc}^{0})\,,
   \label{eq:unprimed}
\end{equation}
and, for comparison with earlier work, we also revisit our previous predictions~\cite{Dulibic:2023jeu} for the doubly charmed baryons
\begin{equation}
   \mathcal{B}_{cc}= (\Xi_{cc}^{++},\, \Xi_{cc}^{+},\, \Omega_{cc}^{+})\,.
\end{equation}

Throughout this work, we adopt the diquark--quark picture of doubly heavy baryons, in which the heavy-quark pair is treated as a compact color-antitriplet diquark interacting with the light quark. In this picture, the $cc$ and $bb$ diquarks in the ground-state $J^P=(1/2)^+$ baryons $\mathcal{B}_{cc}$ and $\mathcal{B}_{bb}$ are in the spin-triplet configuration, $S_{QQ}=1$, with a symmetric spin wave function. The corresponding hyperfine partners are the $J^P=(3/2)^+$ baryons $\mathcal{B}^{\ast}_{cc}$ and $\mathcal{B}^{\ast}_{bb}$, with the same diquark spin as the ground states.

The situation is more involved for doubly heavy $(bcq)$ baryons. For each light flavor, the low-lying spectrum contains two distinct $J^P=(1/2)^+$ states, corresponding to a $bc$ diquark with spin $S_{bc}=0$ or $S_{bc}=1$. The heavier state is expected to decay rapidly to the ground state through an electromagnetic transition, so that only the ground state is weakly decaying and accessible to lifetime measurements. However, current theoretical analyses of the mass spectrum do not conclusively determine whether this weakly decaying state corresponds to $S_{bc}=0$ or $S_{bc}=1$. The lattice-QCD evaluation of Ref.~\cite{Brown:2014ena} yields small mass splittings,
\begin{equation*}
    m_{\Xi'_{bc}}-m_{\Xi_{bc}} = 16(18)(38)\,\mathrm{MeV}\,,\qquad
    m_{\Omega'_{bc}}-m_{\Omega_{bc}} = 35(9)(25)\,\mathrm{MeV}\,,
\end{equation*}
compatible with either sign within uncertainties, so that both orderings remain viable. 
This ambiguity has affected previous lifetime analyses, which relied on specific assumptions about the mass ordering. For instance, Ref.~\cite{Kiselev:1999kh} evaluated $\mathcal{B}_{bc}$ lifetimes assuming a scalar $bc$ diquark in the ground state, whereas the more recent analysis of Ref.~\cite{Cheng:2019sxr} assumed a vector $bc$ diquark.

Given the lack of consensus among different theoretical approaches, we adopt an agnostic standpoint and present lifetime predictions for both the unprimed and primed baryons. Following the naming convention of the lattice study~\cite{Brown:2014ena}, we denote
\begin{equation}
    \Xi_{bc} (S_{bc}=0), \qquad\qquad \Xi'_{bc} (S_{bc}=1)\,,
\end{equation}
and analogously for the $\Omega_{bc}$ states\footnote{Note that the opposite naming convention is used in several papers in the literature.}. We therefore define the primed $S_{bc}=1$ baryons as
\begin{equation}
  \mathcal{B}_{bc}' \equiv
  (\Xi_{bc}^{\prime\, +},\, \Xi_{bc}^{\prime\, 0},\, \Omega_{bc}^{\prime\, 0})\,,
\end{equation}
in contrast to the unprimed $S_{bc}=0$ states in Eq.~\eqref{eq:unprimed}. Since the heavy-diquark spin significantly affects the relevant operator matrix elements through spin-dependent interactions with the light quark, these two sets of lifetime predictions may help distinguish the nature of the $(bcq)$ ground states.

The $\Xi_{cc}^{++}$ remains the only ground-state doubly heavy baryon whose lifetime has been measured so far. Its discovery and lifetime measurement were reported by the LHCb Collaboration~\cite{LHCb:2017iph,LHCb:2018zpl}, and the measured lifetime is in good agreement with our previous prediction in Ref.~\cite{Dulibic:2023jeu}. The $\Xi_{cc}^{+}$ was discovered very recently~\cite{LHCb:2026pxn}, while the $\Omega_{cc}^{+}$ remains the only weakly decaying doubly charmed baryon that has not yet been established experimentally, and existing LHCb searches set limits on its production rate and lifetime~\cite{LHCb:2021rkb}. No states have been observed to date in the doubly bottom  or bottom-charm sectors, although dedicated LHCb studies of $(bcq)$ final states are currently underway~\cite{LHCb:2021xba,LHCb:2022fbu}.

To complete the set of doubly heavy hadron lifetime predictions, we also revisit the lifetime of the $B_c$ meson, extending the analyses of Refs.~\cite{Beneke:1996xe,Aebischer:2021ilm} by including NNLO corrections to the leading decay term and NLO corrections to the chromomagnetic term. Furthermore, we discuss the impact of the Darwin term and compare our predictions with the experimentally measured $B_c$ lifetime~\cite{HFLAV:2024ctg}.

The rest of the paper is organized as follows. In Section~2, we review the HQE framework for doubly heavy hadrons, introduce the relevant operator basis and notation, and discuss the heavy-quark mass schemes used in our analysis. In Section~3, we determine the nonperturbative matrix elements entering the lifetime predictions for doubly heavy baryons. In Section~4, we present our numerical results and discuss the corresponding theoretical uncertainties. In Section~5, we revisit the lifetime of the $B_c$ meson, including corrections that have not been incorporated in earlier analyses. We briefly summarize in Section~6. Additional materials are collected in four appendices. 

\section{OPE for doubly heavy hadrons}
\subsection{Background}
We begin with a brief overview of the HQE framework for inclusive decays of heavy hadrons that underlies our analysis of doubly heavy baryon and $B_c$ meson lifetimes.

The total decay width of a heavy hadron is related via the optical theorem to the imaginary part of the forward matrix element of the transition operator
\begin{equation}
\mathcal{T} 
= i\,\int d^4x\, 
T\,\left[{\cal H}_{\rm eff}(x)\,{\cal H}_{\rm eff}(0)\right]\,
\label{eq:OpticalTheorem}
\end{equation}
as
\begin{equation}
\frac{1}{\Lifetime{H}}
= \Ga(H)
= \frac{1}{2M_H} \Im\, \langle H| \mathcal{T} |H\rangle\,.
\label{Eq:UnitRes}
\end{equation}
Here ${\cal H}_{\rm eff}$ denotes the low-energy weak effective Hamiltonian governing the interactions of the relevant heavy quarks. Since both bottom- and charm-quark decays enter our discussion, we introduce the corresponding $\Delta B=1$ and $\Delta C=1$ Hamiltonians separately.

The effective Hamiltonian for the $\Delta B=1$ transitions (see, e.g., Ref.~\cite{Buchalla:1995vs}) reads
\begin{align}
  {\cal H}_{\rm eff}^{(b)} = 
  \frac{G_F}{\sqrt{2}} \bigg\{\sum_{q_3 = d, s}
  &\bigg[\,
  \sum_{\substack{q_1 = u,c \\ q_2 = u,c} }
   V_{q_1 b}^* V_{q_2 q_3} 
  \Bigl(C_1 \, Q_{1,b}^{q_1 q_2 q_3}  + C_2 \, Q_{2,b}^{q_1 q_2 q_3}  \Bigr)
    -  V_{tb}^* V_{t q_3} 
  \sum \limits_{j=3, \ldots, 6, 8} C_j\, Q_{j,b}^{q_3} 
   \bigg]\nonumber\\
  &
   +\sum_{\substack{q=u,c \\ e,\mu, \tau}}
V_{q b}^* \, Q_{\rm SL,b}^{q \ell}\,\bigg\} + {\rm h.c.}\, ,
   \label{eq:Heff-NL-b}
\end{align}
where $G_F$ is the Fermi constant, $V_{i j}$ are the Cabibbo–Kobayashi–Maskawa (CKM) matrix elements, and $C_i$ are the Wilson coefficients of the $\Delta B=1$ operators evaluated at the renormalization scale $\mu \sim m_b$.
The relevant current–current operators\footnote{Following Ref.~\cite{Buchalla:1995vs}, $Q_2$ denotes the color-singlet operator, which differs from the alternative naming convention in which the color singlet is denoted by $Q_1$.} are 
\begin{equation}
\begin{aligned}
Q_{1,b}^{q_1 q_2 q_3} 
 =&   
\left(\bar b^i \, \gamma_\mu(1-\gamma_5) \, q_1^j \right)
\left(\bar q_2^j \, \gamma^\mu(1-\gamma_5)  \, q_3^i \right),
\\
Q_{2,b}^{q_1 q_2 q_3} 
 =&  
\left(\bar b^i  \, \gamma_\mu(1-\gamma_5)  \, q_1^i \right)
\left(\bar{q}_2^j \, \gamma^\mu(1-\gamma_5)  \, q_3^j \right)\,.
\end{aligned}
\end{equation}
The penguin operators are
\begin{equation}
\begin{aligned}
Q_{3,b}^{q_3} 
&
= (\bar b^i \, \gamma_\mu(1-\gamma_5) \, q_3^i) \sum_{q} ( \bar q^j \, \gamma^\mu(1-\gamma_5) \, q^j)
,  
\quad 
Q_{4,b}^{q_3} = (\bar b^i \, \gamma_\mu(1-\gamma_5) \, q_3^j) \sum_{q} (\bar q^j \, \gamma^\mu(1-\gamma_5) \, q^i)\,, 
\\
Q_{5,b}^{q_3} 
& 
=  (\bar b^i \, \gamma_\mu(1-\gamma_5) \, q_3^i) \sum_{q} (\bar q^j \, \gamma^\mu(1+\gamma_5) \, q^j), 
\quad
Q_{6,b}^{q_3}  = (\bar b^i \, \gamma_\mu(1-\gamma_5) \, q_3^j) \sum_{q} 
(\bar q^j \, \gamma^\mu(1+\gamma_5) \, q^i)\,,
\label{Eq:Q3456b}
\end{aligned}
\end{equation}
while the chromomagnetic operator reads\footnote{For the penguin and chromomagnetic operators we follow the convention in Ref.~ \cite{Krinner:2013cja}.}:
\begin{equation}
Q_{8,b}^{q_3} = \frac{g_s}{8 \pi^2} m_b
\left(\bar b^i \, \sigma^{\mu\nu} (1 - \gamma_5) t^a_{ij} \, q_3^j \right) G^a_{\mu \nu}\,.
\label{Eq:Q8b}
\end{equation}
Here, the color indices are denoted by $i$, and $j$, while $G_{\mu\nu} = G^a_{\mu\nu} t^a$ is the gluon field strength tensor, and $g_s$ is the strong coupling. Finally, the remaining semileptonic operators are:
\begin{equation}
    Q^{q \ell}_{\rm SL, b} =\left(\bar{b}^i\, \gamma_\mu(1-\gamma_5)\, q^i \right)
\left(\bar \nu_\ell \, \gamma^\mu(1-\gamma_5) \, \ell \right).
\end{equation}
In our analysis, the effects of penguin operators and chromomagnetic operator $Q_{8,b}^{q_3}$ are included as part of NLO contributions to the leading Wilson coefficient $\mathcal{C}_3$ in the $1/m_Q$ expansion, as well as in the NLO contributions to the Wilson coefficients of dimension-six four-quark operators.

Similarly, the effective Hamiltonian for $\Delta C = 1$ charm-quark
transitions reads
\begin{equation}
\begin{aligned}
    \mathcal{H}_{\rm eff}^{(c)}
  =\frac{G_F}{\sqrt{2}} & \bigg\{
    \sum_{q_1,q_2=d,s} V^{\vphantom{\ast}}_{cq_1}V^\ast_{uq_2}
    \big(C_1 Q_{1,c}^{q_1q_2}+C_2 Q_{2,c}^{q_1q_2}\big)
    -V^{\vphantom{\ast}}_{ub}V^\ast_{cb}\sum_{j=3}^{6}C_jQ_{j,c}\\
    &+\sum_{\substack{q=d,s \\ e,\mu}}
V_{c q}\,Q_{\text{SL},c}^{q\ell}\bigg\}  +\textrm{h.c.}\,,
    \label{eq:Heff-NL-c}
\end{aligned}
\end{equation}
with the $\Delta C=1$ current–current and semileptonic operators:
\begin{equation}
\begin{split}
    Q_{1,c}^{q_1q_2}&=
    (\bar{c}^i \gamma^\mu(1-\gamma_5) q_1^j)\,
    (\bar{q}_2^{\,j}\gamma_\mu(1-\gamma_5) u^i)\,,\\[0.2em]
    Q_{2,c}^{q_1q_2}&=
    (\bar{c}^i \gamma^\mu(1-\gamma_5) q_1^i)\,
    (\bar{q}_2^{\,j}\gamma_\mu(1-\gamma_5) u^j)\,,\\[0.2em]
    Q_{\text{SL},c}^{q\ell}&=
    (\bar{c}^i\,\gamma^\mu(1-\gamma_5) \,q^i)\,
    (\bar{\ell}\,\gamma_\mu(1-\gamma_5)\,\nu_\ell)\,,
\end{split}
\end{equation}
where the corresponding Wilson coefficients are defined at the scale $\mu\sim m_c$.
The penguin operators $Q_{3\text{--}6,c}$ in the charm sector, defined in analogy with \eqref{Eq:Q3456b}, are suppressed by the small CKM factor $V^{\vphantom{\ast}}_{ub}V^\ast_{cb}$ and by the small values of the Wilson coefficients \cite{Buchalla:1995vs}, and are therefore neglected.

In the HQE framework, the transition operator $\mathcal{T}$ is expanded as a series of local operators with $\Delta B = \Delta C = 0$, organized as a power expansion in $\Lambda_{\text{QCD}}/m_Q$
\begin{equation}
\Im\,\mathcal{T} = \bigg(\mathcal{C}_{3}\,\mathcal{O}_{3} +\frac{\mathcal{C}_{5}}{m_Q^2}\,\mathcal{O}_{5} +\frac{\mathcal{C}_{6}}{m_Q^3}\,\mathcal{O}_{6} +\dots\bigg) + 16\pi^2\bigg(\frac{\tilde{\mathcal{C}}_6}{m_Q^3}\,\tilde{\mathcal{O}}_{6} + \frac{\tilde{\mathcal{C}}_{7}}{m_Q^4}\,\tilde{\mathcal{O}}_{7} + \dots\bigg)\,,\label{Eq:HQEschematic} 
\end{equation} 
where $m_Q$ denotes the relevant heavy-quark mass, $m_b$ or $m_c$, and the expansions in both $1/m_b$ and $1/m_c$ are understood implicitly. The Wilson coefficients $\mathcal{C}_i$ and $\tilde{\mathcal{C}}_{i}$ have been calculated in perturbation theory as expansions in the strong coupling constant $\alpha_s$ and incorporate the phase-space mass effects. The first bracket contains heavy-quark bilinear operators and corresponds to non-spectator contributions, while the second bracket contains so-called spectator four-quark operators, sensitive to light-quark flavors and enhanced by the factor $16\pi^2$ relative to the nonspectator contributions.

\begin{table}[h]
\footnotesize
    \centering
    \renewcommand{\arraystretch}{1.5}
    \begin{tabular}{lcccc}
    \hline\hline
     {\rm Wilson coeff.}&  & $\alpha_s^0$ (LO) & $\alpha_s^1$ (NLO) & $\alpha_s^2$ (NNLO) \\ 
    \hline
    \multirow{2}{*}{$c_{3,Q}$} 
        & SL & \multirow{2}{*}{} & \cite{Fael:2024gyw,Czarnecki:1994bn} & \cite{Fael:2024gyw,Czarnecki:1994bn} \\
        & NL & & \cite{Egner:2024azu} & \cite{Egner:2024azu} \\
    \hline
    \multirow{2}{*}{$c_{G,Q}$} 
        & SL & \multirow{2}{*}{\cite{Bigi:1992su,Bigi:1992ne}} & \cite{Moreno:2022goo} & \multirow{2}{*}{} \\
        & NL & & \cite{Mannel:2023zei,Mannel:2024uar,Mannel:2025fvj} & \\
    \hline
    \multirow{1}{*}{$c_{\rho,c}$} 
        & SL+NL & \cite{Lenz:2020oce,King:2021xqp} & \multirow{1}{*}{} & \multirow{1}{*}{} \\
    \hline
    \multirow{2}{*}{$c_{\rho,b}$} 
        & SL & \cite{Moreno:2022goo,Moreno:2024bgq} & \multirow{2}{*}{}\cite{Moreno:2024bgq,Moreno:2022goo,Mannel:2021zzr}$^*$ & \multirow{2}{*}{} \\
        & NL & \cite{Lenz:2020oce} &  & \\
    \hline
    \multirow{2}{*}{$\tilde{c}_{6,i}$} 
        & SL & \cite{Lenz:2013aua} & \cite{Lenz:2013aua} & \multirow{2}{*}{} \\
        & NL & \cite{Franco:2002fc,Ciuchini:2001vx} & \cite{Franco:2002fc,Ciuchini:2001vx} & \\
    \hline
    \multirow{1}{*}{$\tilde{c}_{6,i}^{bc}$} 
        & SL+NL & \cite{Aebischer:2021ilm,Beneke:1996xe,Cheng:2019sxr,Chang:2000ac} & & \\
    \hline
    \multirow{2}{*}{$\tilde{c}_{7,i}$} 
        & SL & \cite{Lenz:2013aua} & \multirow{2}{*}{} & \multirow{2}{*}{} \\
        & NL & \cite{Lenz:2013aua} & & \\
    \hline\hline
    \end{tabular}
    \caption{Overview of the references with the calculations of the Wilson coefficients used in this analysis. The term $\tilde{c}_{6}^{bc}$ represents the specific dimension-six four-quark contribution for the $bc$ baryon sector involving $b$ and $c$ fields. Note that in our analysis we incorporate only results for which the total coefficient is known. Partial results, such as the semileptonic NLO correction to the Darwin coefficient, are not included; these are denoted by an asterisk.}
    \label{tab:OPEcoefReferences}
\end{table}

Taking the forward matrix element of the expanded transition operator between $\mathcal{B}_{Q_1Q_2}$ states, the HQE of the transition operator $\mathcal{T}$ translates, via Eq.~\eqref{Eq:UnitRes}, into the corresponding expansion of the total width of a doubly heavy baryon,
\begin{equation}
\Gamma=\Gamma_{3+5}+\Gamma_\rho+\widetilde{\Gamma}_6+\widetilde{\Gamma}_7,
\end{equation}
where $\Gamma_{3+5}$ contains the dimension-three and dimension-five two-quark contributions, $\Gamma_\rho$ denotes the two-quark contribution at dimension-six coming from the Darwin term, and $\widetilde{\Gamma}_6$ and $\widetilde{\Gamma}_7$ denote the dimension-six (including penguin contributions) and dimension-seven four-quark contributions, respectively. Each contribution can in turn be decomposed into nonleptonic and semileptonic parts, which we denote by the superscripts $\mathrm{NL}$ and $\mathrm{SL}$, respectively.

Specifically, for the case of doubly bottom baryons we have\footnote{Following the notation in Ref.~\cite{Dulibic:2023jeu}, we use $c_G'$ to distinguish the total contribution to chromomagnetic Wilson coefficient $c_G$ from the contribution that comes from the $1/m_Q$ expansion of $\mathcal{O}_3$ (see Eq.~\ref{eq:QQ} below), that is $c_G' = c_G - c_3/2$.}: 
\begin{align}
    \Gamma(\mathcal{B}_{\scriptscriptstyle b b})&=\Gamma_0(m_b)\bigg[c_{3,b}(\mu_b)\frac{\langle\mathcal{B}_{\scriptscriptstyle bb}\vert \bar{b}b\vert\mathcal{B}_{\scriptscriptstyle bb}\rangle}{2M_{\mathcal{B}_{bb}}}+c^{\prime}_{G,{\scriptscriptstyle b}}(\mu_b)\frac{\hat{\mu}^2_{G,b}}{m_b^2}+c_{\rho,b}(\mu_b)\frac{\hat{\rho}^3_{D,b}}{m_b^3}+\ldots\nonumber\\
    &+\frac{16\pi^2}{2M_{\mathcal{B}_{bb}}}\sum_{q,i}\bigg(\tilde{c}^q_{6,i}(\mu_b)\frac{\langle\mathcal{B}_{\scriptscriptstyle bb}\vert O^{bq}_i\vert \mathcal{B}_{\scriptscriptstyle bb}\rangle}{m_b^3}+\tilde{c}^q_{7,i}(\mu_b)\frac{\langle\mathcal{B}_{\scriptscriptstyle bb}\vert P^{bq}_i\vert \mathcal{B}_{\scriptscriptstyle bb}\rangle}{m_b^4}+\ldots\bigg)\bigg]\,
    \label{Eq:OPEbb}
\end{align}
where 
\begin{equation}
   \Gamma_0(m_Q)= \frac{G_F^2 m_Q^5}{192\pi^3}\,,\label{Eq:Gamma0Def}
\end{equation}
and the chromomagnetic and Darwin matrix elements, $\hat{\mu}^2_{G,Q}$ and $\hat{\rho}^3_{D,Q}$, are defined in Eq.~\eqref{Eq:muDefs} below. The corresponding expression for doubly charmed baryons is obtained by replacing $b$ with $c$.

For the $\mathcal{B}_{bc}$ baryons both $1/m_c$ and $1/m_b$ expansions are present, and the total width can be written as 
\begin{align}
    \Gamma(\mathcal{B}_{\scriptscriptstyle b c})&=\sum_{Q=c,b}\Gamma_0(m_Q)\bigg[c_{3,Q}(\mu_Q)\frac{\langle\mathcal{B}_{\scriptscriptstyle bc}\vert \bar{Q}Q\vert\mathcal{B}_{\scriptscriptstyle bc}\rangle}{2M_{\mathcal{B}_{bc}}}+c^{\prime}_{G,{\scriptscriptstyle Q}}(\mu_Q)\frac{\hat{\mu}^2_{G,Q}}{m_Q^2}+c_{\rho,Q}(\mu_Q)\frac{\hat{\rho}^3_{D,Q}}{m_Q^3}+\ldots\nonumber\\
    &+\frac{16\pi^2}{2M_{\mathcal{B}_{bc}}}\sum_{q,i}\bigg(\tilde{c}^q_{6,i}(\mu_Q)\frac{\langle\mathcal{B}_{\scriptscriptstyle bc}\vert O^{Qq}_i\vert \mathcal{B}_{\scriptscriptstyle bc}\rangle}{m_Q^3}+\tilde{c}^q_{7,i}(\mu_Q)\frac{\langle\mathcal{B}_{\scriptscriptstyle bc}\vert P^{Qq}_i\vert \mathcal{B}_{\scriptscriptstyle bc}\rangle}{m_Q^4}+\ldots\bigg)\bigg]\,.
    \label{Eq:OPEbc}
\end{align}
Table~\ref{tab:OPEcoefReferences} summarizes the references in which the relevant Wilson coefficients have been computed to various orders in $\alpha_s$.

The forward matrix element of the dimension-three operator for the doubly heavy baryon $\Bary$ can be expressed in terms of the kinetic and chromomagnetic terms as
\begin{equation}
    \frac{\langle \mathcal{B} \vert \bar{Q}Q\vert \mathcal{B}\rangle}{2M_{\mathcal{B}}}=n_Q(\mathcal{B})-\frac{\hat{\mu}^2_{\pi,Q}(\mathcal{B})}{2m_Q^2}+\frac{\hat{\mu}^2_{G,Q}(\mathcal{B})}{2m_Q^2}\,,
    \label{eq:QQ}
\end{equation}
with no additional terms at higher orders\footnote{This is a consequence of defining $\hat{\mu}_{\pi,Q}^2$ and $\hat{\mu}_{G,Q}^2$ in terms of the phase-redefined field $Q_v$, rather than the HQET field $h_v$ (see e.g. Ref.~\cite{Dassinger:2006md}).} in $1/m_Q$.
In what follows, we neglect non-valence contributions, which are expected to be small, and therefore set to zero those matrix elements in which the quark field $Q$ within the operator does not match any of the heavy valence quarks within the baryon. For instance, we take quantities such as $\hat{\mu}_{\pi,c}^2(\mathcal{B}_{bb})$ to vanish.
In Eq. \eqref{eq:QQ} the constants are:
\begin{equation}
n_b(\mathcal{B}_{bb})=n_c(\mathcal{B}_{cc})=2\,,\qquad \text{and} \qquad  n_b(\mathcal{B}_{bc})=n_c(\mathcal{B}_{bc})=1\,.
\label{Eq:Factor-n}
\end{equation}

Following Refs.~\cite{Bigi:1993ex,Dassinger:2006md,King:2021xqp}, we define the nonspectator matrix elements as: 
\begin{align}
\hat{\mu}^2_{\pi,Q}(\mathcal{B}) &\equiv -\frac{1}{2M_{\mathcal{B}}} \langle \mathcal{B} |\bar{Q}_v (iD)^2 Q_v | \mathcal{B} \rangle \,, \nonumber \\
\hat{\mu}^2_{G,Q}(\mathcal{B})&\equiv\frac{1}{2M_{\mathcal{B}}} \langle \mathcal{B} |\bar{Q}_v \frac{1}{2} \sig_{\mu\nu} (g_s G^{\mu\nu}) Q_v  |\mathcal{B} \rangle\,, \nonumber \\
\hat{\rho}^3_{D,Q}(\mathcal{B}) &\equiv \frac{1}{2M_{\mathcal{B}}}\langle \mathcal{B} |\bar{Q}_v (i D_\mu) (i v \cdot D) (i D^\mu) Q_v | \mathcal{B} \rangle \,,
\label{Eq:muDefs}
\end{align} 
where $Q_v(x)\equiv e^{im_Q v\cdot x}Q(x)$ denotes the phase-redefined heavy quark field.  Throughout the paper, the operators are defined in terms of the phase-redefined QCD field $Q_v$ and are denoted with hat, while the corresponding operators written in terms of HQET field $h_v$ are denoted without a hat.

In addition to two-quark contributions, the decay widths receive four-quark contributions, which are conveniently organized according to their diagram topologies. The corresponding topologies relevant for $\mathcal{B}_{bb}$, $\mathcal{B}_{cc}$, and $\mathcal{B}_{bc}$ baryons are shown in the Appendix \ref{app:figures}, Figs.~\ref{fig:bb-topologies}, \ref{fig:cc-topologies}, and~\ref{fig:bc-topologies}, respectively. These correspond to weak exchange, constructive, and destructive Pauli interference, denoted by exc, $\mathrm{int}^{+}$, and $\mathrm{int}^{-}$, respectively. For $\mathcal{B}_{bb}$ and $\mathcal{B}_{cc}$, the four-quark contributions involve interactions of the heavy quark with the light valence quark, whereas for $\mathcal{B}_{bc}$ both heavy flavors can participate, leading to topologies with either a heavy or a light spectator, as indicated in Fig.~\ref{fig:bc-topologies}.

With this classification in hand, we decompose the four-quark contributions to the total widths into CKM factors multiplying topology-specific expressions. Keeping only valence contributions, the resulting expressions read for the doubly bottom baryons:
\begin{equation}
\begin{aligned}
\tilde{\Gamma}(\Xi_{bb}^{0})
=|V_{cb}|^2&\left(|V_{ud}|^2+|V_{us}|^2\right)\gammaexc{u}{y_c,0}
\\
+|V_{ub}|^2&\Big( \left (|V_{ud}|^2 + |V_{us}|^2\right ) \left ( \gammaexc{u}{0,0}+\gammaintp{u}{0,0} \right )
\\
&
+\left(|V_{cs}|^2+|V_{cd}|^2\right)\gammaintp{u}{y_c,0}\Big)+|V_{ub}|^2\sum_{l=e,\mu,\tau}\gammaintpSL{u}{y_l,0} 
\,,
\label{eq:Xibbspectator}
\end{aligned}
\end{equation}
\begin{equation}
\hspace*{-4cm}
\begin{aligned}
\tilde{\Gamma}(\Xi_{bb}^{-})
=&|V_{cb}|^2\Big( |V_{ud}|^2\gammaintm{d}{y_c,0}+|V_{cd}|^2\gammaintm{d}{y_c,y_c}\Big)
\\
+&|V_{ub}|^2\Big( |V_{ud}|^2\gammaintm{d}{0,0}+|V_{cd}|^2\gammaintm{d}{0,y_c} \Big)\,,
\label{eq:Xibbmspectator}
\end{aligned}
\end{equation}
\begin{equation}
\hspace*{-4cm}
\begin{aligned}
\tilde{\Gamma}(\Omega_{bb}^{-})
=&|V_{cb}|^2\Big( |V_{cs}|^2\gammaintm{s}{y_c,y_c}+|V_{us}|^2\gammaintm{s}{y_c,0} \Big)
\\
+&|V_{ub}|^2\Big( |V_{cs}|^2\gammaintm{s}{0,y_c}+|V_{us}|^2\gammaintm{s}{0,0} \Big)\,,
\label{eq:Omegabbspectator}
\end{aligned}
\end{equation}
with the notation referring to the topologies in Figure \ref{fig:bb-topologies}, and the corresponding leading order expressions given in Appendix \ref{AppA1}.

The analogous expressions for doubly-charmed baryons are \cite{Dulibic:2023jeu}:
\begin{align}
\tilde{\Gamma}(\Xi_{cc}^{++})
&=
|V_{cs}|^2  \Big ( |V_{ud}|^2
\gammaintm{u}{x_s,0}
+
 |V_{us}|^2
\gammaintm{u}{x_s,x_s} \Big )\nonumber\\
&\quad +|V_{cd}|^2 |V_{ud}|^2
\gammaintm{u}{0,0}
\,,
\label{eq:Xiccppspectator}\\
\tilde{\Gamma}(\Xi_{cc}^{+})
&=
|V_{cs}|^2 |V_{ud}|^2
\gammaexc{d}{x_s,0}
+
|V_{cd}|^2 |V_{ud}|^2
\Big(
\gammaintp{d}{0,0}
+
\gammaexc{d}{0,0}
\Big)
\nonumber\\[2mm]
&\quad
+
|V_{cd}|^2 \sum_{l=e,\mu}
\gammaintpSL{d}{x_l,0}
\,,
\label{eq:Xiccpspectator}\pagebreak[2]\\[2mm]
\tilde{\Gamma}(\Omega_{cc}^{+})
&=
|V_{cs}|^2 |V_{ud}|^2
\gammaintp{s}{0,0}
+
|V_{cs}|^2 |V_{us}|^2
\Big(
\gammaintp{s}{x_s,0}
+
\gammaexc{s}{x_s,0}
\Big)
\nonumber\\
&\quad
+
|V_{cs}|^2 \sum_{l=e,\mu}
\gammaintpSL{s}{x_l,0}
\,,
\label{eq:Omccspectator}
\end{align}
with the corresponding topology diagrams shown in Figure \ref{fig:cc-topologies}.

For the $bc$ baryons, we distinguish between topologies involving the interaction of the two heavy quarks, $b$ and $c$, and those involving a single heavy quark, $b$ or $c$, interacting with the light quark. The corresponding expressions are
\begin{equation}
\begin{aligned}
\tilde{\Gamma}(\Xi_{bc}^{+})
=\ckm{cs}\Big(&\ckm{ud}\gammaintm{cu}{x_s,0}+\ckm{us}\gammaintm{cu}{x_s,x_s}\Big)\\
+\ckm{cd}\Big(&\ckm{ud}\gammaintm{cu}{0,0}+\ckm{us}\gammaintm{cu}{0,x_s}\Big)\\
+\ckm{cb}\Big(& \left( \ckm{cs} + \ckm{cd} \right) \left (\gammaexc{bc}{y_+^c,0}+\gammaintp{bc}{y_-^c,0} \right )  \\ 
&+\left( \ckm{ud} + \ckm{us} \right) 
\left ( \gammaexc{bu}{y_c,0} + \gammaintp{bc}{0,0} \right ) \Big)
+ \ckm{cb} \sum_{l=e,\mu,\tau}\gammaintpSL{bc}{y_-^l,0},
\end{aligned}
\end{equation}
\begin{equation}
\begin{aligned}
\tilde{\Gamma}(\Xi_{bc}^{0})
=\ckm{ud}\Big(&\ckm{cs}\gammaexc{cd}{x_s,0}+\ckm{cd}\gammaexc{cd}{0,0}\Big)\\
+\ckm{cd}\Big(&\ckm{ud}\gammaintp{cd}{0,0}+\ckm{us}\gammaintp{cd}{0,x_s}\Big) + \ckm{cd}\sum_{l=e,\mu}\gammaintpSL{cd}{x_l,0}\\
+\ckm{cb}\Big(&\left ( \ckm{cs} + \ckm{cd} \right ) \left ( \gammaexc{bc}{y_+^c,0}+ \gammaintp{bc}{y_-^c,0} \right ) \\
&+\left (\ckm{ud} + \ckm{us} \right ) \gammaintp{bc}{0,0}
 \\
&+\ckm{ud}\gammaintm{bd}{y_c,0}+\ckm{cd}\gammaintm{bd}{y_c,y_c}\Big)+ \ckm{cb}\sum_{l=e,\mu,\tau}\gammaintpSL{bc}{y^l_-,0},\\
\end{aligned}
\end{equation}
\begin{equation}
\begin{aligned}
\tilde{\Gamma}(\Omega_{bc}^{0})
=\ckm{us}\Big(&\ckm{cs}\gammaexc{cs}{x_s,0}+\ckm{cd}\gammaexc{cs}{0,0}\Big)\\
+\ckm{cs}\Big(&\ckm{ud}\gammaintp{cs}{0,0}+\ckm{us}\gammaintp{cs}{0,x_s}\Big) + \ckm{cs}\sum_{l=e,\mu}\gammaintpSL{cs}{x_l,0}\\
+\ckm{cb}\Big(& \left ( \ckm{cs} + \ckm{cd} \right ) \left ( \gammaexc{bc}{y_+^c,0} + \gammaintp{bc}{y_-^c,0} \right ) \\
&+\left (\ckm{ud} + \ckm{us} \right ) \gammaintp{bc}{0,0} \\
&+\ckm{cs}\gammaintm{bs}{y_c,y_c}+\ckm{us}\gammaintm{bs}{y_c,0}\Big)+\ckm{cb}\sum_{l=e,\mu,\tau}\gammaintpSL{bc}{y_-^l,0},
\label{eq:Ombcspectator}
\end{aligned}
\end{equation}
For the $bc$ baryons, we extend the notation by explicitly writing both the heavy and light quarks which interact in the appropriate four-quark topologies, shown in Figure \ref{fig:bc-topologies}. For example, $\gammaexc{Q_1 Q_2}{x_{q_1},x_{q_2}}$ indicates the matrix element of the weak exchange topology with external quarks $Q_1$ and $Q_2$, with dependence on internal light quark mass ratios $x_{q_1}$ and $x_{q_2}$. The leading-order expressions for the topologies with two massive external quarks can be found, for example, in Refs.~\cite{Aebischer:2021ilm,Cheng:2019sxr,Chang:2000ac}; for convenience, we collect them again in Appendix~\ref{AppA1}. The corresponding penguin contributions are listed in Appendix~\ref{AppA2}. 

For dimension-six four-quark $b$-quark decays, we neglect mass effects of all light quarks and leptons, except for the 
$\tau$-lepton mass. In $c$-quark decays, we retain only the strange-quark mass and, for semileptonic modes, the muon mass. Accordingly, the mass-ratio parameters entering our analysis are
\begin{equation}\label{eq:massratios}
\begin{gathered}
     x_s=\frac{m_s^2}{m_c^2},\quad
     x_\mu=\frac{m_\mu^2}{m_c^2},\quad 
     \\
     y_c=\frac{m_c^2}{m_b^2},\quad
     y_\tau = \frac{m_\tau^2}{m_b^2},\quad 
     y_{\pm}^c =\frac{m_c^2}{p_\pm^2},\quad 
    y_{\pm}^{\tau} =\frac{m_{\tau}^2}{p_\pm^2}\,,
\end{gathered}
\end{equation}
where $p_+^2=(m_b+m_c)^2$ and $p_-^2=(m_b-m_c)^2$. 

The NLO corrections to the heavy-light four-quark contributions are based on the results of Refs.~\cite{Ciuchini:2001vx,Franco:2002fc,Beneke:2002rj}. 
Following the standard counting, we include the effect of the penguin operators to the dimension-six spectator Wilson coefficients as the part of the NLO corrections~\cite{Ciuchini:2001vx}. For the heavy-heavy four-quark operators, the NLO corrections are currently unavailable, but we do include their penguin contributions.

\subsection{Heavy quark mass schemes}\label{sec:massSchemes}

The decay width of a heavy hadron is highly sensitive to the heavy-quark mass, which enters the leading term of the HQE as $m_Q^5$. The choice of mass definition is particularly significant in the charm sector, where an uncertainty of order $\Lambda_{\text{QCD}}$
 constitutes a larger fraction of the charm-quark mass than it does of the bottom-quark mass in the bottom sector\footnote{For a recent attempt to address this issue, see, e.g., Ref.~\cite{Boushmelev:2023kmf}.}.

The analytic expressions in the HQE are commonly expressed in terms of the heavy-quark pole mass. However, the pole mass itself is not a physical parameter, because confinement prevents quarks from appearing as asymptotic on-shell states. Moreover, using the pole mass as a numerical input in the evaluation of total widths is problematic, since the perturbative expansion relating it to a short-distance mass is asymptotic, with coefficients that grow factorially at large orders due to the infrared renormalon. As a consequence, the pole mass carries an irreducible ambiguity of order ${\cal O}(\Lambda_{\rm QCD})$, see, for example, Refs.~\cite{Bigi:1994em,Beneke:1998ui,Beneke:2021lkq} for more detailed discussions. For this reason, we use a short-distance mass as the basic input in the numerical analysis and express the pole mass appearing in the analytic formulas through the corresponding perturbative expansions.

The most common short-distance mass definition is the $\overline{\rm MS}$ mass, denoted by $\overline{m}(\mu)$. In our analysis, we use the relation between the pole and $\overline{\rm MS}$ masses through $\mathcal{O}(\alpha_s^2)$, in order to match the perturbative accuracy of the dimension-three contribution to the total decay widths\footnote{Higher-order terms are known~\cite{Fael:2020iea}, but are not required for our purposes.}. The corresponding expression \cite{Fael:2020iea} through $\mathcal{O}(\alpha_s^2)$ is
\begin{equation}
\begin{aligned}
    \frac{m_0}{\overline{m}(\mu)}=&1+\frac{\alpha_s(\mu)}{\pi}\Bigg(\frac{4}{3}+\ln\frac{\mu^2}{\overline{m}(\mu)^2}\Bigg)\\
    &+\left(\frac{\alpha_s(\mu)}{\pi}\right)^2
    \Bigg[\frac{307}{32} + \frac{509}{72}\ln\frac{\mu^2}{\overline{m}(\mu)^2} + \frac{47 }{24}\ln^2\frac{\mu^2}{\overline{m}(\mu)^2} \\
    &-n_l\left(\frac{71}{144} + \frac{13}{36}\ln\frac{\mu^2}{\overline{m}(\mu)^2} + \frac{1}{12}\ln^2\frac{\mu^2}{\overline{m}(\mu)^2}\right) + \frac{\pi^2}{3}\left(1-\frac{n_l}{6}+\frac{1}{3}\ln 2\right)-\frac{\zeta_3}{6}\Bigg]+\ldots,
\end{aligned}
\end{equation}
where $n_l$ denotes the number of light quarks.

\begin{table}[ht]
\small
    \centering
    \renewcommand{\arraystretch}{1.4}
    \begin{tabular}{cccc}
    \hline\hline
         & Heavy quark mass & Loop quark mass & Renormalization scale\\\hline
        $b$ decay & \makecell[l]{$m_b^{\rm kin}(\mu_{\rm kin}=1\,{\rm GeV})=4.573\,{\rm GeV}$\\$\overline{m}_b(\overline{m}_b)=4.18$ GeV\\ $m_b^{\Upsilon,(0)} =m_\Upsilon/2=4.73\,{\rm GeV}$}
        & $\overline{m}_c(\mu_b)=0.904\,{\rm GeV}$ & \makecell{$\mu_b=4.5\,{\rm GeV}$\\$\mu\in [3,6]$ GeV}
        \\\hline 
        $c$ decay & \makecell[l]{$m_c^{\rm kin}(\mu_{\rm kin}=0.5\,{\rm GeV})=1.4\,{\rm GeV}$\\$\overline{m}_c(\overline{m}_c)=1.28$ GeV\\ $m_c^{\Upsilon,(0)}=m_c^{(0)}=1.30\,{\rm GeV}$} & $\overline{m}_s(\mu_c)=0.093\,{\rm GeV}$& \makecell{$\mu_c=1.5\,{\rm GeV}$\\$\mu\in [1,2]$ GeV}\\\hline\hline
    \end{tabular}
    \caption{Quark masses and central values for renormalization scales used in the numerical analysis, as well as the scale variations used for estimating the scale dependence uncertainties. In evaluating the charm quark kinetic mass we use \cite{Fael:2020iea,Herren:2017osy}. The LO masses in the $\Upsilon$ scheme quoted here,  $m_b^{\Upsilon,(0)}$ and $m_c^{\Upsilon,(0)}$ are defined via Eqs.~\eqref{Eq:UpsilonRelation} and \eqref{Eq:mcExpansion}.}
    \label{tab:massSchemes}
\end{table}

A particularly suitable low-scale short-distance mass definition, especially for inclusive $b$-hadron decays, is the kinetic mass scheme~\cite{Bigi:1994ga}. This scheme is designed to suppress infrared sensitivity by absorbing long-distance fluctuations below a cutoff scale $\mu_\text{kin}$ into the mass definition. The relation between the pole mass and the kinetic mass $m_{\text{kin}}$ is given by~\cite{Fael:2020iea}
\begin{equation}
\begin{split}
    \frac{m_0}{m_{\text{kin}}(\mu_{\text{kin}})} &= 1+\frac{\alpha_s(\mu)}{\pi}\left(\frac{16}{9}\frac{\mu_{\text{kin}}}{m_{\text{kin}}}+\frac{2}{3}\frac{\mu_{\text{kin}}^2}{m_{\text{kin}}^2}\right) \\
    &\quad + \left(\frac{\alpha_s(\mu)}{\pi}\right)^2 \Bigg[ \frac{\mu_{\text{kin}}}{m_{\text{kin}}} \left(
    \frac{860}{27} - \frac{128}{81}n_l - \frac{8}{9}\pi^2 
    - \frac{88}{9} \ln \frac{2\mu_{\text{kin}}}{\mu} 
    + \frac{16}{27}n_l \ln \frac{2\mu_{\text{kin}}}{\mu}
    \right) \\[10pt]
    &\qquad + \left( \frac{\mu_{\text{kin}}}{m_{\text{kin}}} \right)^2 \left(
    \frac{91}{9} - \frac{13}{27}n_l - \frac{\pi^2}{3} 
    - \frac{11}{3} \ln \frac{2\mu_{\text{kin}}}{\mu} 
    + \frac{2}{9}n_l \ln \frac{2\mu_{\text{kin}}}{\mu}
    \right) \\[10pt]
    &\qquad -\frac{32}{27}\left( \frac{\mu_{\text{kin}}}{m_{\text{kin}}} \right)^3  
    -\frac{4}{9}\left( \frac{\mu_{\text{kin}}}{m_{\text{kin}}} \right)^4 \Bigg] + \ldots
\end{split}
\end{equation}

In addition to kinetic and $\overline{\rm MS}$ schemes, we also evaluate lifetimes of $\mathcal{B}_{bc}$ baryons and of the $B_c$ meson using the $\Upsilon$ scheme. In this mass scheme~\cite{Pineda:1997hz,Melnikov:1998ug,Hoang:1998hm,Hoang:1998ng} the $b$-quark pole mass is expressed in terms of the mass of $\Upsilon(1S)$ meson, starting at order\footnote{The $\alpha_s^4$ corrections are also known \cite{Pineda:1997hz,Melnikov:1998ug}, but are beyond the scope of the present work.} $\alpha_s^2$: 
\begin{equation}
\frac{m_\Upsilon/2}{m_b} = 1 - \frac{(\alpha_s C_F)^2}{8}\left\{1\cdot\epsilon + \frac{\alpha_s}{\pi}\left[\left(\ln\frac{\mu}{\alpha_s C_F m_b} + \frac{11}{6}\right)\beta_0 - 4\right]\epsilon^2 + \mathcal{O}(\epsilon^3)\right\}\,,
\label{Eq:UpsilonRelation}
\end{equation}
where $\beta_0 = 11 - 2n_f/3$, $C_F = 4/3$,  and $\epsilon=1$ is a bookkeeping parameter that tracks the power counting so that the $\mathcal{O}(\epsilon^n)$ term in the above expansion corresponds to $\mathcal{O}(\alpha^n)$ correction in the $\alpha_s$-expansion of the Wilson coefficients entering the inclusive decay widths.
To order $\epsilon^2$, we have:
\begin{equation}
m_b = \frac{m_\Upsilon}{2}\Big(1 + \epsilon\, A + \epsilon^2\left(A^2 + AL\right)\Big)
\label{Eq:mbMeson}
\end{equation}
with
\begin{equation}
A \equiv \frac{\alpha_s^2 C_F^2}{8}\,, \qquad L \equiv \frac{\alpha_s}{\pi}\left[\left(\ln\frac{2\mu}{\alpha_s C_F m_{\Upsilon}} + \frac{11}{6}\right)\beta_0 - 4\right]\,.
\end{equation}

In principle, an analogous relation could be used to relate the charm-quark mass to the mass of the 
$J/\psi$ quarkonium; however, nonperturbative effects are large and not well controlled \cite{Pineda:1997hz,Hoang:1998ng}. However, the charm-quark mass can be obtained by using the HQET expression
~\cite{Falk:1992wt, Falk:1992ws},
\begin{equation}
m_b - m_c = \overline{m}_B - \overline{m}_D + \frac{\lambda_1}{2}\left(\frac{1}{m_b} - \frac{1}{m_c}\right) + \mathcal{O}(1/m_Q^2)\,,
\label{Eq:HQETmassdiff}
\end{equation}
given in terms of the spin-averaged meson masses $\overline{m}_B = 1/4(3m_{B^*}+m_B)$ and $\overline{m}_D = 1/4(3m_{D^*}+m_D)$, with the HQET kinetic parameter $\lambda_1 = - 0.362 \pm 0.067$ GeV$^2$ taken from HFLAV \cite{HFLAV:2019otj}.

One can then expand the charm-quark mass to order $\epsilon^2$ as
\begin{equation}
m_c = m_c^{(0)} + \epsilon\,\delta m_c^{(1)} + \epsilon^2\,\delta m_c^{(2)} + \mathcal{O}(\epsilon^3)\,,
\label{Eq:mcExpansion}
\end{equation}
and determine the coefficients by substituting Eqs.~\eqref{Eq:mbMeson} and \eqref{Eq:mcExpansion} into Eq.~\eqref{Eq:HQETmassdiff} and matching order by order in $\epsilon$, to linear order in $\lambda_1$. We obtain:
\begin{align}
m_c^{(0)} &= m_{\Upsilon}/2 - \bar{m}_B+\bar{m}_D - \frac{\lambda_1}{2}\left(\frac{2}{m_{\Upsilon}} - \frac{2}{m_{\Upsilon} - 2\bar{m}_B+2\bar{m}_D}\right)\,,\label{Eq:mc0}\\[2mm]
\delta m_c^{(1)} &= \frac{m_\Upsilon}{2}A\,\bigg[1 + 2\lambda_1\left(\frac{1}{m_{\Upsilon}^2} - \frac{1}{(m_\Upsilon-2\bar{m}_B+2\bar{m}_D)^2}\right)\bigg]\,,\label{Eq:mc1}\\[2mm]
\delta m_c^{(2)} &= \frac{m_{\Upsilon}}{2}(A^2+AL)\left[1 + 2\lambda_1\left(\frac{1}{m_{\Upsilon}^2} - \frac{1}{(m_\Upsilon-2\bar{m}_B+2\bar{m}_D)^2}\right)\right]\\
&-\lambda_1m_{\Upsilon}^2A^2\left[\frac{1}{m_{\Upsilon}^3} - \frac{1}{(m_\Upsilon-2\bar{m}_B+2\bar{m}_D)^3}\right]\,.\label{Eq:mc2}
\end{align}
The zeroth-order and first-order results reproduce Eq.~(12) of Ref.~\cite{Aebischer:2021ilm}. 

In the numerical analysis, the pole masses $m_b$ and $m_c$ entering the $\mathcal{B}_{bc}$ and $B_c$ decay widths are replaced by the above expressions in terms of meson masses. Each contribution is then expanded to the order in $\epsilon$ that matches the perturbative accuracy of its Wilson coefficient, in particular, up to $\mathcal{O}(\epsilon^2)$ for the dimension-three term, matching the perturbative accuracy of the NNLO corrections to the dimension-three Wilson coefficient.

The numerical values of the masses and corresponding renormalization-scale inputs used in the mass schemes are summarized in Table~\ref{tab:massSchemes}.

\section{Matrix elements for doubly heavy baryons}
\label{Sec:MEs}

In this section we determine the nonperturbative matrix elements entering the lifetime predictions for doubly heavy baryons $\mathcal{B}_{Q Q'}$. Our analysis is based on a nonrelativistic diquark picture, in which the heavy-quark pair $(Q_1Q_2)$ is treated as a compact color-antitriplet diquark $\mathcal{D}$\footnote{Several approaches in the literature go beyond this simplified approximation by resolving the finite spatial extent of the heavy diquark~\cite{Soto:2020pfa,Soto:2021cgk,Eichten:2025toq}.}. We express all the matrix elements in terms of the suitable nonrelativistic constituent wave functions. We interpret the resulting values as defined at a low hadronic scale $\mu_h=1\,\mathrm{GeV}$, and evolve them via renormalization-group running to high scales $\mu_b=4.5\,\mathrm{GeV}$ for $b$-quark matrix elements and $\mu_c=1.5\,\mathrm{GeV}$ for $c$-quark matrix elements, unless stated otherwise.

\subsection{Matrix elements of four-quark operators}\label{sec:fourquark}

We begin with the matrix elements of the dimension-six and dimension-seven spectator operators. In the diquark picture, the heavy--heavy contributions are expressed through the $Q_1Q_2$ wave functions at the origin taken from the potential-model analysis of Ref.~\cite{Kiselev:2001fw}, while the diquark-light contributions are determined from hyperfine splittings in the De Rujula--Georgi--Glashow constituent model~\cite{DeRujula:1975qlm}.
\subsubsection{Dimension-six operators}
We work with the following basis of dimension-six four-quark operators 
\begin{equation}
    \begin{aligned}
        O_{1}^{Qq}      &= (\overline{Q}^i\gamma_\mu(1-\gamma^5)q^i)(\overline{q}^j\gamma^\mu(1-\gamma^5)Q^j), & 
        O_2^{Qq}        &= (\overline{Q}^i(1-\gamma^5)q^i)(\overline{q}^j(1+\gamma^5)Q^j), \\
        \tilde{O}_1^{Qq}&= (\overline{Q}^i\gamma_\mu(1-\gamma^5)q^j)(\overline{q}^j\gamma^\mu(1-\gamma^5)Q^i), & 
        \tilde{O}_2^{Qq}&= (\overline{Q}^i(1-\gamma^5)q^j)(\overline{q}^j(1+\gamma^5)Q^i)\,,
        \label{Eq:4qDef}
    \end{aligned}
\end{equation}
where $Q$ and $q$ are defined as the quark fields in QCD. 
For the doubly-bottom and doubly-charmed baryons, the relevant dimension-six spectator terms involve only the heavy-light operators $O_i^{bq},\tilde O_i^{bq}$ and $O_i^{cq},\tilde O_i^{cq}$, respectively. In contrast, for the $\mathcal{B}_{bc}$ baryons one also encounters heavy-heavy contributions induced by the operators $O_i^{bc}$ and $\tilde O_i^{bc}$.

In what follows, we use the abbreviated notation for the matrix elements of four-quark operators:
\begin{equation}
    \Opsix{i}{Qq}_\Bary\equiv\frac{\langle \Bary| O_{i}^{Qq}| \Bary\rangle}{2M_{\Bary}}, \label{Eq:Def}
\end{equation}
with the appropriate replacements for the color-rearranged operators $\Opsixt{i}{Qq}$. 

For the doubly bottom baryons $\Barybb$, the matrix elements are expressed in terms of the heavy-light wave functions at the origin as
\begin{equation}
\begin{aligned}
\Opsix{1}{bq}_{\Barybb}  &= - 6\,|\Psi_{bq}^{\Barybb}(0)|^2 \,, \qquad 
\Opsix{2}{bq}_{\Barybb}  = - \,|\Psi_{bq}^{\Barybb}(0)|^2 \,, \\
\Opsixt{1}{bq}_{\Barybb}  &=  6\tilde{B}\,|\Psi_{bq}^{\Barybb}(0)|^2 \,, \qquad 
\Opsixt{2}{bq}_{\Barybb} =  \tilde{B}\,|\Psi_{bq}^{\Barybb}(0)|^2 \,,
\end{aligned}
\label{eq:dimsixBbb-bq}
\end{equation}
where $q$ denotes the corresponding light valence quark. The corresponding matrix elements for doubly charmed baryons are obtained by replacing $b$ with $c$. The same replacement rule applies to all heavy-light matrix elements discussed below. Furthermore, we note that in the Nonrelativistic Constituent Quark Model (NRCQM), the matrix elements of the color-rearranged operators $\tilde{O}_{1,2}^{bq}$ differ by a relative minus sign from those of $O_{1,2}^{bq}$. Possible deviations from this relation are parametrized by $\tilde B$, which we set to unity at the low scale $\mu_h$. 

For the $\Barybc$ baryons containing a scalar diquark ($S_{\mathcal{D}}=0$), the matrix elements are given by
\begin{equation}
\begin{aligned}
\Opsix{1}{bq}_{\Barybc}  &= - \,|\Psi_{bq}^{\Barybc}(0)|^2 \,, \qquad 
\Opsix{2}{bq}_{\Barybc} = \frac{1}{2} \,|\Psi_{bq}^{\Barybc}(0)|^2 \,, \\
\Opsixt{1}{bq}_{\Barybc}  &=  \tilde{B}\,|\Psi_{bq}^{\Barybc}(0)|^2 \,, \qquad 
\Opsixt{2}{bq}_{\Barybc} =  -\frac{1}{2}\tilde{B}\,|\Psi_{bq}^{\Barybc}(0)|^2 \,,
\end{aligned}\label{eq:dimsixBbc-Qq}
\end{equation}
with the analogous relations for the matrix elements $\Opsix{i}{cq}_\Barybc$ with the appropriate replacements of the $b$ quark with the $c$ quark.
In addition to heavy-light spectator interactions, the $\Barybc$ baryons also receive heavy-heavy spectator contributions involving both the $b$ and the $c$ quark, encoded in the matrix elements:
\begin{equation}
\begin{aligned}
\Opsix{1}{bc}_{\Barybc} &= - 4\,|\Psi_{bc}^{\Barybc}(0)|^2 \,, \qquad 
\Opsix{2}{bc}_{\Barybc}  = - \,|\Psi_{bc}^{\Barybc}(0)|^2 \,, \\
\Opsixt{1}{bc}_{\Barybc}  &=  4\tilde{B}\,|\Psi_{bc}^{\Barybc}(0)|^2 \,, \qquad 
\Opsixt{2}{bc}_{\Barybc}  =  \tilde{B}\,|\Psi_{bc}^{\Barybc}(0)|^2 \,.
\end{aligned}\label{eq:dimsixBbc-bc}
\end{equation}

For the primed baryons $\Barybcprime$, which contain a vector diquark ($S_\mathcal{D}=1$), the spin structure of the matrix elements differs from that of the unprimed $\Barybc$ states. The heavy-light matrix elements are given by: 
\begin{equation}
\begin{aligned}
\Opsix{1}{bq}_{\Barybcprime}  &= - 3\,|\Psi_{bq}^{\Barybcprime}(0)|^2 \,, \qquad 
\Opsix{2}{bq}_{\Barybcprime}  = - \frac{1}{2}\,|\Psi_{bq}^{\Barybcprime}(0)|^2 \,, \\
\Opsixt{1}{bq}_{\Barybcprime}  &=  3\tilde{B}\,|\Psi_{bq}^{\Barybcprime}(0)|^2 \,, \qquad 
\Opsixt{2}{bq}_{\Barybcprime}  =  \frac{1}{2}\tilde{B}\,|\Psi_{bq}^{\Barybcprime}(0)|^2 \,.
\end{aligned}\label{eq:dimsixBbcprime-Qq}
\end{equation}
The remaining heavy-heavy spectator matrix elements involving the $bc$ pair are
\begin{equation}
\label{eq:dimsixBbcprime-bc}
\begin{aligned}
\Opsix{1}{bc}_{\Barybcprime}  &= 0\,, \qquad 
\Opsix{2}{bc}_{\Barybcprime}  =  \,|\Psi_{bc}^{\Barybcprime}(0)|^2 \,, \\
\Opsixt{1}{bc}_{\Barybcprime}  &=  0 \,, \qquad 
\Opsixt{2}{bc}_{\Barybcprime}  =  -\tilde{B}\,|\Psi_{bc}^{\Barybcprime}(0)|^2 \,.
\end{aligned}
\end{equation}
We note that the vanishing of $\Opsix{1}{bc}_{\Barybcprime}$ and $\Opsixt{1}{bc}_{\Barybcprime}$ in Eq.~\eqref{eq:dimsixBbcprime-bc} holds at leading order in the nonrelativistic expansion at which we are working.

Within the diquark picture, we identify the heavy-light wave functions with a common diquark-light quark wave function,
\begin{equation}
    |\Psi_{Qq}^\Bary(0)|^2=|\Psi_{Q'q}^\Bary(0)|^2\equiv|\Psi^\Bary_{\mathcal{D}-q}(0)|^2,\label{Eq:diquarkpicturewf}
\end{equation}
where $Q_1$ and $Q_2$ denote the heavy quarks bound in the diquark $\mathcal{D}$.

We now turn to the extraction of the diquark-light quark wave function $\wavefnsq{\mathcal{D}-q}{\Bary}$. We employ the De Rujula-Georgi-Glashow NRCQM \cite{DeRujula:1975qlm} in which the hyperfine mass splittings are induced by the spin-spin interactions with the corresponding spin-dependent Hamiltonians given for baryons as
\begin{equation}
\begin{aligned}
    \mathcal{H}_{\mathrm{spin, baryons}}=\sum_{i>j}\frac{16\pi\alpha_s}{9}\frac{\vec S_i\cdot  \vec S_j}{m_i^\Bary m_j^\Bary}\delta^3(\vec r_{ij})\,,
    \label{Eq:Hspinbary}
\end{aligned}
\end{equation}
and for mesons as
\begin{equation}
    \mathcal{H}_{\mathrm{spin, mesons}}=\frac{32\pi\alpha_s}{9}\frac{\vec S_1 \cdot \vec S_2}{m_1^\Mes m_2^\Mes}\delta^3(\vec r_{12})\,,
\end{equation} 
where $m_i^{\Bary (M)}$ indicate baryon (meson) constituent masses.
The mass of the baryon can then be expressed as:
\begin{equation}
    M_\Bary=\sum_i m_i^\Bary+\frac{16\pi\alpha_s}{9}\left( \frac{\langle\vec S_1 \cdot \vec S_2\rangle_\Bary}{m_1^\Bary m_2^\Bary}\wavefnsq{12}{\Bary}+\frac{\langle\vec S_2\cdot \vec S_3\rangle_\Bary}{m_2^\Bary m_3^\Bary}\wavefnsq{23}{\Bary}+\frac{\langle\vec S_1\cdot  \vec S_3\rangle_\Bary}{m_1^\Bary m_3^\Bary}\wavefnsq{13}{\Bary} \right)+\ldots\,,\label{eq:RGGBaryons}
\end{equation}
while for the mesons we have 
\begin{equation}
    M_\Mes=\sum_i m_i^\Mes+\frac{32\pi\alpha_s}{9} \frac{\langle\vec S_1 \cdot \vec S_2\rangle_\Mes}{m_1^M m_2^M}\wavefnsq{12}{\Mes}+\ldots\,,\label{eq:RGGMesons}
\end{equation}
where the ellipses represent the remaining spin-independent contributions.
Using the diquark picture relations \eqref{Eq:diquarkpicturewf} and evaluating the matrix elements of the spin-dependent terms in Eq.~\eqref{eq:RGGBaryons}, one can express the hyperfine splittings between spin-$3/2$ and spin-$1/2$ partners, both with diquark spin $S_\mathcal{D}=1$, in terms of the diquark-quark wave functions as: 
\begin{align}
    M_{\Barybbstar}-M_{\Barybb}&=\frac{8\pi\alpha_s}{3}\frac{1}{m_b^\Bary m_q^\Bary}\wavefnsq{\mathcal{D}-q}{\Barybb},\label{Eq:MassSplitbb}
    \\
    M_{\Barybcstar}-M_{\Barybcprime}&=\frac{4\pi\alpha_s}{3}\frac{m_b^\Bary+m_c^\Bary}{m_b^\Bary m_c^\Bary m_q^\Bary}\wavefnsq{\mathcal{D}-q}{\Barybcprime},\label{Eq:MassSplitbc}
\end{align}
where the terms involving the spin interactions between two heavy quarks in the diquark are equal for both hyperfine partners and cancel in the mass differences, and $q$ denotes the corresponding valence light quark in the baryon. We note that the analogous wave functions for the $\Barybc$ cannot be extracted in a similar manner, since these baryons do not possess hyperfine partners. However, we assume they are given by the wave functions of corresponding $\Barybcprime$ baryons.

Following the standard procedure, we express the diquark-light quark baryon wave functions in terms of well-known meson wave functions by normalising  expressions \eqref{Eq:MassSplitbb} and \eqref{Eq:MassSplitbc} to the corresponding meson mass difference relations
\begin{equation}
\begin{aligned}
    M_{B^*_q}-M_{B_q}&=\frac{32\pi\alpha_s}{9}\frac{1}{m_b^\Mes m_q^\Mes}\wavefnsq{bq}{B_q}\,.
\end{aligned}
\end{equation}
This results in
\begin{equation}
    \wavefnsq{\mathcal{D}-q}{\Barybb}=\frac{4}{3}r_{bq}\frac{M_\Barybbstar-M_\Barybb}{M_{B^*_q}-M_{B_q}}\wavefnsq{bq}{B_q},\label{Eq:PsiDqnonrel1}
\end{equation}
and 
\begin{equation}
    \wavefnsq{\mathcal{D}-q}{\Barybcprime}=\frac{8}{3}r_{bq}\frac{m_c^\Bary}{m_b^\Bary+m_c^\Bary}\frac{M_{\Barybcstar}-M_{\Barybcprime}}{M_{B^*_q}-M_{B_q}}\wavefnsq{bq}{B_q},\label{Eq:PsiDqnonrel2}
\end{equation}
where we use the diquark picture relation
\begin{equation}
    \wavefnsq{\mathcal{D}-q}{\Barybc}=\wavefnsq{\mathcal{D}-q}{\Barybcprime}\label{Eq:PsiDqnonrel3}.
\end{equation}
In the above expressions, the difference between the values of the constituent quark masses in mesons and baryons is encoded in the ratios
\begin{equation}
r_{Qq}=\frac{m_Q^\Bary m_q^\Bary}{m_Q^\Mes m_q^\Mes}\,.
\end{equation}
The numerical values of the constituent quark masses used in our analysis are taken from Ref.~\cite{Karliner:2014gca} and are listed in Table~\ref{tab:constituentMasses}.

Finally, the nonrelativistic meson wave functions are related to the corresponding meson masses and decay constants through
\begin{equation}
    \wavefnsq{bq}{B_q}=\frac{1}{12}f_{B_q}^2 M_{B_q}\,,
\end{equation}
which allows one to express the diquark-light quark baryon wave functions in terms of known quantities.

\begin{table}[h]
\small
\centering
\renewcommand{\arraystretch}{1.15}
\begin{tabular}{lc}
\hline\hline
Mass splitting & $\Delta M$ [MeV] \\
\hline
$\Xi_{bb}^{\ast}-\Xi_{bb}$         & $34.6(2.5)(7.4)$ \\
$\Omega_{bb}^{\ast}-\Omega_{bb}$   & $35.7(1.3)(5.5)$ \\
\hline
$\Xi_{cc}^{\ast}-\Xi_{cc}$         & $82.8(7.2)(5.8)$ \\
$\Omega_{cc}^{\ast}-\Omega_{cc}$   & $83.8(1.4)(5.3)$ \\
\hline
$\Xi_{cb}^{\ast}-\Xi_{cb}^{\prime}$       & $26.7(3.3)(8.4)$ \\
$\Omega_{cb}^{\ast}-\Omega_{cb}^{\prime}$ & $27.4(2.0)(6.7)$ \\
\hline\hline
$D^{\ast+}-D^+$& $140.603(15)$\\
\hline 
$D^{\ast 0}-D^0$& $142.016(30)$\\
\hline 
$D_s^\ast-D_s$& $143.8(4)$\\
\hline 
$B^{\ast}-B$& $45.18(20)$\\
\hline
$B_s^{\ast}-B_s$ &$48.5(14)$\\
\hline \hline
\end{tabular}
\caption{The values of the hadronic hyperfine mass splittings used in this work. The baryon mass splittings are taken from the lattice-QCD analysis of Ref.~\cite{Brown:2014ena}. The two quoted uncertainties for the baryon splittings are combined in quadrature for the purposes of our analysis. The meson mass splittings are taken from the PDG averages~\cite{ParticleDataGroup:2024cfk}.}
\label{tab:lattice-hyperfine-splittings}
\end{table}

Using the mass splittings listed in Table~\ref{tab:lattice-hyperfine-splittings} together with the corresponding meson decay constants given in Table~\ref{tab:mes_masses} of Appendix~\ref{app:inputs}, we extract the diquark-light wave functions and obtain the resulting four-quark matrix elements shown in Table~\ref{tab:fourquarkME}. The remaining four-quark matrix elements depend on the $bc$ wave functions, which cannot be determined from the hyperfine splittings. For these, we adopt the corresponding values from the analysis of Ref.~\cite{Kiselev:2001fw}, listed in Table~\ref{Tab:KiselevValues}.

\begin{table}[htbp]
\small
\centering
\renewcommand{\arraystretch}{1.3}
\begin{tabular}{lc}
\hline\hline
Matrix element & Value [GeV$^3$] \\
\hline
$\langle O_{1}^{bu}\rangle_{\Xi_{bb}^{0}} = \langle O_{1}^{bd}\rangle_{\Xi_{bb}^{-}}$ & $-0.120(28)$ \\
$\langle O_{1}^{bs}\rangle_{\Omega_{bb}^{-}}$                                          & $-0.158(32)$ \\[0.2em]
\hline
$\langle O_{1}^{cu}\rangle_{\Xi_{cc}^{++}} = \langle O_{1}^{cd}\rangle_{\Xi_{cc}^{+}}$ & $-0.040(8)\phantom{0}$ \\
$\langle O_{1}^{cs}\rangle_{\Omega_{cc}^{+}}$                                          & $-0.055(11)$ \\[0.2em]
\hline
$\langle O_{1}^{bu}\rangle_{\Xi_{bc}^{+}} = \langle O_{1}^{bd}\rangle_{\Xi_{bc}^{0}}$  & $-0.0077(26)$ \\
$\langle O_{1}^{bs}\rangle_{\Omega_{bc}^{0}}$                                          & $-0.0101(26)$ \\
$\langle O_{1}^{bc}\rangle_{\Xi_{bc}^{+}} = \langle O_{1}^{bc}\rangle_{\Xi_{bc}^{0}}$  & $-0.17(3)\phantom{00}$ \\
$\langle O_{1}^{bc}\rangle_{\Omega_{bc}^{0}}$                                          & $-0.17(3)\phantom{00}$ \\[0.2em]
\hline
$\langle O_{1}^{bu}\rangle_{\Xi_{bc}^{\prime\, +}} = \langle O_{1}^{bd}\rangle_{\Xi_{bc}^{\prime\, 0}}$ & $-0.023(8)\phantom{0}$ \\
$\langle O_{1}^{bs}\rangle_{\Omega_{bc}^{\prime\, 0}}$                                                & $-0.030(8)\phantom{0}$ \\
$\langle O_{2}^{bc}\rangle_{\Xi_{bc}^{\prime\, +}} = \langle O_{2}^{bc}\rangle_{\Xi_{bc}^{\prime\, 0}}$ & $\phantom{-}0.041(8)\phantom{0}$ \\
$\langle O_{2}^{bc}\rangle_{\Omega_{bc}^{\prime\, 0}}$                                                & $\phantom{-}0.041(8)\phantom{0}$ \\
\hline\hline
\end{tabular}
\caption{Numerical values of the four-quark matrix elements at the 
low hadronic scale $\mu_h= 1\,\mathrm{GeV}$. For $\Xi'_{bc}$ and 
$\Omega'_{bc}$ we quote the matrix element of $O_2$, since the matrix 
element of the operator $O_1$ vanishes in the nonrelativistic limit. 
Within the diquark picture employed here, the heavy-light quark matrix 
elements  for the $bc$ baryons 
$\Barybc$ and $\Barybcprime$ satisfy $\Opsix{i}{bq}=\Opsix{i}{cq}$. Note that we take the uncertainties at the order of $20\%$.}
\label{tab:fourquarkME}
\end{table}
\vfill
\subsubsection{Dimension-seven operators}
For the dimension-seven four-quark contributions, we work with the following operator basis:
\begin{align}
& P_{1}^{Qq}=m_q\left(\bar{Q}_i\left(1-\gamma_5\right) q_i\right)\left(\bar{q}_j\left(1-\gamma_5\right) Q_j\right),
\\
& P_2^{Qq}=\frac{1}{m_Q}\left(\bar{Q}_i \overleftarrow{D}_\rho \gamma_\mu\left(1-\gamma_5\right) D^\rho q_i\right)\left(\bar{q}_j \gamma^\mu\left(1-\gamma_5\right) Q_j\right),
\\
& P_3^{Qq}=\frac{1}{m_Q}\left(\bar{Q}_i \overleftarrow{D}_\rho\left(1-\gamma_5\right) D^\rho q_i\right)\left(\bar{q}_j\left(1+\gamma_5\right) Q_j\right),
\\
& \tilde{P}_1^{Qq}=m_q\left(\bar{Q}_i\left(1-\gamma_5\right) q_j\right)\left(\bar{q}_j\left(1-\gamma_5\right) Q_i\right),
\\
& \tilde{P}_2^{Qq}=\frac{1}{m_Q}\left(\bar{Q}_i \overleftarrow{D}_\rho \gamma_\mu\left(1-\gamma_5\right) D^\rho q_j\right)\left(\bar{q}_j \gamma^\mu\left(1-\gamma_5\right) Q_i\right),
\\
& \tilde{P}_3^{Qq}=\frac{1}{m_Q}\left(\bar{Q}_i \overleftarrow{D}_\rho\left(1-\gamma_5\right) D^\rho q_j\right)\left(\bar{q}_j\left(1+\gamma_5\right) Q_i\right)\,,\label{Eq:dim7Ops}
\end{align}
where, as for the case of dimension-six four-quark operators, we take the quarks as full QCD fields.

For the $\Barybb$ and $\Barycc$ baryons, spectator effects involve interactions between a heavy quark and the light valence quark, and we retain these contributions up to and including dimension seven.
For the $\Barybc$ baryons, on the other hand, three classes of four-quark spectator contributions arise, coming from the charm-light quark, the bottom-light quark, and the heavy-heavy $bc$ spectator interactions. Of these, we retain dimension-seven corrections only for the charm-light quark contributions. For the bottom-light quark spectator contributions, the dimension-seven corrections are suppressed by $\Lambda_{\rm QCD}/m_b$ relative to dimension six. Since $\Lambda_{\rm QCD}/m_b \sim (\Lambda_{\rm QCD}/m_c)^2$, these corrections are expected to be of the order of the neglected dimension-eight charm-light quark contributions and therefore within the theoretical uncertainties of the present analysis.
Furthermore, the four-quark operators involving $b$ and $c$ fields are treated at leading order in the nonrelativistic expansion, with the heavy fields understood as two-component Pauli spinors. In the NRQCD velocity counting, the leading four-quark contribution enters at order $v^6$, while the next parity-allowed corrections appear at order $v^8$, where $v$ denotes the relative velocity of the heavy quarks inside the diquark. Given that $\langle v^2 \rangle$ is small for the $bc$ system (see Table~\ref{tab:v2Values}), it is sufficient for our purposes to work with the leading-order heavy-heavy contribution.

Regarding the matrix elements of the dimension-seven heavy-light operators, we follow Ref.~\cite{Gratrex:2022xpm} and relate them to the matrix elements of the dimension-six operators using simple scaling relations\footnote{Note that a somewhat different parametrization of the dimension-seven matrix elements was employed in Refs.~\cite{Cheng:2018mwu,Cheng:2019sxr}. In particular, in their scheme, the matrix elements of the $P_1$
operator are larger; however, this choice has little impact on the final predictions, as this operator enters with a small Wilson coefficient.}:
\begin{equation}
    \OpsevenP{1}{Qq}_\Bary=m_q^\mathcal{B}\Opsix{2}{Qq}_\Bary,\quad \OpsevenP{2}{Qq}_\Bary=\Lambda_{\mathrm{QCD}}\Opsix{1}{Qq}_\Bary,\quad \OpsevenP{3}{Qq}_\Bary=\Lambda_{\mathrm{QCD}}\Opsix{2}{Qq}_\Bary,
\end{equation}
where $\OpsevenP{i}{Qq}_\Bary$ is normalized in the same way as the matrix elements of the dimension-six operators in
Eq.~\eqref{Eq:Def}. For the central value we take $\Lambda_{\rm QCD}=0.33~\mathrm{GeV}$~\cite{Herren:2017osy}.
The matrix elements of the color-rearranged operators are related to those of the corresponding operators $P_i^{Qq}$ as
\begin{equation}
\OpsevenPt{i}{Qq}_\Bary = -\,\tilde{B}\,\OpsevenP{i}{Qq}_\Bary\,,
\end{equation}
and we set $\tilde{B}=1$.

\subsection{Matrix elements of two-quark operators}
\label{Sec:ME2quark}

The NRQCD expansion of $\bar{Q}Q$ in terms of the non-relativistic Pauli spinor field $\Psi_Q$ is obtained by implementing the standard Foldy--Wouthuysen transformation for a heavy-quark field $Q$, and reads~\cite{Beneke:1996xe,Lepage:1992tx,Brambilla:2005yk}:
\begin{align}
\bar{Q}Q
&=
\Psi_Q^\dagger \Psi_Q
-\frac{1}{2m_Q^2}\Psi_Q^\dagger (i\vec{D})^2 \Psi_Q
+\frac{3}{8m_Q^4}\Psi_Q^\dagger (i\vec{D})^4 \Psi_Q
\nonumber\\
&\quad
-\frac{1}{2m_Q^2}\Psi_Q^\dagger\,g_s\,\vec{\sigma}\cdot\vec{B}\,\Psi_Q
-\frac{1}{4m_Q^3}\Psi_Q^\dagger\,g_s\,(\vec{\mathcal{D}}\cdot\vec{E})\,\Psi_Q
+\dots \,,
\label{eq:QQNRQCDexp}
\end{align}
up to ${\cal O}(v_Q^7)$ according to the NRQCD velocity power counting~\cite{Lepage:1992tx}:
\begin{equation}
\Psi_Q \sim (m_Q v_Q)^{3/2},\qquad 
\vec{D}\sim m_Q v_Q,\qquad 
g_s\vec{B}\sim m_Q^2 v_Q^4,\qquad 
g_s\vec{E}\sim m_Q^2 v_Q^3\,.
\end{equation}

Similarly, for the  corresponding expansion of the chromomagnetic operator we obtain:
\begin{align}
\bar{Q}\,\frac{1}{2}\, g_s\sigma_{\mu\nu}G^{\mu\nu}Q
&=
-\Psi_Q^\dagger\, g_s\,\vec{\sigma}\,\cdot\,\vec{B}\,\Psi_Q
-\frac{1}{2m_Q}\Psi_Q^\dagger\, g_s\,(\vec{\mathcal{D}}\,\cdot\,\vec{E})\,\Psi_Q
+\dots \,.
\label{eq:chromoNR}
\end{align}
We omit the spin–orbit operator whose matrix elements vanish in the non-relativistic limit for all 
S-wave ground-state baryons considered here.

\subsubsection{\texorpdfstring{Darwin $\hat{\rho}^3_{D}$ terms}{Darwin rhoD3 terms}}
\label{Sec:Darwin}
We begin our discussion of the matrix elements of the two-quark operators with the matrix element of the Darwin operator, which enters the HQE at dimension six. This choice is convenient because the same nonrelativistic Darwin parameter also appears as a higher-order correction in the expansion of the chromomagnetic operator in Eq.~\eqref{eq:chromoNR}, so the results derived here will be used again in the determination of $\hat{\mu}_{G,Q}^2$ below.

Within the diquark picture, the Darwin matrix element receives two separate contributions, one from the diquark-light interaction and one from the interaction of the two heavy quarks inside the diquark:
\begin{equation}
\hat{\rho}_{D,Q}^3(\Bary)=\hat{\rho}_{D,Q}^3(\Bary)|_{\diq}+\hat{\rho}_{D,Q}^3(\Bary)|_{Q-Q'}.
\end{equation}
The diquark-light quark contribution can be obtained by relating the Darwin operator to four-quark operators using the equation of motion for the gluon field \cite{Bigi:1993ex} resulting with
\begin{equation}
\hat{\rho}_{D,Q}^3(\mathcal{B})\vert_{\diq}=
   g_s^2 \left(-\frac{1}{8}\Opsix{1}{Qq}+\frac{1}{24}\Opsixt{1}{Qq}+\frac{1}{4}\Opsix{2}{Qq}-\frac{1}{12}\Opsixt{2}{Qq}\right)\,+\mathcal{O}(1/m_Q)\,.\label{eq:Darwin4qbar}
\end{equation}
For the remaining $\hat{\rho}_D^3|_{Q-Q'}$ piece, arising from the interactions between two heavy quarks within the heavy diquark, the leading nonrelativistic term reads
\begin{equation}
    \rho_{D,Q}^3(\Bary)|_{Q-Q'}=\left. \frac{1}{2}\frac{\langle\Bary|\Psi_Q^\dagger\, g_s\,(\vec{\mathcal{D}}\,\cdot\,\vec{E})\,\Psi_Q |\Bary\rangle}{2M_\Bary}\right|_{Q-Q'},
\end{equation}
and in the NRCQM we obtain
\begin{equation}\label{eq:darwinQQnonrel}
    \rho_{D,Q}^3(\Bary)|_{Q-Q'}=-\frac{g_s^2}{2}n_Q(\Bary)\langle t^a_Q t^a_{Q'}\rangle_\Bary |\Psi_{QQ'}^\Bary(0)|^2\,.
\end{equation}

In particular, for the doubly bottom baryons we obtain 
\begin{equation}
    \rho_{D,b}^3(\Barybb)|_{b-b}=\frac{2}{3}g_s^2\wavefnsq{bb}{\Barybb},
\end{equation}
with the analogous expression for the doubly charmed baryons with the straightforward replacement of labels. For the $\Barybc$ and $\Barybcprime$ baryons we have
\begin{equation}
    \rho_{D,b}^3(\Barybc)|_{b-c}=\rho_{D,c}^3(\Barybc)|_{b-c}=\frac{1}{3}g_s^2\wavefnsq{bc}{\Barybc}\,,\label{Eq:rhoDbc1}
\end{equation}
and
\begin{equation}
\hat{\rho}_{D,b}^3(\Barybcprime)|_{b-c}=\rho_{D,c}^3(\Barybcprime)|_{b-c}=\frac{1}{3}g_s^2\wavefnsq{bc}{\Barybcprime}\,,\label{Eq:rhoDbc2}
\end{equation}
respectively.

\begin{table}[t]
\small
\centering
\renewcommand{\arraystretch}{1.15}
\setlength{\tabcolsep}{10pt}
\begin{tabular}{l c c}
\hline\hline
Baryon $\mathcal{B}$ & $Q$ &
$\hat{\rho}_{D,Q}^3(\mathcal{B})\;[\mathrm{GeV}^3]$ \\
\hline
$\Xi_{bb}$      & $b$ & $0.36(7)$ \\
$\Omega_{bb}$   & $b$ & $0.38(8)$ \\
\hline
$\Xi_{cc}$      & $c$ & $0.084(17)$ \\
$\Omega_{cc}$   & $c$ & $0.093(12)$ \\
\hline
$\Xi_{bc}$      & $b$ & $0.054(11)$ \\
$\Omega_{bc}$   & $b$ & $0.057(11)$ \\
$\Xi_{bc}$      & $c$ & $0.076(15)$ \\
$\Omega_{bc}$   & $c$ & $0.080(16)$ \\
\hline
$\Xi'_{bc}$     & $b$ & $0.054(11)$ \\
$\Omega'_{bc}$  & $b$ & $0.057(11)$ \\
$\Xi'_{bc}$     & $c$ & $0.076(15)$ \\
$\Omega'_{bc}$  & $c$ & $0.080(16)$ \\
\hline\hline
\end{tabular}
\caption{Numerical values of $\hat{\rho}_{D,Q}^3(\mathcal{B})$ in units of GeV$^3$ evaluated at the scale $\mu_b$ for $\hat{\rho}_{D,b}^3(\mathcal{B})$ and at the scale $\mu_c$ for $\hat{\rho}_{D,c}^3(\mathcal{B})$.}
\label{tab:rhoDhat_values}
\end{table}

The quantities $\vert \Psi_{QQ'}^{\mathcal{B}}(0)\vert^2$ represent the squares of the corresponding nonrelativistic wave functions of heavy quarks at the origin of a $QQ'$ diquark system. The corresponding values are derived from heavy baryon spectroscopy using potential models~\cite{Bagan:1994dy,Kiselev:2001fw}. We adopt the results from Ref.~\cite{Kiselev:2001fw} as listed in Table~\ref{Tab:KiselevValues}.

We interpret the corresponding Darwin terms as defined at the low hadronic scale $\mu_h=1\,\mathrm{GeV}$ and evolve them using the anomalous dimension of the Darwin operator~\cite{Shifman:1984wx,Pirjol:1998ur}, taking matrix elements involving a $b$ quark to the scale $\mu_b$ and those involving a $c$ quark to the scale $\mu_c$.

\begin{table}[t]
\centering
\begin{tabular}{ccc}
\hline\hline
$|\Psi_{bb}^{\Barybb}(0)|^2$ &
$|\Psi_{cc}^{\Barycc}(0)|^2$ &
$|\Psi_{bc}^{\Barybc}(0)|^2$ \\
\hline
0.14 & 0.02 & 0.04 \\
\hline \hline
\end{tabular}
\caption{Values of the heavy-heavy wave functions at the origin, in units of GeV$^3$ taken from Ref.~\cite{Kiselev:2001fw}.}
\label{Tab:KiselevValues}
\end{table}

\subsubsection{\texorpdfstring{Chromomagnetic $\hat{\mu}_G^2$ terms}{Chromomagnetic muG2 terms}}
For the determination of the chromomagnetic matrix elements, we use the nonrelativistic expansion in Eq.~\eqref{eq:chromoNR}, which gives
\begin{equation}
\hat{\mu}_{G,Q}^2(\Bary)=
-\frac{\langle\Bary|\Psi_Q^\dagger \,g_s\,\vec{\sigma}\cdot\vec{B}\,\Psi_Q|\Bary\rangle}{2M_\Bary}
-\frac{1}{2m_Q}\frac{\langle\Bary|\Psi_Q^\dagger \,g_s\,(\vec{\mathcal{D}}\cdot\vec{E})\,\Psi_Q |\Bary\rangle}{2M_\Bary}
+\dots \,.
\label{eq:chromoNR2}
\end{equation}
The two terms on the right-hand side are identified with $\mu_{G,Q}^2$ and the nonrelativistic Darwin parameter $\rho_{D,Q}^3$, respectively:
\begin{equation}
\begin{aligned}
\mu_{G,Q}^2(\Bary)&=-\frac{\langle\Bary|\Psi_Q^\dagger\, g_s\,\vec{\sigma}\cdot\vec{B}\,\Psi_Q|\Bary\rangle}{2M_\Bary},
\\
\rho_{D,Q}^3(\Bary)&=\frac{1}{2}\frac{\langle\Bary|\Psi_Q^\dagger\, g_s\,(\vec{\mathcal{D}}\,\cdot\,\vec{E})\,\Psi_Q |\Bary\rangle}{2M_\Bary}.
\end{aligned}
\end{equation}
In the diquark picture, each of these terms receives two contributions, one from the diquark-light quark interaction and the other from the interaction of the two heavy quarks within the diquark. We therefore write
\begin{equation}\label{eq:muGhatDQdecompose}
\hat{\mu}_{G,Q}^2(\Bary)=\mu_{G,Q}^2(\Bary)|_{\diq}+\mu_{G,Q}^2(\Bary)|_{Q-Q'}-\frac{1}{m_Q}\rho_{D,Q}^3(\Bary)|_{\diq}-\frac{1}{m_Q}\rho_{D,Q}^3(\Bary)|_{Q-Q'}\,.
\end{equation}

We first extract the diquark-light quark contribution $\mu_{G,Q}^2(\Bary)|_{\diq}$ from the mass splitting between the $\mathcal{B}^\ast$ and $\mathcal{B}$ baryons for which we use the NRCQM expression
\begin{equation}
    \mu_{G,Q}^2(\Bary)|_{\mathcal{D}-q}= n_Q(\Bary) \frac{4}{3} \frac{g_s^2}{m^\Bary_{q}}  \, \langle t_Q^a t_{q}^a \rangle_\Bary \, \langle \vec{S}_Q \cdot \vec{S}_{q} \rangle_\Bary \,|\Psi_{\mathcal{D}-q}^\Bary(0) |^2\,.\label{eq:muGdiqNRCQM}
\end{equation}

For doubly bottom case we insert $\langle \vec{S}_b \cdot \vec{S}_{q} \rangle=-1/2$, and use the relation $|\Psi_{\mathcal{D}-q}^\Barybb(0) |=|\Psi_{bq}^\Barybb(0) |$ from Eq.~\eqref{Eq:MassSplitbb}, to obtain:
\begin{equation}
\left.\mu_{G,b}^2(\mathcal{B}_{bb})\right|_{\mathcal{D}\text{--}q}
=
\frac{4}{3}\,m^\Bary_b\left(M_{\mathcal{B}_{bbq}^{\ast}}-M_{\mathcal{B}_{bbq}}\right)\,.
\label{Eq: Diquarkbb-q}
\end{equation}
For the $\Barybc$ baryons, the diquark spin vanishes and therefore 
\begin{equation}
    \mu_{G,Q}^2(\Barybc)|_{\diq}=0\,.
\end{equation}

For the $\Barybcprime$ baryons, using $\langle \vec{S}_Q \cdot \vec{S}_{q} \rangle=-1/2$ and expressing the wave function through the mass splitting in Eq.~\eqref{Eq:MassSplitbc}, we obtain
\begin{equation}
\mu_{G,Q}^2(\Barybcprime)|_{\diq}=\frac{4}{3}\frac{m^\Bary_c m^\Bary_b}{m^\Bary_c + m^\Bary_b}\left( M_\Barybcqstar-M_\Barybcqprime\right)\,.
\end{equation}

We next consider the contribution to $\mu_G^2$ arising from the interaction of the two heavy quarks inside the diquark. This term is evaluated from the same expression, Eq.~\eqref{eq:muGdiqNRCQM}, after replacing the light quark by the heavy quark $Q'$. For the $\Barybb$ baryons, using $\langle \vec{S}_b \cdot \vec{S}_b \rangle = 1/4$, one finds
\begin{equation}
    \mu_{G,b}^2(\Barybb)|_{b-b}=-\frac{4}{9}\frac{g_s^2}{m^\Bary_b}\wavefnsq{bb}{\Barybb}\,.
\end{equation}
For the $\Barybc$ baryons, using $\langle \vec{S}_b \cdot \vec{S}_c \rangle = -3/4$, one obtains
\begin{equation}
    \mu_{G,b}^2(\Barybc)|_{b-c}=\frac{2}{3}\frac{g_s^2}{m^\Bary_c}\wavefnsq{bc}{\Barybc},\qquad
    \mu_{G,c}^2(\Barybc)|_{b-c}=\frac{2}{3}\frac{g_s^2}{m^\Bary_b}\wavefnsq{bc}{\Barybc}\,.
\end{equation}
Similarly, for the $\Barybcprime$ baryons, where $\langle \vec{S}_b \cdot \vec{S}_c \rangle = 1/4$, we have
\begin{equation}
    \mu_{G,b}^2(\Barybcprime)|_{b-c}=-\frac{2}{9}\frac{g_s^2}{m^\Bary_c}\wavefnsq{bc}{\Barybcprime},\qquad
    \mu_{G,c}^2(\Barybcprime)|_{b-c}=-\frac{2}{9}\frac{g_s^2}{m^\Bary_b}\wavefnsq{bc}{\Barybcprime}\,.
\end{equation}

It remains to collect the nonrelativistic Darwin corrections entering the chromomagnetic terms. The diquark-light quark Darwin contribution $\rho_{D,Q}^3|_{\diq}$ is obtained from the right-hand side of Eq.~\eqref{eq:darwinQQnonrel} by replacing $Q'$ with the light quark $q$ and using the diquark-picture relation
$|\Psi_{Q q}^\Bary(0)|^2=|\Psi_{Q' q}^\Bary(0)|^2\equiv|\Psi^\Bary_{\mathcal{D}-q}(0)|^2$.
For doubly bottom baryons this yields\footnote{Substituting in Eq.~\eqref{eq:Darwin4qbar} the nonrelativistic expressions for the four-quark matrix elements at the low hadronic scale $\mu_h$, where we assume $\tilde{B}=1$, in terms of the corresponding diquark-light quark wave functions reproduces the results for the nonrelativistic Darwin term in Eqs.~\eqref{eq:darwinBbb-Dq}--\eqref{eq:darwinBbcprime-Dq}.}
\begin{gather}
    \rho_{D,b}^3(\Barybb)|_{\diq}=\frac{2}{3}g_s^2\wavefnsq{\diq}{\Barybb}\,,
    \label{eq:darwinBbb-Dq}
\end{gather}
while for $\Barybc$ and $\Barybcprime$ we obtain
\begin{gather}
\rho_{D,b}^3(\Barybc)|_{\diq}=\rho_{D,c}^3(\Barybc)|_{\diq}=\frac{1}{3}g_s^2\wavefnsq{\diq}{\Barybc},
\label{eq:darwinBbc-Dq}
\\
\rho_{D,b}^3(\Barybcprime)|_{\diq}=\rho_{D,c}^3(\Barybcprime)|_{\diq}=\frac{1}{3}g_s^2\wavefnsq{\diq}{\Barybcprime}\,.
\label{eq:darwinBbcprime-Dq}
\end{gather}
The corresponding heavy--heavy $bc$ contributions are given in Eqs.~\eqref{Eq:rhoDbc1} and \eqref{Eq:rhoDbc2}.
The last remaining piece entering $\hat{\mu}_{G,Q}^2$ is the heavy--heavy Darwin contribution $\rho_{D,Q}^3|_{Q-Q'}$, which is evaluated in the same way as the standalone Darwin term in Eq.~\eqref{eq:darwinQQnonrel}.

Collecting all contributions, including the heavy--heavy Darwin terms from Eq.~\eqref{eq:darwinQQnonrel}, we obtain the following expressions for the total chromomagnetic matrix elements $\hat{\mu}_{G,Q}^2$:
\begin{align}
    \hat{\mu}_{G,b}^2(\Barybb)&=\frac{4}{3}\,m^\Bary_b\left(M_{\mathcal{B}_{bb}^{\ast}}-M_{\mathcal{B}_{bb}}\right)-\frac{10}{9}\frac{g_s^2}{m^\Bary_b}\wavefnsq{bb}{\Barybb}-\frac{2}{3}\frac{g_s^2}{m^\Bary_b}\wavefnsq{\diq}{\Barybb},
    \\
    \hat{\mu}_{G,b}^2(\Barybc)&=\left(\frac{2}{m^\Bary_c}-\frac{1}{m^\Bary_b}\right)\frac{g_s^2}{3}\wavefnsq{bc}{\Barybc}-\frac{1}{3}\frac{g_s^2}{m^\Bary_b}\wavefnsq{\diq}{\Barybc},
    \\
    \hat{\mu}_{G,c}^2(\Barybc)&=\left(\frac{2}{m^\Bary_b}-\frac{1}{m^\Bary_c}\right)\frac{g_s^2}{3}\wavefnsq{bc}{\Barybc}-\frac{1}{3}\frac{g_s^2}{m^\Bary_c}\wavefnsq{\diq}{\Barybc},
    \\
    \hat{\mu}_{G,b}^2(\Barybcprime)&=\frac{4}{3}\frac{m^\Bary_c m^\Bary_b}{m^\Bary_c + m^\Bary_b}\left(M_\Barybcstar-M_\Barybcprime\right)-\left( \frac{2}{m^\Bary_c}+\frac{3}{m^\Bary_b} \right)\frac{g_s^2}{9}\wavefnsq{bc}{\Barybcprime}-\frac{1}{3}\frac{g_s^2}{m^\Bary_b}\wavefnsq{\diq}{\Barybcprime},
    \\
    \hat{\mu}_{G,c}^2(\Barybcprime)&=\frac{4}{3}\frac{m^\Bary_c m^\Bary_b}{m^\Bary_c + m^\Bary_b}\left(M_\Barybcstar-M_\Barybcprime\right)-\left( \frac{2}{m^\Bary_b}+\frac{3}{m^\Bary_c} \right)\frac{g_s^2}{9}\wavefnsq{bc}{\Barybcprime}-\frac{1}{3}\frac{g_s^2}{m^\Bary_c}\wavefnsq{\diq}{\Barybcprime}.
\end{align}

As for the case of Darwin terms in Sec.~\eqref{Sec:Darwin}, we assume the values extracted for chromomagnetic matrix elements correspond to low scale $\mu_h= 1$ GeV, and evolve them towards the corresponding higher scales using RG evolution from Ref.~\cite{Grozin:2007fh}.
\begin{table}[t]
\centering
\renewcommand{\arraystretch}{1.15}
\setlength{\tabcolsep}{10pt}
\begin{tabular}{l c c}
\hline\hline
Baryon $\mathcal{B}$ & $Q$ &
$\hat{\mu}_{G,Q}^2(\mathcal{B})\;[\mathrm{GeV}^2]$ \\
\hline
$\Xi_{bb}$      & $b$ & $0.024(5)$ \\
$\Omega_{bb}$   & $b$ & $0.026(5)$ \\
\hline
$\Xi_{cc}$      & $c$ & $0.086(17)$ \\
$\Omega_{cc}$   & $c$ & $0.083(17)$ \\
\hline
$\Xi_{bc}$      & $b$ & $\phantom{}0.057(11)$ \\
$\Omega_{bc}$   & $b$ & $\phantom{}0.056(11)$ \\
$\Xi_{bc}$      & $c$ & $-0.022(4)$ \\
$\Omega_{bc}$   & $c$ & $-0.024(5)$ \\
\hline
$\Xi'_{bc}$     & $b$ & $-0.004(11)$ \\
$\Omega'_{bc}$  & $b$ & $-0.003(11)$ \\
$\Xi'_{bc}$     & $c$ & $-0.019(4)$ \\
$\Omega'_{bc}$  & $c$ & $-0.020(4)$ \\
\hline\hline
\end{tabular}
\caption{Numerical values of $\hat{\mu}_{G,Q}^2(\mathcal{B})$ in units of GeV$^2$ evaluated at the scale $\mu_b$ for $\hat{\mu}_{G,b}^2(\mathcal{B})$ and at the scale $\mu_c$ for $\hat{\mu}_{G,c}^2(\mathcal{B})$.}
\label{tab:muGhat_values}
\end{table}

\newpage
\subsubsection{\texorpdfstring{Kinetic $\hat{\mu}_\pi^2$ terms}{Kinetic mupi2 terms}}
We next turn to a nonrelativistic estimate of the matrix elements of the operators $\Psi_Q^\dagger (i\vec D)^2 \Psi_Q$ and $\Psi_Q^\dagger (i\vec D)^4 \Psi_Q$ entering the kinetic term.
The quadratic term can be written as
\begin{equation}
\frac{1}{m_Q^2}\,
\frac{\langle \mathcal{B} \vert \Psi_Q^\dagger (i\vec{D})^2 \Psi_Q \vert \mathcal{B} \rangle}{2M_{\mathcal{B}}}
= n_Q(\mathcal{B})\,\langle v_Q^2\rangle_{\mathcal{B}}\,,
\end{equation}
where the multiplicity factor $n_Q(\mathcal{B})$ is given in Eq.~\eqref{Eq:Factor-n}, and $\langle v_Q^2\rangle_{\mathcal{B}}$ denotes the mean squared three-velocity of the heavy quark $Q$ in the baryon.
This term can be decomposed into the contribution from the motion of $Q$ inside the diquark and the motion of the diquark inside the baryon,
\begin{equation} 
\langle v_Q^2\rangle=\langle v_Q^2(\mathcal{D})\rangle+\langle v_\mathcal{D}^2\rangle\,. 
\end{equation} 
We further use $m_{\mathcal D}\,|\mathbf v_{\mathcal D}|=m_q\,|\mathbf v_q|$ and $m_{\mathcal D}\simeq m_Q+m_{Q'}$, where $Q$ and $Q'$ denote the heavy quarks and $q$ is the light constituent quark. We take the internal kinetic energy of the diquark subsystem to be one half of $T$, where $T$ denotes the kinetic energy associated with the relative motion of the diquark and the light quark in the baryon rest frame. The factor of $1/2$ is motivated by virial-theorem scaling together with the ratio of color factors for the $QQ'$ interaction in the color-antitriplet channel relative to the diquark-light quark interaction in the color-singlet channel~\cite{Kiselev:1998sy}. This yields 
\begin{equation} 
\langle v_Q^2\rangle_{\mathcal{B}}=\bigg(\frac{m_{Q'}}{m_Q(m_Q+m_{Q'})}+\frac{2m_q}{(m_Q+m_{Q'})(m_q+m_Q+m_{Q'})}\bigg)T\,,
\label{Eq:vQ2}
\end{equation} 
where the quark masses are understood as constituent masses, with numerical values collected in Table \ref{tab:constituentMasses}.

For the $\Psi_Q^\dagger (i\vec{D})^4 \Psi_Q$ term, we approximate the matrix element by the square of the quadratic contribution,
\begin{equation}
   \frac{1}{m_Q^4} \frac{\langle \mathcal{B} \vert \Psi_Q^\dagger (i\vec{D})^4\Psi_Q\vert \mathcal{B} \rangle}{2M_{\mathcal{B}}}\simeq n_Q(\mathcal{B}) \langle v_Q^2\rangle^2_{\mathcal{B}}\,.
\end{equation}

The general expression for the kinetic matrix element is therefore
\begin{equation}
\hat{\mu}_{\pi,Q}^2(\mathcal{B})=m_Q^2n_Q(\mathcal{B})\left(\langle v_Q^2\rangle_{\mathcal{B}}-\frac{3}{4}\langle v_Q^2\rangle_\mathcal{B}^2\right)\,,
\end{equation}
with $\langle v_Q^2\rangle_\mathcal{B}$ given in Eq.~\eqref{Eq:vQ2}. We also note that within our approximation
\begin{equation}
\hat{\mu}_{\pi,Q}^2(\Barybcprime)=\hat{\mu}_{\pi,Q}^2(\Barybc)\,.
\end{equation}
\begin{table}[t]
\centering
\renewcommand{\arraystretch}{1.25}
\setlength{\tabcolsep}{12pt}
\begin{tabular}{ccc}
\hline\hline
Baryon $\mathcal{B}$ & $Q$ & $\langle v^2_Q\rangle_\mathcal{B}$ \\
\hline
$\mathcal{B}_{bb}$    & $b$ & $0.042(8)$ \\
\hline
$\mathcal{B}_{cc}$ & $c$ & $0.139(28)$ \\
\hline
\multirow{2}{*}{$\Barybc, \Barybcprime$}
  & $b$ & $0.026(5)$ \\
  & $c$ & $0.18(4)$ \\
\hline\hline
\end{tabular}
\caption{Values of the average squared velocities $\langle v^2_Q\rangle_\mathcal{B}$ of a heavy quark $Q$ for doubly heavy baryons, obtained using NRCQM.}
\label{tab:v2Values}
\end{table}
Following Refs.~\cite{Likhoded:1999yv,Kiselev:1999kh}, we use numerical value $T=0.40\pm 0.08\,\text{GeV}$ for all doubly heavy baryons obtained using several potential models \cite{Gershtein:1998sx}, and assign a conservative $20\%$ uncertainty. For the quark masses we use the constituent values listed in Table~\ref{tab:constituentMasses}.
The resulting numerical values for the average square velocities are collected in Table \ref{tab:v2Values}.

\section{Numerical predictions for doubly heavy baryons}
\label{Sec:Numerical Estimates}

We now turn to the numerical analysis and present our predictions for the lifetimes, lifetime ratios, and inclusive semileptonic decay widths of doubly heavy baryons.

Before turning to the numerical discussion, we briefly summarize the set of short-distance contributions retained in our analysis. For the two-quark contributions, we include the leading dimension-three decay term up to NNLO, the dimension-five chromomagnetic term up to NLO, and the dimension-six Darwin term at LO. In the heavy-light spectator sector, we include the dimension-six four-quark terms at LO and NLO, with the leading penguin contributions accounted for as part of the NLO corrections, together with the dimension-seven terms at LO. In the heavy-heavy spectator sector, relevant for the $bc$ hadrons, we include the dimension-six four-quark terms only at LO, since the corresponding NLO corrections for two massive external quarks are currently unavailable. The associated penguin contributions are nevertheless retained.

The theoretical uncertainties are estimated from two sources, 
namely, by propagation of the uncertainties of the hadronic matrix 
elements, and by estimate of the 
residual renormalization-scale dependence. For the latter, we vary 
the renormalization scale in the ranges $\mu\in[3,6]~\GeV$ for the 
terms involving a $b$ quark decay, and $\mu\in[1,2]~\GeV$ for the 
terms involving the $c$ quark decay. The central values of our predictions correspond to the central 
values of the hadronic parameters and the scale choices 
$\mu_b=4.5~\GeV$ and $\mu_c=1.5~\GeV$. For the $B_c$ meson we take the central scale choice $\mu_{bc}=\sqrt{\mu_b\mu_c}= 2.6$ GeV and for the uncertainties vary it in the range $\mu\in [1.5,3.5]$ GeV. We present the results for the $\Barybb$ and $\Barycc$ 
baryons in the kinetic and the 
$\overline{\rm MS}$ mass schemes, while in the $\Barybc$ sector we 
additionally consider the $\Upsilon$ scheme, as well as for the $B_c$ meson. The corresponding numerical inputs 
for the mass-scheme definitions are listed in 
Table~\ref{tab:massSchemes}.

Lifetime ratios are particularly useful observables, since the two-quark contributions largely cancel in these combinations. As a result, the corresponding theoretical uncertainties are reduced, and the ratios provide a more direct probe of the spectator effects. For the doubly bottom baryons, where no experimental lifetime information is currently available, we quote ratios normalized to the negatively charged state, $\tau(\Barybb)/\tau(\Xi_{bb}^{-})$. In the bottom--charm sector, we analogously normalize the lifetime ratios to the positively charged baryon, $\tau(\Barybc)/\tau(\Xi_{bc}^{+})$ and $\tau(\Barybcprime)/\tau(\Xi_{bc}^{\prime\, +})$. For the doubly charmed baryons, by contrast, we normalize the ratios to the $\Xi_{cc}^{++}$ baryon lifetime according to
\begin{equation}
    \frac{\tau(\Barycc)}{\tau(\Xiccpp)}
    =
    \frac{1}{1+\big[\Gamma(\Barycc)-\Gamma(\Xiccpp)\big]^{\rm theory}\,\tau(\Xiccpp)^{\rm exp}}\,,
\end{equation}
using the experimental input provided by the LHCb measurement
\begin{equation}
    \tau(\Xiccpp)^{\rm exp}=0.256^{+0.024}_{-0.022}\pm0.014\,\mathrm{ps}\,.
\end{equation}

The leading dimension-three decay term deserves special attention, since its overall $m_Q^5$ scaling makes it particularly sensitive to the choice of heavy-quark mass scheme. This sensitivity is especially pronounced for the charm-quark contributions. To illustrate the behaviour of the perturbative expansion and the residual scheme dependence, we display the decomposition of this term for a single charm decay in all three mass schemes considered\footnote{The $\Upsilon$ scheme is used here for $\mathcal{B}_{bc}$ baryons and $B_c$ meson.} here:
\begin{equation}
\begin{aligned}
    \Gamma_{3}[m_c^{\rm kin}]/\mathrm{ps}^{-1}
    &= 1.09 + 0.45 + 0.48\,,\\
    \Gamma_{3}[\overline{m}_c]/\mathrm{ps}^{-1}
    &= 0.47 + 0.50 + 0.54\,,\\
    \Gamma_{3}[m_c^{\Upsilon}]/\mathrm{ps}^{-1}
    &= 0.73 + 0.41 + 0.79\,,
\end{aligned}
\label{Eq:Gamma3schemes}
\end{equation}
where the three terms in each sum correspond to the LO, NLO, and NNLO contributions. While the series does not show an order-by-order convergence in any of these mass schemes, higher-order corrections substantially reduce the relative scheme dependence of the total result. This suggests that the perturbative calculation captures the expected cancellation of mass-scheme dependence, even though sizable higher-order effects can still be expected. 

\subsection{\texorpdfstring{Predictions for $\Barybb$ and $\Barycc$ baryons}{Predictions for bb and cc baryons}}

We begin with the discussion of the lifetime predictions for the 
doubly bottom and doubly charmed baryons, $\Barybb$ and $\Barycc$.
To illustrate the relative sizes of the various terms entering the 
HQE, we display the numerical expressions for the decay widths 
in the kinetic mass scheme. For the $\Barybb$ baryons 
we obtain
\begin{align}
\Gamma(\Xi_{bb}^{0})/\mathrm{ps}^{-1}
=&2(\underbrace{0.728}_{\rm LO}-\underbrace{0.048}_{\rm NLO}-\underbrace{0.009}_{\rm NNLO})
\nonumber\\
&-0.027\frac{\mu_\pi^2}{1.70\,{\rm GeV}^2}
-3\times 10^{-4}\frac{\mu_G^2}{0.024\,{\rm GeV}^2}
-0.060\frac{\rho_D^3}{0.36\,{\rm GeV}^3}
\nonumber\\
&+(\undermarker{0.363}{LO}+\undermarker{0.024}{NLO})\frac{\Opsix{1}{bu}}{-0.120\,{\rm GeV}^3}
\nonumber\\
&+0.055\frac{\OpsevenP{2}{bu}}{-0.056\,{\rm GeV}^4}\,,
\label{Eq:semiNumXibb0}
\end{align}
\begin{align}
\Gamma(\Xi_{bb}^{-})/\mathrm{ps}^{-1}
=&2(\underbrace{0.728}_{\rm LO}-\underbrace{0.048}_{\rm NLO}-\underbrace{0.009}_{\rm NNLO})
\nonumber\\
&-0.027\frac{\mu_\pi^2}{1.70\,{\rm GeV}^2}
-3\times 10^{-4}\frac{\mu_G^2}{0.024\,{\rm GeV}^2}
-0.060\frac{\rho_D^3}{0.36\,{\rm GeV}^3}
\nonumber\\
&-(\undermarker{0.094}{LO}-\undermarker{0.024}{NLO})\frac{\Opsix{1}{bd}}{-0.120\,{\rm GeV}^3}
\nonumber\\
&+0.016\frac{\OpsevenP{2}{bd}}{-0.056\,{\rm GeV}^4}
-2\times 10^{-5}\frac{\OpsevenP{3}{bd}}{-0.009\,{\rm GeV}^4}\,,
\label{Eq:semiNumXibbm}
\end{align}
\begin{align}
\Gamma(\Omega_{bb}^{-})/\mathrm{ps}^{-1}
=&2(\underbrace{0.728}_{\rm LO}-\underbrace{0.048}_{\rm NLO}-\underbrace{0.009}_{\rm NNLO})
\nonumber\\
&-0.027\frac{\mu_\pi^2}{1.70\,{\rm GeV}^2}
-3\times 10^{-4}\frac{\mu_G^2}{0.024\,{\rm GeV}^2}
-0.063\frac{\rho_D^3}{0.38\,{\rm GeV}^3}
\nonumber\\
&-(\undermarker{0.117}{LO}-\undermarker{0.034}{NLO})\frac{\Opsix{1}{bs}}{-0.158\,{\rm GeV}^3}
\nonumber\\
&-0.004\frac{\OpsevenP{1}{bs}}{-0.014\,{\rm GeV}^4}+0.015\frac{\OpsevenP{2}{bs}}{-0.052\,{\rm GeV}^4}
-5\times 10^{-5}\frac{\OpsevenP{3}{bs}}{-0.009\,{\rm GeV}^4}\,,
\label{Eq:semiNumOmegabbm}
\end{align}
where the dimension-six matrix elements are defined at the low 
hadronic scale $\mu_h=1$\,GeV. The resulting numerical predictions 
for the lifetimes of the $\Barybb$ baryons and the corresponding 
ratios are collected in Table~\ref{tab:fin_bb}, while the breakdown of various contributions is listed in Appendix \ref{sec:appendixBreakdown} in Table \ref{tab:bbBreakdown}.
The absolute widths and lifetimes exhibit a mild dependence on the choice of heavy-quark mass scheme. The lifetime ratios are remarkably stable across the two schemes, which simply reflects the fact that the dominant scheme dependence is largely common to all three baryons.

The dominant lifetime splittings originate from the dimension-six 
spectator contributions leading to the predicted lifetime hierarchy
\begin{equation}
   \tau(\Xi_{bb}^{0})< \tau(\Xi_{bb}^{-}) \simeq \tau(\Omega_{bb}^{-})\,,
\end{equation}
which is primarily driven by the sizable weak-exchange contribution to the 
$\Xi_{bb}^{0}$ decay width. This can 
be traced to the combination of the sizable Wilson coefficient of 
the weak exchange topology and the factor of $6$ in the 
corresponding four-quark matrix element, which enters due to the 
spin-1 $bb$ diquark, see Eq.~\eqref{eq:dimsixBbb-bq}. This 
situation contrasts with the case of singly bottom baryons, where an 
analogous spin-dependent enhancement factor appears only for the 
$\Omega_b$ baryon, as a consequence of its spin-1 $ss$ diquark. However, in the valence approximation the exchange topology does not contribute to the $\Omega_b$ width, so that the 
corresponding enhancement has no phenomenological effect, see e.g.~\cite{Gratrex:2023pfn}.

\begin{table}[htbp]
\small
\centering
\renewcommand{\arraystretch}{1.3}
\begin{tabular}{lcc}
\hline\hline
Observable & $\overline{\rm MS}$ & Kinetic \\
\hline
$\Gamma(\Xi_{bb}^0)$    [$10^{-12}$\,GeV] & $1.02\pm0.06\,^{+0.08}_{-0.05}$ & $1.12\pm0.06\,^{+0.01}_{-0.01}$ \\
$\Gamma(\Xi_{bb}^-)$    [$10^{-12}$\,GeV] & $0.72\pm0.01\,^{+0.06}_{-0.04}$ & $0.79\pm0.01\,^{+0.001}_{-0.003}$ \\
$\Gamma(\Omega_{bb}^-)$ [$10^{-12}$\,GeV] & $0.70\pm0.01\,^{+0.06}_{-0.04}$ & $0.78\pm0.02\,^{+0.002}_{-0.003}$ \\[0.2em]
\hline
$\tau(\Xi_{bb}^0)$    [ps] & $0.64\pm0.04\,^{+0.03}_{-0.05}$ & $0.59\pm0.03\,^{+0.005}_{-0.004}$ \\
$\tau(\Xi_{bb}^-)$    [ps] & $0.92\pm0.02\,^{+0.05}_{-0.07}$ & $0.83\pm0.02\,^{+0.003}_{-0.001}$ \\
$\tau(\Omega_{bb}^-)$ [ps] & $0.94\pm0.02\,^{+0.05}_{-0.07}$ & $0.85\pm0.02\,^{+0.004}_{-0.002}$ \\[0.2em]
\hline
$\tau(\Xi_{bb}^{0})/\tau(\Xi_{bb}^{-})$    & $0.70\pm0.04^{+0.004}_{-0.000}$ & $0.71\pm0.04^{+0.003}_{-0.004}$ \\
$\tau(\Omega_{bb}^{-})/\tau(\Xi_{bb}^{-})$ & $1.02\pm0.03^{+0.001}_{-0.001}$ & $1.02\pm0.03^{+0.000}_{-0.001}$ \\[0.2em]
\hline
$\Gamma_{\rm SL}(\Xi_{bb}^0)$    [$10^{-12}$\,GeV] & $0.190\pm0.002\,^{+0.002}_{-0.002}$ & $0.202\pm0.003\,^{+0.005}_{-0.011}$ \\
$\Gamma_{\rm SL}(\Xi_{bb}^-)$    [$10^{-12}$\,GeV] & $0.190\pm0.002\,^{+0.002}_{-0.002}$ & $0.202\pm0.003\,^{+0.010}_{-0.011}$ \\
$\Gamma_{\rm SL}(\Omega_{bb}^-)$ [$10^{-12}$\,GeV] & $0.190\pm0.002\,^{+0.002}_{-0.002}$ & $0.201\pm0.003\,^{+0.010}_{-0.011}$ \\
\hline\hline
\end{tabular}
\caption{Final predictions for the decay widths, lifetimes, and 
lifetime ratios of the doubly bottom baryons $\Barybb$ in two 
$b$-quark mass schemes. The first uncertainty is parametric, while 
the second reflects the residual renormalization-scale dependence 
obtained by varying $\mu_b\in[3,6]$\,GeV.}
\label{tab:fin_bb}
\end{table}

For the doubly charmed baryons, the lifetime hierarchy is more 
pronounced, reflecting the large  spectator 
contributions. We obtain the following numerical expressions in the kinetic mass scheme:
\begin{align}
\Gamma(\Xi_{cc}^{++})/\mathrm{ps}^{-1}
=&2(\underbrace{1.09}_{\rm LO}+\underbrace{0.45}_{\rm NLO}+\underbrace{0.48}_{\rm NNLO})
\nonumber\\
&-0.25\frac{\mu_\pi^2}{0.5\,{\rm GeV}^2}
+0.04\frac{\mu_G^2}{0.086\,{\rm GeV}^2}
+0.37\frac{\rho_D^3}{0.084\,{\rm GeV}^3}
\nonumber\\
&-(\underbrace{1.71}_{\rm LO}-\underbrace{0.42}_{\rm NLO})\frac{\Opsix{1}{u}}{-0.04\,{\rm GeV}^3}
\nonumber\\
&+0.83\frac{\OpsevenP{2}{u}}{-0.013\,{\rm GeV}^4}\,,
\label{Eq:semiNumXiccpp}
\end{align}
\begin{align}
\Gamma(\Xi_{cc}^{+})/\mathrm{ps}^{-1}
=&2(\underbrace{1.09}_{\rm LO}+\underbrace{0.45}_{\rm NLO}+\underbrace{0.48}_{\rm NNLO})
\nonumber\\
&-0.25\frac{\mu_\pi^2}{0.5\,{\rm GeV}^2}
+0.04\frac{\mu_G^2}{0.086\,{\rm GeV}^2}
+0.37\frac{\rho_D^3}{0.084\,{\rm GeV}^3}
\nonumber\\
&+(\underbrace{8.38}_{\rm LO}+\underbrace{2.00}_{\rm NLO})\frac{\Opsix{1}{d}}{-0.04\,{\rm GeV}^3}
\nonumber\\
&+3.29\frac{\OpsevenP{2}{d}}{-0.013\,{\rm GeV}^4}\,,
\label{Eq:semiNumXiccp}
\end{align}
\begin{align}
\Gamma(\Omega_{cc}^{+})/\mathrm{ps}^{-1}
=&2(\underbrace{1.09}_{\rm LO}+\underbrace{0.45}_{\rm NLO}+\underbrace{0.48}_{\rm NNLO})
\nonumber\\
&-0.25\frac{\mu_\pi^2}{0.5\,{\rm GeV}^2}
+0.04\frac{\mu_G^2}{0.083\,{\rm GeV}^2}
+0.41\frac{\rho_D^3}{0.093\,{\rm GeV}^3}
\nonumber\\
&+(\underbrace{6.11}_{\rm LO}-\underbrace{0.97}_{\rm NLO})\frac{\Opsix{1}{s}}{-0.055\,{\rm GeV}^3}
\nonumber\\
&+0.75\frac{\OpsevenP{1}{s}}{-0.005\,{\rm GeV}^4}
-2.51\frac{\OpsevenP{2}{s}}{-0.018\,{\rm GeV}^4}-2\times10^{-5}\frac{\OpsevenP{3}{s}}{-0.003\,{\rm GeV}^4}\,,
\label{Eq:semiNumOmegaccp}
\end{align}
where the dimension-six matrix elements are defined at the low 
hadronic scale\footnote{Note that we fixed a typo in the code of our previous analysis \cite{Dulibic:2023jeu} which influenced the running of four-quark matrix elements resulting in small numerical difference with respect to corrected results in equations \eqref{Eq:semiNumXiccpp}-\eqref{Eq:semiNumOmegaccp}.} $\mu_h=1$\,GeV.

Among the two-quark terms, the Darwin operator provides the largest contribution, with part of it offset by the kinetic term. In the spectator part of the expansion, the width of $\Xi_{cc}^{+}$ is enhanced predominantly by the exchange topology, whereas the width of $\Xi_{cc}^{++}$ is reduced by the destructive interference. This qualitative picture is unchanged once the NLO corrections to the dimension-six coefficients and the LO dimension-seven terms are taken into account.
These results imply 
the lifetime hierarchy
\begin{equation}
    \tau(\Xiccp) < \tau(\Omega_{cc}^{+}) < \tau(\Xiccpp)\,.
\end{equation}

We summarize our numerical predictions for the lifetimes of the 
$\Barycc$ baryons and the corresponding semileptonic widths in Table~\ref{tab:fin_cc}, obtained in the 
$\overline{\rm MS}$ and kinetic mass schemes, while the breakdown of various contributions is listed in Appendix \ref{sec:appendixBreakdown} in Table \ref{tab:ccBreakdown}. The absolute decay widths and lifetimes show a noticeable dependence on the charm-quark mass scheme, driven primarily by the leading dimension-three contribution. However, the results across the schemes remain marginally compatible within the theoretical uncertainties. The predicted lifetime ratios remain rather stable across the two schemes, indicating the cancellation of the dominant scheme dependence entering the universal leading charm quark decay term.

Compared with our previous analysis of the doubly charmed 
baryons~\cite{Dulibic:2023jeu}, the present analysis also includes 
the NNLO correction to the leading decay term and 
the NLO correction to the chromomagnetic term, with the former providing the largest difference which increases the total decay widths by 
approximately $30$\%, $5\%$, $15\%$, for the $\Xiccpp$, $\Xiccp$, and $\Omega_{cc}^{+}$, respectively. For the $\Xi_{cc}^{++}$ baryon, this 
reduces our previous central prediction\footnote{The abstract of the published version of our previous paper on doubly charmed baryon lifetimes~\cite{Dulibic:2023jeu} contains a misprint: $\tau(\Xi_{cc}^{++}) = 0.32 \pm 0.5^{+0.8}_{-0.7}$~ps should read $\tau(\Xi_{cc}^{++}) = 0.32 \pm 0.05^{+0.08}_{-0.07}$~ps.} for the lifetime from 
$\tau(\Xi_{cc}^{++})\simeq 0.32$ ps in the kinetic scheme 
to the present value $\tau(\Xi_{cc}^{++})\simeq 0.27$ ps, 
moving the prediction closer to the LHCb measurement \cite{LHCb:2018zpl} 
$\tau(\Xi_{cc}^{++})^{\rm exp}=0.256^{+0.024}_{-0.022}\pm 0.014$\,ps.

\begin{table}[htbp]
\small
\centering
\renewcommand{\arraystretch}{1.3}
\begin{tabular}{lcc}
\hline\hline
Observable & $\overline{\rm MS}$ & Kinetic \\
\hline
$\Gamma(\Xi_{cc}^{++})$ [$10^{-12}$\,GeV] & $1.8\pm0.2\,^{+2.0}_{-0.6}$   & $2.5\pm0.2\,^{+0.9}_{-0.3}$ \\
$\Gamma(\Xi_{cc}^{+})$  [$10^{-12}$\,GeV] & $10.0\pm1.3\,^{+5.2}_{-2.0}$  & $11.8\pm1.4\,^{+2.8}_{-1.1}$ \\
$\Gamma(\Omega_{cc}^+)$ [$10^{-12}$\,GeV] & $4.2\pm0.7\,^{+1.8}_{-0.6}$   & $5.0\pm0.8\,^{+0.5}_{-0.2}$ \\[0.2em]
\hline
$\tau(\Xi_{cc}^{++})$ [ps] & $0.37\pm0.04\,^{+0.18}_{-0.20}$     & $0.27\pm0.02\,^{+0.04}_{-0.07}$ \\
$\tau(\Xi_{cc}^{+})$  [ps] & $0.066\pm0.008\,^{+0.016}_{-0.023}$ & $0.056\pm0.007\,^{+0.006}_{-0.011}$ \\
$\tau(\Omega_{cc}^+)$ [ps] & $0.16\pm0.03\,^{+0.03}_{-0.05}$     & $0.13\pm0.02\,^{+0.005}_{-0.012}$ \\[0.2em]
\hline
$\tau(\Xi_{cc}^{+})/\tau(\Xi_{cc}^{++})$    & $0.32\pm0.04^{+0.04}_{-0.07}$ & $0.30\pm0.03^{+0.02}_{-0.04}$ \\
$\tau(\Omega_{cc}^{+})/\tau(\Xi_{cc}^{++})$ & $0.62\pm0.07^{+0.02}_{-0.00}$ & $0.61\pm0.08^{+0.04}_{-0.01}$ \\[0.2em]
\hline
$\Gamma_{\rm SL}(\Xi_{cc}^{++})$ [$10^{-12}$\,GeV] & $0.46\pm0.01^{+0.09}_{-0.08}$ & $0.55\pm0.01^{+0.01}_{-0.03}$ \\
$\Gamma_{\rm SL}(\Xi_{cc}^{+})$  [$10^{-12}$\,GeV] & $0.47\pm0.01^{+0.09}_{-0.08}$ & $0.56\pm0.02^{+0.01}_{-0.04}$ \\
$\Gamma_{\rm SL}(\Omega_{cc}^+)$ [$10^{-12}$\,GeV] & $0.84\pm0.20^{+0.00}_{-0.13}$ & $0.90\pm0.22^{+0.09}_{-0.23}$ \\
\hline\hline
\end{tabular}
\caption{Final predictions for the decay widths, lifetimes, and 
lifetime ratios of the doubly charmed baryons $\Barycc$ in two 
$c$-quark mass schemes. The first uncertainty is parametric, while 
the second reflects the residual renormalization-scale dependence 
obtained by varying $\mu_c\in[1,2]$\,GeV.}
\label{tab:fin_cc}
\end{table}

\subsection{\texorpdfstring{Predictions for $\Barybc$ baryons}{Predictions for bc baryons}}
We next turn to the bottom-charm baryons. In this sector, both $b$- and $c$-quark decays contribute to the total width, while the resulting lifetime pattern is shaped by the interplay of heavy-light ($bq$ and $cq$) and heavy-heavy ($bc$) spectator contributions. The latter are CKM suppressed relative to the heavy-light ones by the factor $V_{cb}$.
This suppression is partially compensated by the enhanced sizes of the corresponding nonperturbative matrix elements, see Table~\ref{tab:fourquarkME}.

The corresponding numerical expressions for the $\Barybc$ 
baryons in the kinetic mass scheme read:
\begin{equation}
\begin{aligned}
\Gamma(\Xi_{bc}^{+})/\mathrm{ps}^{-1}
=&(\undermarker{1.09}{LO}+\undermarker{0.45}{NLO}+\undermarker{0.48}{NNLO})+(\undermarker{0.73}{LO}-\undermarker{0.05}{NLO}-\undermarker{0.01}{NNLO})
\\
&-0.16\frac{\mu_{\pi,c}^2}{0.31\,{\rm GeV}^2}
-0.001\frac{\mu_{G,c}^2}{-0.022\,{\rm GeV}^2}
+0.33\frac{\rho_{D,c}^3}{0.076\,{\rm GeV}^3}
\\
&-0.009\frac{\mu_{\pi,b}^2}{0.53\,{\rm GeV}^2}
-0.001\frac{\mu_{G,b}^2}{0.057\,{\rm GeV}^2}
-0.009\frac{\rho_{D,b}^3}{0.054\,{\rm GeV}^3}
\\
&-(\undermarker{0.55}{LO}-\undermarker{0.12}{NLO})\frac{\Opsix{1}{cu}}{-0.008\,{\rm GeV}^3}
+(0.024+0.001)\frac{\Opsix{1}{bu}}{-0.008\,{\rm GeV}^3}
+0.83\frac{\Opsix{1}{bc}}{-0.17\,{\rm GeV}^3}
\\
&+0.16\frac{\OpsevenP{2}{cu}}{-0.003\,{\rm GeV}^4}
-3\times 10^{-5}\frac{\OpsevenP{3}{cu}}{-0.003\,{\rm GeV}^4}\,,
\end{aligned}
\label{Eq:semiNumXibcp}
\end{equation}
\begin{equation}
\begin{aligned}
\Gamma(\Xi_{bc}^{0})/\mathrm{ps}^{-1}
=&(\undermarker{1.09}{LO}+\undermarker{0.45}{NLO}+\undermarker{0.48}{NNLO})+(\undermarker{0.73}{LO}-\undermarker{0.05}{NLO}-\undermarker{0.01}{NNLO})
\\
&-0.16\frac{\mu_{\pi,c}^2}{0.31\,{\rm GeV}^2}
-0.001\frac{\mu_{G,c}^2}{-0.022\,{\rm GeV}^2}
+0.33\frac{\rho_{D,c}^3}{0.076\,{\rm GeV}^3}
\\
&-0.009\frac{\mu_{\pi,b}^2}{0.53\,{\rm GeV}^2}
-0.001\frac{\mu_{G,b}^2}{0.057\,{\rm GeV}^2}
-0.009\frac{\rho_{D,b}^3}{0.054\,{\rm GeV}^3}
\\
&+(\undermarker{1.65}{LO}+\undermarker{0.37}{NLO})\frac{\Opsix{1}{cd}}{-0.008\,{\rm GeV}^3}
-(\undermarker{0.009}{LO}-\undermarker{0.002}{NLO})\frac{\Opsix{1}{bd}}{-0.008\,{\rm GeV}^3}
+0.83\frac{\Opsix{1}{bc}}{-0.17\,{\rm GeV}^3}
\\
&+0.63\frac{\OpsevenP{2}{cd}}{-0.003\,{\rm GeV}^4}
-9\times 10^{-7}\frac{\OpsevenP{3}{cd}}{-0.003\,{\rm GeV}^4}\,,
\end{aligned}
\label{Eq:semiNumXibc0}
\end{equation}
\begin{equation}
\begin{aligned}
\Gamma(\Omega_{bc}^{0})/\mathrm{ps}^{-1}
=&(\undermarker{1.09}{LO}+\undermarker{0.45}{NLO}+\undermarker{0.48}{NNLO})+(\undermarker{0.73}{LO}-\undermarker{0.05}{NLO}-\undermarker{0.01}{NNLO})
\\
&-0.16\frac{\mu_{\pi,c}^2}{0.31\,{\rm GeV}^2}
-0.001\frac{\mu_{G,c}^2}{-0.024\,{\rm GeV}^2}
+0.35\frac{\rho_{D,c}^3}{0.08\,{\rm GeV}^3}
\\
&-0.009\frac{\mu_{\pi,b}^2}{0.53\,{\rm GeV}^2}
-0.001\frac{\mu_{G,b}^2}{0.056\,{\rm GeV}^2}
-0.009\frac{\rho_{D,b}^3}{0.057\,{\rm GeV}^3}
\\
&+(\undermarker{2.07}{LO}-\undermarker{0.33}{NLO})\frac{\Opsix{1}{cs}}{-0.010\,{\rm GeV}^3}
-(\undermarker{0.011}{LO}-\undermarker{0.003}{NLO})\frac{\Opsix{1}{bs}}{-0.010\,{\rm GeV}^3}
+0.80\frac{\Opsix{1}{bc}}{-0.17\,{\rm GeV}^3}
\\
&-0.41\frac{\OpsevenP{1}{cs}}{0.003\,{\rm GeV}^4}
-0.46\frac{\OpsevenP{2}{cs}}{-0.003\,{\rm GeV}^4}
+2\times 10^{-5}\frac{\OpsevenP{3}{cs}}{-0.003\,{\rm GeV}^4}\,.
\end{aligned}
\label{Eq:semiNumOmegabc0}
\end{equation}
In these expressions, the first line 
of each decay width displays the dimension-three contribution, with 
the $c$-quark partonic term in the first bracket and the $b$-quark 
partonic term in the second bracket, each separated into LO, NLO, 
and NNLO pieces. 
These expressions make the hierarchy of contributions in the $\Barybc$ sector transparent. The dominant part of the total width comes from the leading dimension-three term, with the charm-quark contribution substantially larger than the bottom-quark one. A similar pattern persists among the power-suppressed two-quark terms. The charm-sector corrections, in particular the Darwin term, are numerically much more important than corresponding bottom-sector contributions. The lifetime splittings are governed mainly by the dimension-six heavy-light $cq$ spectator terms, whose size and sign differ significantly among the three baryons. This leads to the predicted hierarchy for the $\Barybc$ baryons with $S_{\cal D}=0$,
\begin{equation}
    \tau(\Xi_{bc}^0) \lesssim \tau(\Omega_{bc}^0) < \tau(\Xi_{bc}^+)\,.
\end{equation}

The analogous expressions for the $\Barybcprime$ baryons, 
corresponding to a spin-1 $bc$ diquark, read:
\begin{equation}
\begin{aligned}
\Gamma(\Xi_{bc}^{\prime\, +})/\mathrm{ps}^{-1}
=&(\undermarker{1.09}{LO}+\undermarker{0.45}{NLO}+\undermarker{0.48}{NNLO})+(\undermarker{0.73}{LO}-\undermarker{0.05}{NLO}-\undermarker{0.01}{NNLO})
\\
&-0.16\frac{\mu_{\pi,c}^2}{0.31\,{\rm GeV}^2}
-0.01\frac{\mu_{G,c}^2}{-0.022\,{\rm GeV}^2}
+0.33\frac{\rho_{D,c}^3}{0.076\,{\rm GeV}^3}
\\
&-0.009\frac{\mu_{\pi,b}^2}{0.53\,{\rm GeV}^2}
-0.001\frac{\mu_{G,b}^2}{0.057\,{\rm GeV}^2}
-0.009\frac{\rho_{D,b}^3}{0.054\,{\rm GeV}^3}
\\
&-(\undermarker{0.98}{LO}-\undermarker{0.26}{NLO})\frac{\Opsix{1}{cu}}{-0.023\,{\rm GeV}^3}
+(\undermarker{0.07}{LO}+\undermarker{0.00}{NLO})\frac{\Opsix{1}{bu}}{-0.023\,{\rm GeV}^3}
+0.45\frac{\Opsix{2}{bc}}{0.041\,{\rm GeV}^3}
\\
&+0.47\frac{\OpsevenP{2}{cu}}{-0.008\,{\rm GeV}^4}
-8\times 10^{-5}\frac{\OpsevenP{3}{cu}}{-0.008\,{\rm GeV}^4}\,,
\end{aligned}
\label{Eq:semiNumXibcprimep}
\end{equation}
\begin{equation}
\begin{aligned}
\Gamma(\Xi_{bc}^{\prime\, 0})/\mathrm{ps}^{-1}
=&(\undermarker{1.09}{LO}+\undermarker{0.45}{NLO}+\undermarker{0.48}{NNLO})+(\undermarker{0.73}{LO}-\undermarker{0.05}{NLO}-\undermarker{0.01}{NNLO})
\\
&-0.16\frac{\mu_{\pi,c}^2}{0.31\,{\rm GeV}^2}
-0.01\frac{\mu_{G,c}^2}{-0.022\,{\rm GeV}^2}
+0.33\frac{\rho_{D,c}^3}{0.076\,{\rm GeV}^3}
\\
&-0.009\frac{\mu_{\pi,b}^2}{0.53\,{\rm GeV}^2}
-0.001\frac{\mu_{G,b}^2}{0.057\,{\rm GeV}^2}
-0.009\frac{\rho_{D,b}^3}{0.054\,{\rm GeV}^3}
\\
&+(\undermarker{4.82}{LO}+\undermarker{1.15}{NLO})\frac{\Opsix{1}{cd}}{-0.023\,{\rm GeV}^3}
-(\undermarker{0.02}{LO}+\undermarker{0.00}{NLO})\frac{\Opsix{1}{bd}}{-0.023\,{\rm GeV}^3}
+0.45\frac{\Opsix{2}{bc}}{0.041\,{\rm GeV}^3}
\\
&+1.89\frac{\OpsevenP{2}{cd}}{-0.008\,{\rm GeV}^4}
+3\times 10^{-6}\frac{\OpsevenP{3}{cd}}{-0.008\,{\rm GeV}^4}\,,
\end{aligned}
\label{Eq:semiNumXibcprime0}
\end{equation}
\begin{equation}
\begin{aligned}
\Gamma(\Omega_{bc}^{\prime\, 0})/\mathrm{ps}^{-1}
=&(\undermarker{1.09}{LO}+\undermarker{0.45}{NLO}+\undermarker{0.48}{NNLO})+(\undermarker{0.73}{LO}-\undermarker{0.05}{NLO}-\undermarker{0.01}{NNLO})
\\
&-0.16\frac{\mu_{\pi,c}^2}{0.31\,{\rm GeV}^2}
-0.01\frac{\mu_{G,c}^2}{-0.022\,{\rm GeV}^2}
+0.35\frac{\rho_{D,c}^3}{0.076\,{\rm GeV}^3}
\\
&-0.009\frac{\mu_{\pi,b}^2}{0.53\,{\rm GeV}^2}
-0.001\frac{\mu_{G,b}^2}{0.057\,{\rm GeV}^2}
-0.009\frac{\rho_{D,b}^3}{0.054\,{\rm GeV}^3}
\\
&+(\undermarker{3.33}{LO}-\undermarker{0.55}{NLO})\frac{\Opsix{1}{cs}}{-0.03\,{\rm GeV}^3}
-(\undermarker{0.02}{LO}+\undermarker{0.01}{NLO})\frac{\Opsix{1}{bs}}{-0.03\,{\rm GeV}^3}
+0.42\frac{\Opsix{2}{bc}}{0.041\,{\rm GeV}^3}
\\
&+0.41\frac{\OpsevenP{1}{cs}}{-0.003\,{\rm GeV}^4}
-1.36\frac{\OpsevenP{2}{cs}}{-0.010\,{\rm GeV}^4}
+6\times 10^{-5}\frac{\OpsevenP{3}{cs}}{-0.010\,{\rm GeV}^4}\,,
\end{aligned}
\label{Eq:semiNumOmegabcprime0}
\end{equation}
organized in the same way as the unprimed expressions. As in the unprimed case, the widths are dominated by the leading charm contribution, while the lifetime splittings are governed mainly by the heavy-light $cq$ spectator terms, which are generally more pronounced for the vector-diquark states.

The final numerical predictions for the $\Barybc$ and $\Barybcprime$ baryons are collected in Tables~\ref{tab:fin_bc} and~\ref{tab:fin_bc_prime}, respectively, in the $\overline{\rm MS}$, kinetic, and $\Upsilon$ mass schemes, while the breakdown of various contributions is listed in Appendix \ref{sec:appendixBreakdown} in Table \ref{tab:breakdownBARYbc}. Also in this sector, the dependence on the choice of heavy-quark mass scheme remains relatively mild and is largely covered by the quoted theoretical uncertainties.

\begin{table}[htbp]
\small
\centering
\renewcommand{\arraystretch}{1.3}
\begin{tabular}{lccc}
\hline\hline
Observable & $\overline{\rm MS}$ & Kinetic & $\Upsilon$ \\
\hline
$\Gamma(\Xi_{bc}^{+})$    [$10^{-12}$\,GeV] & $1.8\pm0.1^{+1.1}_{-0.4}$ & $2.3\pm0.2^{+0.5}_{-0.2}$ & $2.3\pm0.2^{+0.4}_{-0.2}$ \\
$\Gamma(\Xi_{bc}^{0})$    [$10^{-12}$\,GeV] & $3.5\pm0.4^{+1.7}_{-0.6}$ & $4.2\pm0.5^{+0.8}_{-0.3}$ & $4.0\pm0.4^{+0.9}_{-0.4}$ \\
$\Gamma(\Omega_{bc}^{0})$ [$10^{-12}$\,GeV] & $2.4\pm0.3^{+1.0}_{-0.4}$ & $3.0\pm0.3^{+0.3}_{-0.1}$ & $2.9\pm0.3^{+0.4}_{-0.2}$ \\[0.2em]
\hline
$\tau(\Xi_{bc}^{+})$    [ps] & $0.37\pm0.03\,^{+0.09}_{-0.14}$ & $0.29\pm0.02\,^{+0.02}_{-0.05}$ & $0.29\pm0.02\,^{+0.03}_{-0.04}$ \\
$\tau(\Xi_{bc}^{0})$    [ps] & $0.19\pm0.02\,^{+0.04}_{-0.06}$ & $0.16\pm0.02\,^{+0.01}_{-0.03}$ & $0.17\pm0.02\,^{+0.02}_{-0.03}$ \\
$\tau(\Omega_{bc}^{0})$ [ps] & $0.27\pm0.03\,^{+0.05}_{-0.08}$ & $0.22\pm0.03\,^{+0.01}_{-0.02}$ & $0.23\pm0.02\,^{+0.02}_{-0.03}$ \\[0.2em]
\hline
$\tau(\Xi_{bc}^{0})/\tau(\Xi_{bc}^{+})$    & $0.51\pm0.07^{+0.04}_{-0.01}$ & $0.54\pm0.07^{+0.01}_{-0.00}$ & $0.56\pm0.07^{+0.01}_{-0.04}$ \\
$\tau(\Omega_{bc}^{0})/\tau(\Xi_{bc}^{+})$ & $0.72\pm0.10^{+0.10}_{-0.05}$ & $0.76\pm0.10^{+0.07}_{-0.03}$ & $0.78\pm0.10^{+0.01}_{-0.02}$ \\[0.2em]
\hline
$\Gamma_{\rm SL}(\Xi_{bc}^{+})$    [$10^{-12}$\,GeV] & $0.41\pm0.01^{+0.06}_{-0.05}$ & $0.46\pm0.01^{+0.01}_{-0.02}$ & $0.49\pm0.02^{+0.01}_{-0.04}$ \\
$\Gamma_{\rm SL}(\Xi_{bc}^{0})$    [$10^{-12}$\,GeV] & $0.41\pm0.01^{+0.06}_{-0.05}$ & $0.47\pm0.01^{+0.01}_{-0.02}$ & $0.49\pm0.02^{+0.01}_{-0.04}$ \\
$\Gamma_{\rm SL}(\Omega_{bc}^{0})$ [$10^{-12}$\,GeV] & $0.49\pm0.09^{+0.01}_{-0.03}$ & $0.53\pm0.09^{+0.03}_{-0.09}$ & $0.54\pm0.08^{+0.00}_{-0.06}$ \\
\hline\hline
\end{tabular}
\caption{Final predictions for the decay widths, lifetimes, and 
lifetime ratios of the doubly heavy baryons $\Barybc$ (scalar $bc$ 
diquark, $S_{\mathcal{D}}=0$) in three heavy-quark mass schemes. Uncertainties 
are parametric.}
\label{tab:fin_bc}
\end{table}

\begin{table}[htbp]
\small
\centering
\renewcommand{\arraystretch}{1.3}
\begin{tabular}{lccc}
\hline\hline
Observable & $\overline{\rm MS}$ & Kinetic & $\Upsilon$ \\
\hline
$\Gamma(\Xi_{bc}^{\prime\, +})$    [$10^{-12}$\,GeV] & $1.5\pm0.2^{+1.0}_{-0.3}$ & $2.0\pm0.2^{+0.4}_{-0.2}$ & $2.0\pm0.2^{+0.3}_{-0.2}$ \\
$\Gamma(\Xi_{bc}^{\prime\, 0})$    [$10^{-12}$\,GeV] & $6.2\pm1.3^{+3.0}_{-1.1}$ & $7.3\pm1.4^{+1.6}_{-0.6}$ & $6.8\pm1.3^{+1.8}_{-0.7}$ \\
$\Gamma(\Omega_{bc}^{\prime\, 0})$ [$10^{-12}$\,GeV] & $2.8\pm0.5^{+0.9}_{-0.4}$ & $3.3\pm0.6^{+0.2}_{-0.1}$ & $3.1\pm0.5^{+0.4}_{-0.2}$ \\[0.2em]
\hline
$\tau(\Xi_{bc}^{\prime\, +})$    [ps] & $0.43\pm0.05\,^{+0.11}_{-0.17}$ & $0.32\pm0.03\,^{+0.03}_{-0.06}$ & $0.33\pm0.03\,^{+0.03}_{-0.04}$ \\
$\tau(\Xi_{bc}^{\prime\, 0})$    [ps] & $0.11\pm0.02\,^{+0.02}_{-0.03}$ & $0.09\pm0.02\,^{+0.01}_{-0.02}$ & $0.10\pm0.02\,^{+0.01}_{-0.02}$ \\
$\tau(\Omega_{bc}^{\prime\, 0})$ [ps] & $0.24\pm0.04\,^{+0.03}_{-0.06}$ & $0.20\pm0.03\,^{+0.00}_{-0.01}$ & $0.21\pm0.03\,^{+0.01}_{-0.02}$ \\[0.2em]
\hline
$\tau(\Xi_{bc}^{\prime\, 0})/\tau(\Xi_{bc}^{\prime\, +})$    & $0.25\pm0.06^{+0.03}_{-0.01}$ & $0.28\pm0.06^{+0.001}_{-0.001}$ & $0.29\pm0.06^{+0.01}_{-0.03}$ \\
$\tau(\Omega_{bc}^{\prime\, 0})/\tau(\Xi_{bc}^{\prime\, +})$ & $0.56\pm0.12^{+0.14}_{-0.05}$ & $0.61\pm0.12^{+0.08}_{-0.04}$ & $0.64\pm0.12^{+0.02}_{-0.02}$ \\[0.2em]
\hline
$\Gamma_{\rm SL}(\Xi_{bc}^{\prime\, +})$    [$10^{-12}$\,GeV] & $0.42\pm0.02^{+0.06}_{-0.05}$ & $0.51\pm0.02^{+0.01}_{-0.02}$ & $0.53\pm0.02^{+0.01}_{-0.04}$ \\
$\Gamma_{\rm SL}(\Xi_{bc}^{\prime\, 0})$    [$10^{-12}$\,GeV] & $0.43\pm0.02^{+0.06}_{-0.05}$ & $0.51\pm0.02^{+0.01}_{-0.02}$ & $0.53\pm0.02^{+0.01}_{-0.04}$ \\
$\Gamma_{\rm SL}(\Omega_{bc}^{\prime\, 0})$ [$10^{-12}$\,GeV] & $0.6\pm0.2^{+0.01}_{-0.08}$   & $0.7\pm0.2^{+0.05}_{-0.13}$   & $0.7\pm0.2^{+0.01}_{-0.08}$ \\
\hline\hline
\end{tabular}
\caption{Final predictions for the decay widths, lifetimes, and 
lifetime ratios of the doubly heavy baryons $\Barybcprime$ (vector 
$bc$ diquark, $S_{\mathcal{D}}=1$) in three heavy-quark mass schemes. 
Uncertainties are parametric.}
\label{tab:fin_bc_prime}
\end{table}

For the $\Barybcprime$ baryons with $S_{\cal D}=1$, we predict the hierarchy
\begin{equation}
    \tau(\Xi_{bc}^{\prime\, 0}) < \tau(\Omega_{bc}^{\prime\, 0}) < \tau(\Xi_{bc}^{\prime\, +})\,.
\end{equation}
We can now address the question of whether the predicted lifetime 
pattern can help distinguish between the two possible heavy-diquark 
spin assignments in the $\mathcal{B}_{bc}$ sector, namely the 
unprimed states with $S_{\mathcal{D}}=0$ and the primed states with $S_{\mathcal{D}}=1$. 
Comparing the two sets of states at fixed light flavor, a clear 
separation is found for the electrically neutral baryons 
$\Xi_{bc}^{0}$ and $\Xi_{bc}^{\prime\, 0}$. In all three mass schemes, 
the predicted lifetime of $\Xi_{bc}^{0}$ is significantly larger 
than that of $\Xi_{bc}^{\prime\, 0}$, with the difference exceeding 
the quoted theoretical uncertainties. By contrast, the corresponding 
differences between the charged states and between the $\Omega_{bc}$ 
states remain compatible within the present theoretical precision. 
The stronger sensitivity of the neutral baryons can be traced 
directly to the spectator sector. In particular, their total widths 
receive a sizable weak-exchange contribution which is substantially 
enhanced in the $S_{\mathcal{D}}=1$ case. The reason is that the exchange 
topology depends directly on the heavy-light dimension-six matrix 
elements, in particular on those associated with the $cq$ spectator 
contribution, and the corresponding Wilson coefficients are larger 
for the vector-diquark baryons than for the scalar-diquark ones; 
cf.~Eqs.~\eqref{eq:dimsixBbc-Qq} and~\eqref{eq:dimsixBbcprime-Qq}. 
As a result, the exchange contribution to $\Xi_{bc}^{\prime\, 0}$ is 
significantly larger than to $\Xi_{bc}^{0}$, which increases the 
total width and correspondingly shortens the lifetime. For the 
charged and $\Omega_{bc}$ states, the same spin dependence is 
present, but the spectator contributions are distributed differently 
among exchange and interference topologies, so that the resulting 
shifts partially compensate each other.

\section{\texorpdfstring{Lifetime of the $B_c$ meson}{Lifetime of the Bc meson}}
\label{sec:Bc}

Following our previous discussion of the doubly heavy baryon lifetimes, we are now well equipped to extend our analysis to the lifetime of the doubly heavy $B_c$ meson. Its lifetime has been precisely measured~\cite{HFLAV:2024ctg},
\begin{equation}
    \tau(B_c)=0.510\pm0.009~\mathrm{ps}\,,
    \label{Eq:BcExp}
\end{equation}
providing an important test of the HQE description of heavy-hadron lifetimes.

The lifetime of the $B_c$ meson has been analyzed previously within the HQE and NRQCD in Refs.~\cite{Beneke:1996xe,Aebischer:2021ilm}. In particular, Ref.~\cite{Aebischer:2021ilm} organized the calculation as an expansion in the nonrelativistic velocity $v$ of the heavy constituents and included the leading contributions together with the higher-order terms available at the time. In the present work, we revisit the $B_c$ lifetime within the same general framework, but using updated short-distance input. Compared with Ref.~\cite{Aebischer:2021ilm}, we include NNLO corrections to the leading dimension-three contribution, NLO corrections to the chromomagnetic term, and discuss the impact of the Darwin term. We also incorporate presently available penguin contributions to the dimension-six four-quark terms. One of our main goals is to assess how these additional corrections affect the mass-scheme dependence of the $B_c$ lifetime prediction, which was found in Ref.~\cite{Aebischer:2021ilm} to be rather pronounced.

As in the case of doubly heavy baryons, the expansion for the $B_c$ system is naturally organized according to NRQCD power counting. We refer to the discussion at the beginning of Sec.~\ref{Sec:ME2quark} and only note here that the relevant nonrelativistic velocities are not especially small, in particular for the charm quark $v_c^2\sim0.4$, which is about twice the corresponding value in the case of $\mathcal{B}_{bc}$ baryons (see Table \ref{tab:v2Values}), while for the bottom quark $v_b^2 \sim 0.04$. 

We now collect the nonperturbative matrix elements needed for the $B_c$ lifetime prediction. More detailed discussions can be found in Refs.~\cite{Beneke:1996xe,Aebischer:2021ilm}. In NRQCD, the matrix elements of the two-quark operators can be expressed in terms of the nonrelativistic kinetic-energy parameter $T$ and the wave function at the origin. For the kinetic operator we have
\begin{equation}
 \frac{\langle B_c \vert \Psi_{\bar b}^\dagger (i\vec{D})^2\Psi_{\bar b}\vert B_c \rangle}{2M_{B_c}}
 =
 \frac{\langle B_c \vert \Psi_{c}^\dagger (i\vec{D})^2\Psi_{c}\vert B_c \rangle}{2M_{B_c}}
 =
 \frac{2\,m_bm_c}{m_b+m_c}\,T\,.
\end{equation}

The chromomagnetic matrix elements are determined by the $B_c$ wave function at the origin according to 
\begin{equation}
\mu_{G,b}^2(B_c)
=
-\frac{\langle B_c|\Psi_{\bar b}^\dagger\, g_s\,\vec{\sigma}\!\cdot\!\vec{B}\,\Psi_{\bar b}|B_c\rangle}{2M_{B_c}}
=
\frac{4}{3}g_s^2\,\frac{|\Phi(0)|^2}{m_c}\,,
\end{equation}
and
\begin{equation}
\mu_{G,c}^2(B_c)
=
-\frac{\langle B_c|\Psi_c^\dagger\, g_s\,\vec{\sigma}\!\cdot\!\vec{B}\,\Psi_c|B_c\rangle}{2M_{B_c}}
=
\frac{4}{3}g_s^2\,\frac{|\Phi(0)|^2}{m_b}\,.
\end{equation}

We next consider the dimension-six four-quark contributions. For these terms we use the same operator basis as in Eq.~\eqref{Eq:4qDef}, with the light-quark field replaced by the charm field. The corresponding expressions can be obtained from the formulas for the $bc$ baryons by translating the baryonic topology labels $\mathrm{int}^+$ and $\mathrm{exc}$, corresponding to constructive Pauli interference and weak exchange, respectively, into the standard mesonic nomenclature 
\begin{equation*}
\mathrm{int}^+ \to \mathrm{WA}\,,
\qquad
\mathrm{exc} \to \mathrm{PI}\,,
\end{equation*}
where WA and PI denote the weak annihilation and Pauli interference, respectively, see e.g., Figs.~2 and 3 in Ref.~\cite{Gratrex:2022xpm}. The corresponding leading-order expressions are collected in Eqs.~\eqref{Eq:bcintplustree}--\eqref{Eq:bcSL}. In the numerical analysis we also include the penguin contributions to the dimension-six four-quark terms~\cite{Chang:2000ac,Aebischer:2021ilm} with the corresponding expressions collected in Appendix~\ref{AppA2} given in terms of bag parameters $B_{B_c}$ and $B^\prime_{B_c}$, see Eqs.~\eqref{IntPlusPenguin}-\eqref{Eq:excPenguin}. 
In the vacuum insertion approximation one has
\begin{equation}
B_{B_c}=B'_{B_c}=1\,.
\end{equation}
We use this choice for the central values and estimate the associated hadronic uncertainty from variations of the bag parameters by $20\%$ around unity.

The seminumerical expression for the total decay width of the $B_c$ meson in the kinetic mass scheme reads
\begin{equation}
\begin{aligned}
    \Gamma(B_c)/\text{ps}^{-1}=&(\undermarker{1.01}{LO}+\undermarker{0.32}{NLO}+\undermarker{0.37}{NNLO})+(\undermarker{0.68}{LO}-\undermarker{0.03}{NLO}-\undermarker{0.01}{NNLO})
    \\
    &-0.21\frac{\mu_{\pi,c}^2}{0.49\,{\rm GeV}^2}+0.005\frac{\mu_{G,c}^2}{0.13\,{\rm GeV}^2}+1.25\frac{\rho_{D,c}^3}{0.34\,{\rm GeV}^3}
    \\
    &-0.01\frac{\mu_{\pi,b}^2}{0.75\,{\rm GeV}^2}+0.003\frac{\mu_{G,b}^2}{0.41\,{\rm GeV}^2}-0.05\frac{\rho_{D,b}^3}{0.34\,{\rm GeV}^3}
    \\
    &+0.13B'_{B_c}-0.04B_{B_c},
\end{aligned}
\label{Eq:SeminumBc}
\end{equation}
where the first and second brackets in the first line correspond to charm and bottom dimension-three decay contributions, respectively, and the common renormalization scale $\mu_{bc}=\sqrt{\mu_b\mu_c}= 2.6$ GeV is used. 

To test the sensitivity of the $B_c$ lifetime to the heavy-quark mass definition, we collect the breakdown of various contributions in Table \ref{tab:BcmesonResults} in three different mass schemes. As expected, in the case of the contributions coming from $b$ decays, both the scheme dependence, as well as scale variations, are considerably more stable than for the $c$ sector. By including the NNLO term in the charm sector we find that the difference between the kinetic and $\Upsilon$ mass schemes reduces, while the value in the $\overline{\rm MS}$ scheme shows significantly weaker improvement, which is also reflected in the large scale dependence in this scheme. 
In comparison to the previous predictions in the $\overline{\rm MS}$ and the $\Upsilon$ schemes\footnote{We note that our "$\Upsilon$ scheme" corresponds to "meson" scheme in Ref.~\cite{Aebischer:2021ilm}.} in Ref.~\cite{Aebischer:2021ilm}, despite our inclusion of the NNLO correction to the leading decay term, this scheme dependence remains sizable.

\begin{table}
\centering\footnotesize
\begin{tabular}{cccc}
\hline\hline
Decay widths &  $\overline{\mathrm{MS}}$ & Kinetic & $\Upsilon$ \\
\hline
$\bar{b}\to \bar{c}u(\bar{s}+\bar{d})$ & $0.31^{+0.25}_{-0.02}$ & $0.31^{+0.03}_{-0.01}$ & $0.31^{+0.01}_{-0.01}$ \\
\hline
$\bar{b}\to \bar{c}c(\bar{s}+\bar{d})$ & $0.15^{+0.07}_{-0.01}$ & $0.15^{+0.02}_{-0.00}$ & $0.15^{+0.02}_{-0.00}$  \\
\hline
$\bar{b}\to c(e+\mu+\tau)\nu$ & $0.12^{+0.10}_{-0.00}$ & $0.13^{+0.01}_{-0.01}$ & $0.12^{+0.01}_{-0.02}$ \\
\hline
$\sum \bar{b}\to \bar{c}$ & $0.58^{+0.42}_{-0.02}$ & $0.59^{+0.04}_{-0.01}$ & $0.57^{+0.01}_{-0.00}$ \\
\hline\hline
$c\to (s+d)u(\bar{d}+\bar{s})$ & $0.55^{+0.56}_{-0.15}+({0.75^{+0.31}_{-0.10}})\big\vert_{\rho_{D,c}^3}$ & $1.16^{+0.35}_{-0.12}+({0.92^{+0.24}_{-0.08}})\big\vert_{\rho_{D,c}^3}$ & $1.00^{+0.39}_{-0.13}+({0.90^{+0.22}_{-0.08}})\big\vert_{\rho_{D,c}^3}$ 
\\[1ex]
\hline
$c\to (s+d)(e+\mu)\nu$ & $0.18^{+0.09}_{-0.03}+({0.29^{+0.03}_{-0.01}})\big\vert_{\rho_{D,c}^3}$ & $0.33^{+0.00}_{-0.01}+({0.32^{+0.00}_{-0.00}})\big\vert_{\rho_{D,c}^3}$ & $0.32^{+0.03}_{-0.02}+({0.33^{+0.00}_{-0.00}})\big\vert_{\rho_{D,c}^3}$ 
\\[1ex]
\hline
$\sum c\to s$ & $0.72^{+0.65}_{-0.18}+({1.05^{+0.33}_{-0.12}})\big\vert_{\rho_{D,c}^3}$ & $1.48^{+0.35}_{-0.12}+({1.25^{+0.24}_{-0.08}})\big\vert_{\rho_{D,c}^3}$ & $1.32^{+0.22}_{-0.08}+({1.22^{+0.43}_{-0.15}})\big\vert_{\rho_{D,c}^3}$
\\[1ex]
\hline\hline
WA: $\bar{b}c \to c(\bar{s}+\bar{d})$ & $0.12^{+0.13}_{-0.03}$ & $0.13^{+0.09}_{-0.03}$ & $0.13^{+0.09}_{-0.03}$ \\
\hline
WA: $\bar{b}c \to \tau\nu$ & $0.03$ & $0.04$ & $0.04$ \\
\hline
PI & $-0.11^{+0.04}_{-0.17}$ & $-0.08^{+0.03}_{-0.06}$ & $-0.10^{+0.03}_{-0.07}$ \\
\hline\hline
$\Gamma_{B_c}$ & \makecell{\rule{0pt}{3ex}$1.35\pm0.02^{+1.03}_{-0.19}$\\[0.5ex]$\quad+{(1.05\pm0.21^{+0.33}_{-0.12})}\big\vert_{\rho_{D,c}^3}$} & \makecell{\rule{0pt}{3ex}$2.16\pm0.04^{+0.42}_{-0.13}$\\[0.5ex]$\quad+{(1.25\pm0.25^{+0.24}_{-0.08})}\big\vert_{\rho_{D,c}^3}$} & \makecell{\rule{0pt}{3ex}$1.97\pm0.04^{+0.47}_{-0.15}$\\[0.5ex]$\quad+{(1.22\pm0.24^{+0.22}_{-0.08})}\big\vert_{\rho_{D,c}^3}$}
\\[3ex]
\hline\hline
\end{tabular}
\caption{Results for separate two-quark and four-quark decay width contributions given in units of ps$^{-1}$. The uncertainty from renormalization scale running is given for each decay mode, while in the final result we give the hadronic uncertainties, followed by the scale uncertainties. The renormalization scale is varied in the range $\mu\in[1.5,3.5]$ GeV with the central value $\mu_{bc}=\sqrt{\mu_b\mu_c}= 2.6$ GeV. In the charm sector, we separate the contribution from the Darwin term.
}\label{tab:BcmesonResults}
\end{table}

Furthermore, as seen from the Table \ref{tab:BcmesonResults}, the impact of the Darwin term in the $B_c$ system is numerically highly significant. As it is known from previous lifetimes analyses, the Wilson coefficient of the Darwin operator is particularly large, which can be also observed in our numeric expression for the total decay width in Eq.~\eqref{Eq:SeminumBc}, in particular, the corresponding contribution proportional to $\rho_{D,c}^3(B_c)$. 
In addition, this matrix element turns out to be unusually large.  
In the nonrelativistic picture, the matrix element of the Darwin operator is given in terms of the $B_c$ wave function as
\begin{equation}
\rho_{D,Q}^3(B_c)
=
\frac{1}{2}\frac{\langle B_c|\Psi_Q^\dagger g_s\,(\vec{\mathcal D}\!\cdot\!\vec E)\,\Psi_Q|B_c\rangle}{2M_{B_c}}
=
\frac{2}{3}g_s^2\,|\Phi(0)|^2\,,
\end{equation}
where
\begin{equation}
   |\Phi(0)|^2=\frac{f_{B_c}^2M_{B_c}}{12}\,, 
   \label{Eq:PhiBc}
\end{equation}
and the hyperfine splitting relation  
\begin{equation}
   M_{B_c^\ast}-M_{B_c}
   =
   \frac{8}{9}g_s^2\,\frac{|\Phi(0)|^2}{m_bm_c}\,
   \label{eq:BcDarwinMassDiff}
\end{equation}
is used. 
We obtain
\begin{equation}
\rho_{D,Q}^3(B_c)\simeq 0.34~\mathrm{GeV}^3\,.
\end{equation} 
This value is significantly larger than the corresponding estimate for heavy-light $B_q$ mesons. For comparison, Ref.~\cite{Lenz:2022rbq} quotes the result for the $B_q$ meson $\rho_D^3(B_q)\simeq0.061~\mathrm{GeV}^3$ at $\mu=1~\mathrm{GeV}$ obtained in vacuum insertion approximation, so that the Darwin matrix element for $B_c$ meson is larger by approximately a factor of five. Similarly, the value of the Darwin matrix element for the $B_c$ meson is significantly larger than for the $\Barybc$ baryons, where $\rho_{D,c}^3(\Barybc) \simeq 0.08$ GeV$^3$. This enhancement in the Darwin matrix element for the $B_c$ meson can be traced back to the significantly larger wave function reflecting the compactness of the $B_c$ state. Indeed, the value 
$|\Phi(0)|^2\simeq 0.095~\mathrm{GeV}^3\,$
is about a factor of $2.4$ larger than the estimated value $|\Psi_{bc}^{\mathcal B_{bc}}(0)|^2=0.04~\mathrm{GeV}^3$ used for the $\mathcal{B}_{bc}$ baryons, see Table~\ref{Tab:KiselevValues}. 

Our final prediction for the lifetime of the $B_c$ meson in {\it the kinetic mass scheme} is 
\begin{equation}
    \tau^{\rm kin}(B_c)=0.293\pm0.022^{+0.019}_{-0.047}\, {\rm ps},
\end{equation}
where the first error represents the hadronic uncertainties and the second uncertainty represents the renormalization scale variation in the range $\mu\in [1.5,3.5]$ GeV. 
This result is in tension with the measured value in \eqref{Eq:BcExp}. Interestingly, dropping the Darwin contribution results with the value $\tau^{\rm kin}_{{\rm excl. }\, \rho_{D,c}}(B_c)=0.463\pm0.009^{+0.029}_{-0.075}$, compatible with experiment.

\section{Summary and conclusions}
\label{sec:conclusion}
In this work we have presented updated predictions for the lifetimes of 
all weakly decaying doubly heavy baryons, $\Barybb$, $\Barycc$, and 
$\Barybc$, as well as for the $B_c$ meson, within the framework of the 
heavy-quark expansion. Compared with previous analyses 
\cite{Kiselev:1999kh,Likhoded:1999yv,Guberina:1999mx,Kiselev:2001fw,Berezhnoy:2018bde,
Cheng:2018mwu,Cheng:2019sxr,Yang:2022nps,Aebischer:2021ilm,
Beneke:1996xe,Dulibic:2023jeu}, the present analysis incorporates a 
substantially updated set of short-distance contributions. In particular, 
we include NNLO corrections to the leading dimension-three contribution, 
the NLO corrections to the chromomagnetic term, the full set of presently 
available NLO corrections to the dimension-six heavy-light spectator contribution, the effect of 
subleading dimension-seven spectator operators, as well as penguin effects. For the $\Barybb$ and 
$\Barycc$ baryons we present the results in the kinetic and 
$\overline{\rm MS}$ schemes, while for the $\Barybc$ baryons and the 
$B_c$ meson we additionally employ the $\Upsilon$ scheme, in order to 
assess the residual scheme dependence of the predictions.

The nonperturbative matrix elements entering the lifetime predictions 
have been determined within the constituent diquark--quark picture, with the 
heavy-light contributions extracted from hyperfine mass splittings via 
a nonrelativistic constituent quark model, and the heavy--heavy 
contributions evaluated using wave functions at the origin taken from 
potential-model analyses. Using the same setup for all doubly heavy baryons, and applying the analogous NRQCD treatment to the $B_c$ meson, enables a consistent comparison across the different doubly heavy sectors.

Our main findings can be summarized as follows. For the doubly bottom 
baryons we predict the lifetime hierarchy 
$\tau(\Xi_{bb}^{0}) < \tau(\Xi_{bb}^{-}) \simeq \tau(\Omega_{bb}^{-})$, 
with the splitting between the $\Xi_{bb}^{0}$ and the negatively charged 
states driven primarily by the sizable weak-exchange contribution to 
the $\Xi_{bb}^{0}$ decay width. For the doubly charmed baryons we obtain 
the hierarchy 
$\tau(\Xi_{cc}^{+}) < \tau(\Omega_{cc}^{+}) < \tau(\Xi_{cc}^{++})$, in 
agreement with our previous analysis~\cite{Dulibic:2023jeu} and with 
earlier works in the literature. The updated prediction for the $\Xi_{cc}^{++}$ lifetime moves even closer to 
the experimental central value, as a consequence of the newly included NNLO correction to the 
leading dimension-three contribution. We note, however, that 
sizable uncertainties remain, dominated by the residual 
renormalization-scale dependence, and the nonperturbative input for the 
four-quark matrix elements.

\begin{figure}[h]
    \centering
    \includegraphics[width=1\linewidth]{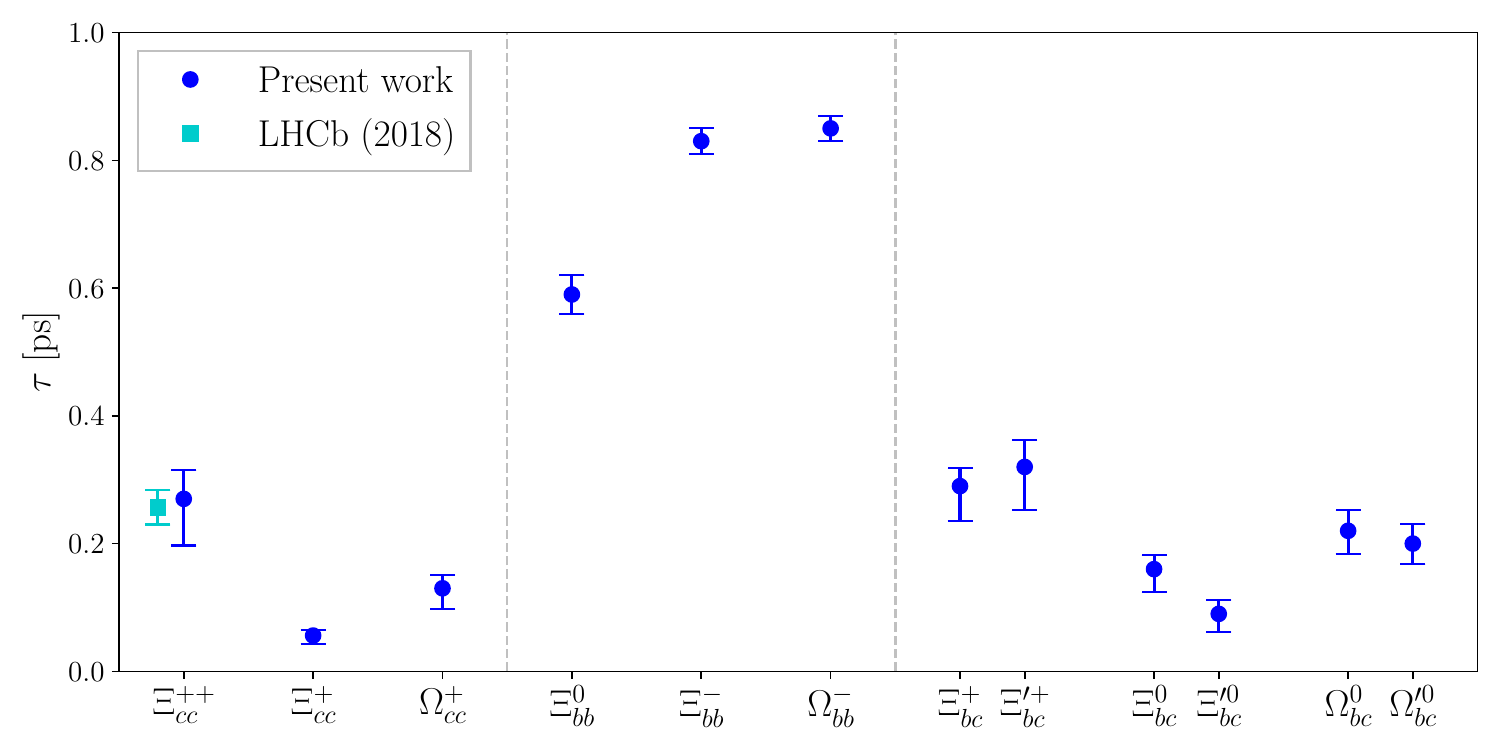}
    \caption{Summary of our predictions for the lifetimes of all doubly heavy baryons, evaluated in {\it the kinetic scheme}, alongside the only measured lifetime of the $\Xi_{cc}^{++}$ baryon by the LHCb Collaboration~\cite{LHCb:2018zpl}.}
    \label{fig:summary}
\end{figure}

For the $\Barybc$ baryons, where the ground-state heavy-diquark spin is 
not firmly established, we have presented predictions for both the 
$S_{bc}=0$ (unprimed) and $S_{bc}=1$ (primed) assignments. 
The possible 
distinction between the two scenarios is visible in the predicted lifetimes of the neutral 
$\Xi_{bc}^{0}$ and $\Xi_{bc}^{\prime\, 0}$ states, with a difference that exceeds the estimated theoretical uncertainty.  

Our final predictions for the lifetimes of doubly heavy baryons are summarized in Fig.~\ref{fig:summary} and they are broadly consistent with 
earlier HQE-based analyses 
\cite{Kiselev:1999kh,Likhoded:1999yv,Kiselev:2001fw,Berezhnoy:2018bde,Guberina:1999mx,
Cheng:2018mwu,Cheng:2019sxr,Yang:2022nps}, although the central values differ due to differences in the treatment of the 
perturbative expansion and of the hadronic matrix elements. Relative to 
the works in Refs.~\cite{Kiselev:1999kh,Kiselev:2001fw}, which relied on the diquark picture 
combined with potential-model inputs, our predictions include the full 
set of known NLO and NNLO perturbative corrections. Compared with the 
more recent analyses in Refs.
\cite{Cheng:2018mwu,Cheng:2019sxr,Cheng:2026mlv}, which rely on 
bag-model estimates of the four-quark matrix elements, we extract the 
heavy-light matrix elements from hyperfine mass splittings. 
The resulting lifetime hierarchies in the $\Barybb$ and $\Barycc$ 
sectors are nevertheless in broad agreement across these analyses, 
providing a useful consistency check between different nonperturbative 
frameworks.

Turning to the $B_c$ meson, our updated prediction, which includes 
NNLO corrections to the leading decay term, NLO corrections to the 
chromomagnetic operator, 
agrees reasonably well across all three mass schemes, within uncertainties, with the precisely measured lifetime 
$\tau(B_c)^{\rm exp}=0.510\pm0.009$\,ps~\cite{HFLAV:2024ctg} only if the Darwin contribution is excluded. We  
emphasize the numerically sizable impact of the Darwin term, whose 
matrix element is significantly larger in the $B_c$ system than in
heavy-light $B_q$ mesons, owing to the compact nature of the 
$B_c$ meson which exhibits a large wave function at the origin.

We also find that the inclusion of higher-order 
perturbative corrections reduces the spread between predictions obtained 
in different heavy-quark mass schemes, although a sizable residual scheme 
dependence remains.

Overall, our analysis demonstrates that the present status of the HQE 
provides a reasonably controlled description of the lifetimes of doubly 
heavy hadrons. The dominant theoretical uncertainties originate from 
the hadronic input parameters, in particular the four-quark matrix 
elements, and from the residual renormalization-scale dependence. It is expected that future lattice calculations of four-quark matrix elements could provide significant improvement to the hadronic uncertainties. 
The ongoing experimental 
efforts at LHCb and other facilities to measure the lifetimes of doubly 
heavy baryons will provide 
valuable tests of the heavy-quark expansion and of the current theoretical framework used to determine the required hadronic matrix elements.

\section*{Note added}
While this work was being finalized, Ref.~\cite{Cheng:2026mlv} appeared. That paper presents an HQE analysis of the $\Barybb$ and $\Barycc$ lifetimes, overlapping partially with our study. The present work, in addition, also covers the $\Barybc$ baryons and the $B_c$ meson not discussed in \cite{Cheng:2026mlv}. In the overlapping sectors, the main differences concern the treatment of the four-quark matrix elements and the perturbative input for the leading dimension-three contribution. In particular, Ref.~\cite{Cheng:2026mlv} employs a bag-model description of the four-quark matrix elements, whereas our analysis uses the nonrelativistic constituent quark model and relates these matrix elements to mass splittings. Moreover, we include the NNLO correction to the leading dimension-three contribution, which is not taken into account there. Overall, we find broad agreement with their numerical results.

\subsection*{Acknowledgments}
We thank Matteo Fael and Matthias Steinhauser for the code and useful discussions on the NNLO corrections. We also acknowledge a useful correspondence with Maria Laura Piscopo. The support of the Croatian Science Foundation (HRZZ) under the project “Nonperturbative QCD in heavy flavour physics” (IP-2024-05-4427) is acknowledged.

\newpage
\appendix

\section{Contributions from four-quark operators}
\label{AppA}
In this appendix we collect the analytic expressions for the dimension-six and dimension-seven spectator contributions to the inclusive decay widths. The leading-order results are presented in subsection~\ref{AppA1}, while the corresponding dimension-six expressions including penguin contributions are given in subsection~\ref{AppA2}.
\subsection{Leading four-quark contributions}
\label{AppA1}

The leading-order dimension-six contributions to the total decay widths arising from the topologies relevant to $\mathcal{B}_{bb}$ and $\mathcal{B}_{cc}$, involving interactions with the light spectator $q$, are given by:
\begin{eqnarray}
\Ga^{q}_{6,\intp}(x_1, x_2)
&=\Ga_0(m_Q)\displaystyle{\frac{16\pi^2\sqrt{\la}}{m_Q^3}}
\Bigg\{
\Big[\big((x_1-x_2)^2+x_1+x_2-2\big)(2C_1 C_2+\NC C_2^2)\Big]\Opsix{1}{q}
\nonumber\\ 
&\quad
-\Big[2\big(2(x_1-x_2)^2-x_1-x_2-1\big)(2C_1 C_2+\NC C_2^2)\Big]\Opsix{2}{q}
\nonumber\\
&\quad
+\Big[\big((x_1-x_2)^2+x_1+x_2-2\big)C_1^2\Big]\Opsixt{1}{q}
\nonumber\\
&\quad
-2\Big[\big(2(x_1-x_2)^2-x_1-x_2-1\big)C_1^2\Big]\Opsixt{2}{q}
\Bigg\}\,,
\label{Eq:A1}
\end{eqnarray}
\begin{equation}
\Ga^q_{6,\exc}(x_1,x_2)
=
\Gamma_0(m_Q)\displaystyle{\frac{16\pi^2}{m_Q^3}}
\left(2 \NC \sqrt{\la}\,(1-x_1-x_2)\right)
\Big\{(2 C_1 C_2)\Opsix{1}{q}+(C_1^2+C_2^2)\Opsixt{1}{q}\Big\}\,,
\end{equation}
\begin{equation}
\Ga^{q}_{6,\intm}(x_1,x_2)
=
\Ga^{q}_{6,\intp}(x_1, x_2)\big|_{C_1\longleftrightarrow C_2}\,,
\end{equation}
and, for the semileptonic decays
\begin{equation}
\Ga^{q,\text{SL}}_{6,\intp}(x_\ell,0)
=
\Ga^{q}_{6,\intp}(x_\ell,0)\big|_{C_1 \to 0,\, C_2 \to 1,\, N_C \to 1}\,.
\end{equation}
Here, $x_i$ denote the dimensionless mass ratios, defined in Eq.~\eqref{eq:massratios}, corresponding to the internal masses in the loop, and $\la\equiv \la(x_1,x_2)=1+x_1^2+x_2^2-2(x_1+x_2+x_1x_2)$ denotes the K\"all\'en function with the first argument set to unity. Once again, we note that the matrix elements $\langle O_i\rangle$ in the above expressions are defined using the normalization in Eq.~\eqref{Eq:Def}.

For the $bc$ baryons we have dimension-six contributions from four-quark operators involving interactions between the $b$ and $c$ quarks with a light-quark spectator. The corresponding transition operators for the relevant $\text{int}^+$ and $\text{exc}$ topologies, shown in Fig.~3 (d,e,i,j), are
\begin{equation}
\begin{aligned}
\Gamma^{bc}_{6,\text{int}+, \text{tree}}&(y_-,0)
=
-\frac{G_F^2}{6\pi}\,(1-y_-)^2\,p_-^{2}
\Bigg[
\left(N_c C_1^2 + 2C_1 C_2\right)
\Bigg(
\left(1 + \frac{y_-}{2}\right) \langle O_1^{bc}\rangle
\\
&\hspace*{2.3cm}
-(1 + 2y_-)\frac{m_b^2}{p_-^2}
\left(
\langle O_2^{bc} \rangle + \frac{m_c}{m_b} \langle S_{--}^{bc}\rangle
+ \frac{m_c}{m_b} \langle S_{++}^{bc}\rangle
+ \frac{m_c^2}{m_b^2} \langle S_{+-}^{bc}\rangle
\right)
\Bigg)
\\
&\hspace*{-1.5cm}
+C_2^2
\Bigg(
\left(1 + \frac{y_-}{2}\right) \langle \widetilde{O}_1^{bc}\rangle
-(1 + 2y_-)\frac{m_b^2}{p_-^2}
\left(
\langle \widetilde{O}_2^{bc}\rangle + \frac{m_c}{m_b} \langle\widetilde{S}_{--}^{bc}\rangle
+ \frac{m_c}{m_b} \langle\widetilde{S}_{++}^{bc}\rangle
+ \frac{m_c^2}{m_b^2} \langle\widetilde{S}_{+-}^{bc}\rangle
\right)
\Bigg)
\Bigg],
\label{Eq:bcintplustree}
\end{aligned}
\end{equation}
and
\begin{equation}
\Gamma^{bc}_{6,\text{exc}, \text{tree}}(y_+,0)
=
\frac{G_F^2}{2\pi}\,(1-y_+)^2\,p_+^{2}
\Big[ 2C_1 C_2 \langle O_1^{bc}\rangle+(C_1^2+C_2^2)\langle\widetilde{O}_1^{bc}\rangle \Big]\,,
\label{Eq:bcexctree}
\end{equation}
respectively.

Here we denote $p_{\pm}^2=(m_b\pm m_c)^2$ and $y^i_{\pm}=m_i^2/p_{\pm}^2$.
For the semileptonic contribution, we have
\begin{equation}
\Gamma^{bc,\text{SL}}_{6,\intp}(y^\ell_-,0)
=
\Ga^{bc}_{6,\intp}(y^\ell_-,0)\big|_{C_1 \to 0,\, C_2 \to 1,\, N_C \to 1}\,.
\label{Eq:bcSL}
\end{equation}

The operators $O^{bc}_{1,2}$ and $\widetilde{O}^{bc}_{1,2}$ are defined in Eq.~\eqref{Eq:4qDef}, while the additional operators arising in the above expressions are defined as
\begin{equation}
\begin{aligned}
\Spp^{bc}&=(\overline{b}^i(1+\gamma^5)c^i)(\overline{c}^j(1+\gamma^5)b^j),
&\qquad
\Opp^{bc}&=(\overline{b}^i\gamma^\mu(1+\gamma^5)c^i)(\overline{c}^j\gamma_\mu(1+\gamma^5)b^j),
\\
\Smm^{bc}&=(\overline{b}^i(1-\gamma^5)c^i)(\overline{c}^j(1-\gamma^5)b^j),
&\qquad
\Omp^{bc}&=(\overline{b}^i\gamma^\mu(1-\gamma^5)c^i)(\overline{c}^j\gamma_\mu(1+\gamma^5)b^j),
\\
\Spm^{bc}&=(\overline{b}^i(1+\gamma^5)c^i)(\overline{c}^j(1-\gamma^5)b^j),
&\qquad
\Opm^{bc}&=(\overline{b}^i\gamma^\mu(1+\gamma^5)c^i)(\overline{c}^j\gamma_\mu(1-\gamma^5)b^j).
\end{aligned}
\label{Eq:PlusMinOps}
\end{equation}
Using the same normalization as in Eq.~\eqref{Eq:Def}, we find for the matrix elements of $bc$ baryons with diquark spin $S_\mathcal{D}=0$ in the leading nonrelativistic limit, in terms of the corresponding diquark wave functions at the origin,
\begin{equation}
\label{eq:scalarRLdimsixBbc-bc}
\begin{aligned}
\ME{\Spp^{bc}}{\Barybc} &=\frac{3}{2}\,|\Psi_{bc}^{\Barybc}(0)|^2\,,
&\qquad
\ME{\Opp^{bc}}{\Barybc} &=-4\,|\Psi_{bc}^{\Barybc}(0)|^2\,,
\\
\ME{\Smm^{bc}}{\Barybc} &=\frac{3}{2}\,|\Psi_{bc}^{\Barybc}(0)|^2\,,
&\qquad
\ME{\Omp^{bc}}{\Barybc} &=2\,|\Psi_{bc}^{\Barybc}(0)|^2\,,
\\
\ME{\Spm^{bc}}{\Barybc} &=-\,|\Psi_{bc}^{\Barybc}(0)|^2\,,
&\qquad
\ME{\Opm^{bc}}{\Barybc} &=2\,|\Psi_{bc}^{\Barybc}(0)|^2\,,
\end{aligned}
\end{equation}
while for $\mathcal{B}'_{bc}$ baryons with diquark spin $S_\mathcal{D}=1$ we obtain
\begin{equation}
\label{eq:vectorRLdimsixBbc-bc}
\begin{aligned}
\ME{\Spp^{bc}}{\Barybc} &=\frac{7}{2}\,|\Psi_{bc}^{\Barybc}(0)|^2\,,
&\qquad
\ME{\Opp^{bc}}{\Barybc} &=0,
\\
\ME{\Smm^{bc}}{\Barybc} &=\frac{7}{2}\,|\Psi_{bc}^{\Barybc}(0)|^2\,,
&\qquad
\ME{\Omp^{bc}}{\Barybc} &=2\,|\Psi_{bc}^{\Barybc}(0)|^2\,,
\\
\ME{\Spm^{bc}}{\Barybc} &=\,|\Psi_{bc}^{\Barybc}(0)|^2\,,
&\qquad
\ME{\Opm^{bc}}{\Barybc} &=2\,|\Psi_{bc}^{\Barybc}(0)|^2\,,
\end{aligned}
\end{equation}
The color-rearranged operators are given by the relation
\begin{equation}
\langle \widetilde{\mathcal O}^{bc} \rangle_\Barybc=-\widetilde{B}\langle \mathcal O^{bc}\rangle_\Barybc,
\end{equation}
where $\tilde{B}(\mu_h)=1$ at the low hadronic scale $\mu_h=1\,\text{GeV}$.

For the case of $B_c$ meson, 
we parametrize the relevant matrix elements in terms of bag parameters $B_{B_c}$ and $B_{B_c}^\prime$.

The full set of required matrix elements for the $B_c$ meson is given by
\begin{equation}
\begin{aligned}
\Opsix{1}{bc}_{B_c}=\Opsix{2}{bc}_{B_c}=\ME{\Spm^{bc}}{B_c}
&=\frac{f_{B_c}^2 M_{B_c}}{2}B_{B_c},
\\
\ME{\Smm^{bc}}{B_c}=\ME{\Spp^{bc}}{B_c}=\ME{\Omp^{bc}}{B_c}=\ME{\Opm^{bc}}{B_c}
&=-\frac{f_{B_c}^2M_{B_c}}{2}B_{B_c},
\end{aligned}
\end{equation}
with the analogous expressions for color-rearranged operators
\begin{equation}
\begin{aligned}
\Opsixt{1}{bc}_{B_c}=\Opsixt{2}{bc}_{B_c}=\ME{\Spmt^{bc}}{B_c}
&=\frac{f_{B_c}^2 M_{B_c}}{2}\frac{1}{3}B'_{B_c},
\\
\ME{\Smmt^{bc}}{B_c}=\ME{\Sppt^{bc}}{B_c}=\ME{\Ompt^{bc}}{B_c}=\ME{\Opmt^{bc}}{B_c}
&=-\frac{f_{B_c}^2M_{B_c}}{2}\frac{1}{3}B'_{B_c}\,.
\end{aligned}
\end{equation}

The expressions for the dimension-seven contributions to the decay widths for the heavy-light quark topologies are given by
\begin{equation}
\begin{split}
\Ga^{q}_{7,\intp}(x_1,x_2)
&=\Ga_0(m_Q)\frac{16\pi^2\sqrt{\la}}{m_Q^4}(2 C_1 C_2 + N_C C_2^2)
\bigg\{
2 \left[2(x_1- x_2)^2  - x_1 -x_2 -1\right]
\left(\OpsevenP{1}{q} +{\OpsevenP{1}{q}}^{\dagger}\right)
\\
&\quad
+\frac{2}{\lambda}\bigg[
(x_1+x_2-1)\Big((x_1-x_2)^2+x_1+x_2-2\Big)
+\lambda\Big(2(x_1-x_2)^2+x_1+x_2\Big)
\bigg]\OpsevenP{2}{q}
\\
&\quad
+\frac{4}{\lambda}\bigg[
(1-x_1-x_2)\Big(\lambda+(x_1-x_2)^2+x_1+x_2-2\Big)
\\
&\qquad\qquad
+\lambda\Big(1+2x_1+2x_2-6(x_1-x_2)^2\Big)
\bigg]\OpsevenP{3}{q}
\bigg\}
\\
&\quad
+\bigg\{
\OpsevenP{i}{q}\to \OpsevenPt{i}{q},
\,
(2 C_1 C_2 + N_C C_2^2)\to C_1^2
\bigg\}\,,
\end{split}
\end{equation}
\begin{equation}
\Ga^q_{7,\exc}(x_1,x_2)
=
\Gamma_0(m_Q)\frac{16\pi^2}{m_Q^4}
\Bigg[
\frac{12\left((1-x_1-x_2)^2+(x_1+x_2)\la\right)}{\sqrt{\la}}
\Bigg]
\Big\{(2 C_1 C_2)\OpsevenP{2}{q}+(C_1^2+C_2^2)\OpsevenPt{2}{q}\Big\}\,.
\end{equation}
\begin{equation}
\Ga^{q}_{7,\intm}(x_1,x_2)
=
\Ga^{q}_{7,\intp}(x_1,x_2)\big|_{C_1\longleftrightarrow C_2}\,,
\end{equation}
\begin{equation}
\Ga^{q,\text{SL}}_{7,\intp}(x_\ell,0)
=
\Ga^{q}_{7,\intp}(x_\ell,0)\big|_{C_1 \to 0,\, C_2 \to 1,\, N_C \to 1}\,.
\label{eq:explicitDim7Bar}
\end{equation}
\vfill

\subsection{\texorpdfstring{Penguin contributions to four-quark operators for $\Barybc$ baryons and the $B_c$ meson}{Penguin contributions to four-quark operators for bc baryons and the Bc meson}}
\label{AppA2}
The expressions for the dimension-six four-quark penguin contributions to the transition operators corresponding to the $\text{int}^+$ and $\text{exc}$ topologies for the $\mathcal{B}_{bc}$ baryons read
\begin{equation}\hspace{-2.2cm}
\begin{aligned}
\Gamma^{bc}_{6,\text{int}+,\text{penguin}}(y_-,0)
=
\frac{G_F^2}{\pi} (1-y_-)^2 p_-^{2}\Bigg\{ \frac{-1}{6}
\Bigg[
\big(&N_c C_4^2 +2N_c C_1 C_4+2C_3 C_4+2C_1 C_3+2C_2 C_4\big)
\\
\cdot\Bigg(
\left(1 + \frac{y_-}{2}\right) \langle O_1^{bc} \rangle
-(1 + 2y_-)\frac{m_b^2}{p_-^2}
&\left(
\langle O_2^{bc} \rangle + \frac{m_c}{m_b} \langle S_{--}^{bc} \rangle
+ \frac{m_c}{m_b} \langle S_{++}^{bc} \rangle
+ \frac{m_c^2}{m_b^2} \langle S_{+-}^{bc} \rangle
\right)
\Bigg)
\\
+\left(C_3^2+2C_2 C_3\right)
\Bigg(
\left(1 + \frac{y_-}{2}\right) \langle\widetilde{O}_1^{bc}\rangle
-(1 + 2y_-)\frac{m_b^2}{p_-^2}
&\left(
\langle\widetilde{O}_2^{bc}\rangle + \frac{m_c}{m_b} \langle\widetilde{S}_{--}^{bc}\rangle
+ \frac{m_c}{m_b} \langle\widetilde{S}_{++}^{bc}\rangle
+ \frac{m_c^2}{m_b^2} \langle\widetilde{S}_{+-}^{bc}\rangle
\right)
\Bigg)
\Bigg]
\\
\hspace*{1cm}
+\frac{m_b m_c}{2 p_-^2}
\Bigg[
\big(N_c C_6(C_1+C_4)+C_5(C_1+C_4)+C_6(C_2+C_3)\big)
&\left( \langle S_{++}^{bc}\rangle+\langle S_{--}^{bc}\rangle+\frac{m_c}{m_b}\langle S_{+-}^{bc}\rangle+\frac{m_c}{m_b}\langle S_{-+}^{bc}\rangle\right)
\\
+\left(C_2 C_5+C_3 C_5\right)
&\left( \langle\widetilde{S}_{++}^{bc}\rangle+\langle\widetilde{S}_{--}^{bc}\rangle+\frac{m_c}{m_b}\langle\widetilde{S}_{+-}^{bc}\rangle+\frac{m_c}{m_b}\langle\widetilde{S}_{-+}^{bc}\rangle\right)
\Bigg]
\\
&\hspace{-3cm}+
\left(N_c C_6^2+2C_5 C_6\right)\langle S_{+-}^{bc}\rangle+C_5^2\langle\widetilde{S}_{+-}^{bc}\rangle
\Bigg\},
\end{aligned}
\label{IntPlusPenguin}
\end{equation}
and
\begin{equation}
\begin{aligned}
\Gamma^{bc}_{6,\text{exc},\text{penguin}}(y_+,0)
=
\frac{G_F^2}{2\pi} (1-y_+)^2 p_+^{2}\Bigg\{
\Bigg[
&\left(2C_1 C_3+2C_2 C_4+2C_3 C_4\right)\langle O_1^{bc}\rangle
\\
+&\left(C_3^2+C_4^2+2C_1 C_4+2C_2 C_3\right)\langle\widetilde{O}_1^{bc}\rangle
\Bigg]
\\
-\frac{m_b m_c}{2 p_+^2}
\Bigg[
\Big(C_2 C_6+C_3 C_6+C_1 C_5+C_4 C_5\Big)
&\left(2\langle S_{--}^{bc}\rangle-\langle O_{-+}^{bc}\rangle+\langle O_{+-}^{bc}\rangle+\frac{m_c}{m_b}(2\langle S_{+-}^{bc}\rangle-2\langle O_1^{bc}\rangle)\right)
\\
+\Big(C_2 C_5+C_3 C_5+C_1 C_6+C_4 C_6\Big)
&\left(
2\langle\widetilde{S}_{--}^{bc}\rangle-\langle\widetilde{O}_{-+}^{bc}\rangle+\langle\widetilde{O}_{+-}^{bc}\rangle
+\frac{m_c}{m_b}(2\langle\widetilde{S}_{+-}^{bc}\rangle-2\langle\widetilde{O}_1^{bc}\rangle)
\right)
\Bigg]\Bigg\},
\end{aligned}
\label{Eq:excPenguin}
\end{equation}
respectively. The definitions of the corresponding operators and the expressions for their matrix elements are given in Appendix subsection~\ref{AppA1}. As in the case of the leading-order expressions, the corresponding formulas for the $B_c$ meson can be obtained from those for the $bc$ baryons after translating the baryonic topology labels $\mathrm{int}^+\to \mathrm{WA}$ and $\mathrm{exc}\to \mathrm{PI}$, together with the exchanges $p_+ \leftrightarrow p_-$ and $y^c_+ \leftrightarrow y^c_-$.

\section{Decomposition of contributions to doubly-heavy baryon decay widths}\label{sec:appendixBreakdown}
In this appendix, we present our results for the decay widths in the kinetic scheme, decomposed into the nonleptonic dimension-three and five contribution, the nonleptonic Darwin contribution, the four-quark contributions, given separately for each topology, and the semi-leptonic contributions.

\begin{table}[ht]
    \centering
    \begin{tabular}{ccccccc}
    \hline\hline
    & $\Gamma_{3+5}^{\rm NL}$ & $\Gamma^{\rm NL}_\rho$ & $\Gamma_{\rm exch}$ & $\Gamma_{\rm int-}$ & $\Gamma_{\rm 3+5+\rho}^{\rm SL}$ \\\hline
    $\Xi_{bb}^0$ & $0.671-0.024+0.003$ & $-0.027$ & \makecell{$0.239+0.016$\\$0.036$} & $0$ & $0.202$\\\hline
    $\Xi_{bb}^{-}$ & $0.671-0.024+0.003$ & $-0.027$ & $0$ & \makecell{$-0.062+0.016$\\$0.010$} & $0.202$\\\hline
    $\Omega_{bb}^{-}$ & $0.671-0.024+0.003$ & $-0.029$ & $0$ & \makecell{$-0.077+0.022$\\$0.007$} & $0.201$\\\hline\hline
    \end{tabular}
    \caption{Different contributions to the decay widths of $\Barybb$ baryons in units of $10^{-12}$ GeV, calculated in \textit{the kinetic mass scheme}. The dimension-three and five contributions are given in three terms corresponding to LO, NLO, and NNLO, respectively, while their semileptonic counterpart is summed into one term. The dimension-six LO and NLO contributions for the four-quark decay widths are given in the first line of their respective cells, with the dimension-seven contributions in the second line.}
    \label{tab:bbBreakdown}
\end{table}
\begin{table}[h]
    \centering\small
    \begin{tabular}{ccccccccc}
    \hline\hline
    & $\Gamma^{\rm NL}_{\rm 3+5}$ & $\Gamma^{\rm NL}_{\rho}$ & $\Gamma_{\rm exch}$ & $\Gamma_{\rm int-}$ & $\Gamma^{\rm NL}_{\rm int+}$ & $\Gamma_{\rm int+}^{\rm SL}$ & $\Gamma_{\rm 3+5+\rho}^{\rm SL}$ \\\hline
    $\Xi_{cc}^{++}$ & $0.92+0.54+0.55$ & $0.19$ & $0$ & \makecell{$-1.12+0.29$\\$0.54$} & $0$ & $0$ & $0.55$ \\\hline
    $\Xi_{cc}^{+}$ & $0.92+0.54+0.55$ & $0.19$ & \makecell{$5.38+1.35$\\$2.23$} & $0$ & \makecell{$0.08-0.00$\\$-0.04$} & \makecell{$0.06-0.02$\\$-0.03$} & $0.55$ \\\hline
    $\Omega_{cc}^{+}$ & $0.92+0.54+0.55$ & $0.21$ & \makecell{$0.37+0.09$\\$0.16$} & $0$ & \makecell{$2.18-0.11$\\$-0.81$} & \makecell{$1.47-0.62$\\$-0.51$} & $0.56$ \\\hline\hline
    \end{tabular}
    \caption{Different contributions to the decay widths of $\Barycc$ baryons in units of $10^{-12}$ GeV, calculated in \textit{the kinetic mass scheme}. The dimension-three and five nonleptonic contributions are given in three terms corresponding to LO, NLO, and NNLO, respectively, while their semileptonic counterpart is summed into one term. The dimension-six LO and NLO contributions for the four-quark decay widths are given in the first line of their respective cells, with the dimension-seven contributions in the second line.}
    \label{tab:ccBreakdown}
\end{table}

\begin{table}[h]
    \centering\small
    \begin{tabular}{ccccccccc}
    \hline\hline
    & $\Gamma_{\rm 3+5}^{\rm NL}$ & $\Gamma^{\rm NL}_{\rho}$ & $\Gamma_{\rm exch}$ & $\Gamma_{\rm int-}$ & $\Gamma^{\rm NL}_{\rm int+}$ & $\Gamma_{\rm int+}^{\rm SL}$ & $\Gamma_{\rm 3+5+\rho}^{\rm SL}$ \\\hline
    $\Xi_{bc}^{+}$ & $0.78+0.24+0.27$ & $0.17$ & \makecell{$0.43-0.02$\\} & \makecell{$-0.36+0.08$\\$0.10$} & \makecell{$0.11-0.01$\\} & \makecell{$0.05$\\} & $0.41$ \\\hline
    $\Xi_{bc}^{0}$ & $0.78+0.24+0.27$ & $0.17$ & \makecell{$1.49+0.23$\\$0.43$} & \makecell{$-0.01+0.00$\\} & \makecell{$0.15-0.01$\\$-0.01$} & \makecell{$0.07-0.01$\\$-0.01$} & $0.41$ \\\hline
    $\Omega_{bc}^{0}$ & $0.78+0.24+0.27$ & $0.18$ & \makecell{$0.49-0.01$\\$0.03$} & \makecell{$-0.01+0.00$\\} & \makecell{$0.90-0.06$\\$-0.37$} & \makecell{$0.54-0.19$\\$-0.23$} & $0.42$ \\\hline\hline
    $\Xi_{bc}^{'+}$ & $0.78+0.24+0.27$ & $0.17$ & \makecell{$0.05+0.00$\\} & \makecell{$-0.65+0.17$\\$0.31$} & \makecell{$0.21-0.01$} & $0.10$ & $0.41$ \\\hline
    $\Xi_{bc}^{'0}$ & $0.78+0.24+0.27$ & $0.17$ & \makecell{$3.09+0.77$\\$1.28$} & \makecell{$-0.01+0.00$\\} & \makecell{$0.26-0.02$\\$-0.03$} & \makecell{$0.13-0.01$\\$-0.02$} & $0.41$ \\\hline
    $\Omega_{bc}^{'0}$ & $0.78+0.24+0.27$ & $0.18$ & \makecell{$0.22+0.05$\\$0.09$} & \makecell{$-0.02+0.00$\\} & \makecell{$1.39-0.09$\\$-0.44$} & \makecell{$0.88-0.34$\\$-0.28$} & $0.42$ \\\hline\hline
    \end{tabular}
    \caption{Different contributions to the decay widths of $\Barybc$ and $\Barybcprime$ baryons in units of $10^{-12}$ GeV, calculated in \textit{the kinetic mass scheme}. The dimension-three and five contributions are given in three terms corresponding to LO, NLO, and NNLO, respectively, while their semileptonic counterpart is summed into one term. The dimension-six LO and NLO contributions for the four-quark decay widths are given in the first line of their respective cells, with the dimension-seven contributions in the second line.}
    \label{tab:breakdownBARYbc}
\end{table}

\newpage
\section{Numerical inputs}
\label{app:inputs}
In this appendix, we collect the numerical inputs used in this work if they are not already specified in the text or in the appendix of \cite{Dulibic:2023jeu}.

The constituent quark masses used in evaluation of nonperturbative matrix elements are given in Table \ref{tab:constituentMasses}.
\newcolumntype{C}{>{$}c<{$}} 
\begin{table}[h]
    \centering
    \begin{tabular}{cccc}
         \hline\hline
         $m_b^\Bary$ & $m_c^\Bary$ & $m_q^\Bary$ & $m_s^\Bary$
         \\\hline
         5.0435\text{ GeV} & 1.7105\text{ GeV} & 0.363\text{ GeV} & 0.538\text{ GeV}
         \\\hline\hline
         $m_b^\Mes$ & $m_c^\Mes$ & $m_q^\Mes$ & $m_s^\Mes$
         \\\hline
         5.0038\text{ GeV} & 1.6633\text{ GeV} & 0.310\text{ GeV} & 0.483\text{ GeV}
         \\\hline\hline
    \end{tabular}
    \caption{Values of constituent quark masses in mesons and baryons, denoted by subscripts $\mathcal{B}$ and $M$, respectively, used in evaluation of nonperturbative matrix elements \cite{Karliner:2014gca}.}
    \label{tab:constituentMasses}
\end{table}

\begin{table}[ht]
\small
    \centering
    \begin{tabular}{cccccccc} \hline\hline
         & $D^\pm$ & $D^0$ & $D_s$ & $B^+$ & $B^0$ & $B_s$ & $B_c$ \\ \hline
    $m_M$     & $1.86966(5)$ & $1.86484(5)$ & $1.96835(7)$ & $5.27941(7)$ & $5.27972(8)$ & $5.36691(11)$ & $6.2749(8)$ \\
    $f_M$     & $0.2120(7)$  & $0.2120(7)$  & $0.2499(5)$  & $0.1920(43)$ & $0.1920(43)$ & $0.2284(37)$ & $0.427(6)$ \\\hline\hline
    \end{tabular}
    \caption{\small Masses and decay constants \cite{FermilabLattice:2011njy,McNeile:2011ng,Na:2012kp,Aoki:2014nga,Christ:2014uea,McNeile:2012qf} of $D_{(s)}$, $B_{(s)}$, and $B_c$ mesons in units of GeV, taken from the latest PDG \cite{ParticleDataGroup:2024cfk} and FLAG \cite{FlavourLatticeAveragingGroupFLAG:2024oxs} averages, with the $B_c$ decay constant taken from the lattice-QCD evaluation in Ref.~\cite{McNeile:2012qf}.}
    \label{tab:mes_masses}
\end{table}

\newpage
\section{Figures}\label{app:figures}
In this appendix, we display the decay topologies contributing to the four-quark spectator terms for doubly heavy baryons.

\begin{figure}[h]
    \centering
    \begin{minipage}{0.32\textwidth}
    \vspace*{0.13cm}
        \centering
        \begin{subfigure}{1\textwidth}
            \centering
        \includegraphics[width=\linewidth]{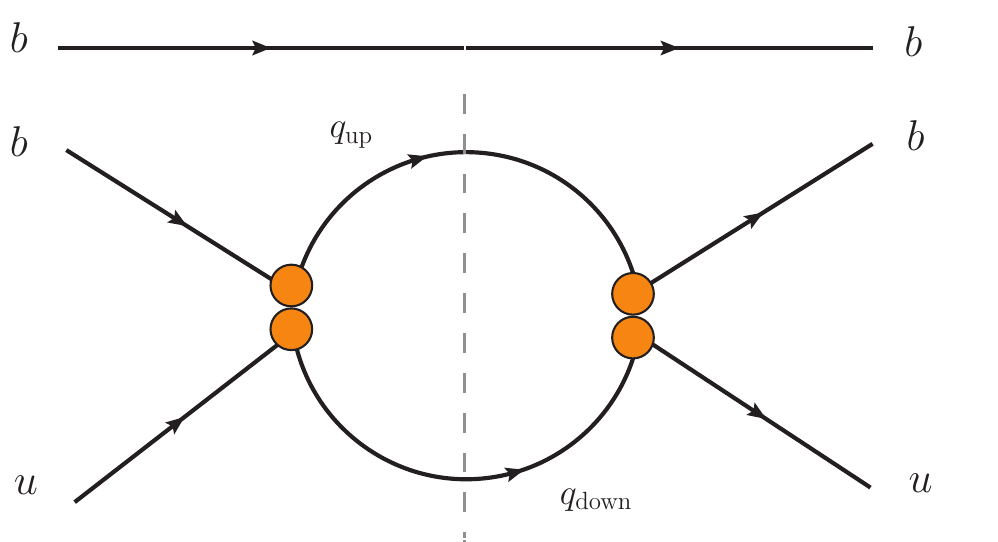}
        \end{subfigure}
        \subcaption*{\footnotesize{(a) WE in $\Xi_{bb}^0$$(bbu)$.}}
    \end{minipage}%
    \hfill
    \begin{minipage}{0.32\textwidth}
    \vspace*{-0.15cm}
        \centering
        \includegraphics[width=\linewidth]{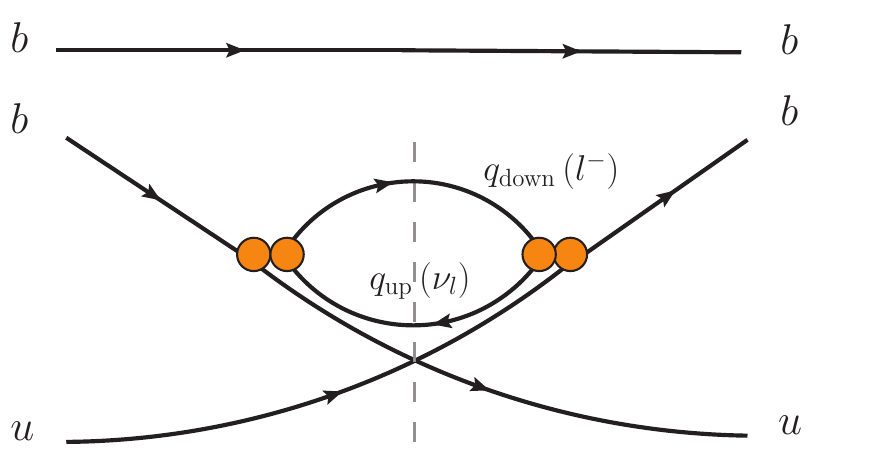}
        \subcaption*{\footnotesize{(b) int$^+$ in $\Xi_{bb}^0$$(bbu)$.}}
    \end{minipage}
    \hfill
    \begin{minipage}{0.28\textwidth}
    \vspace*{0.362cm}
        \centering
        \includegraphics[width=\linewidth]{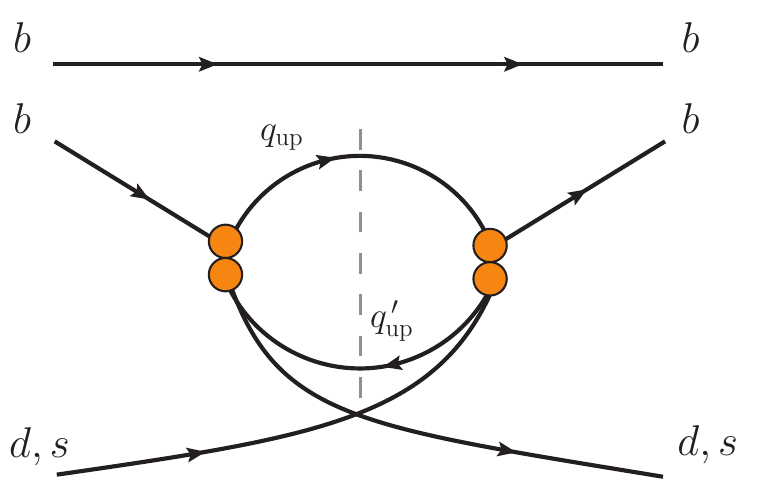}
        \subcaption*{\footnotesize{(c) int$^-$ in $\Xi_{bb}^-$$(bbd)$ and $\Omega_{bb}^-$$(bbs)$.}}
    \end{minipage}
    \caption{\footnotesize Topologies contributing to the decay of $\mathcal{B}_{bb}$ baryons via spectator interactions. Internal lines correspond to $q_{\rm up} = \{u, c\}$, $q_{\rm down} = \{d,s\}$, and, for the semileptonic decays $\ell^- = \{e,\mu,\tau\}$ with the corresponding neutrinos $\nu_\ell$.}  
    \label{fig:bb-topologies}
\end{figure}

\begin{figure}[h]
    \centering
    \begin{minipage}{0.315\textwidth}
        \centering
        \begin{subfigure}{1\textwidth}
            \centering
      \includegraphics[width=\linewidth]{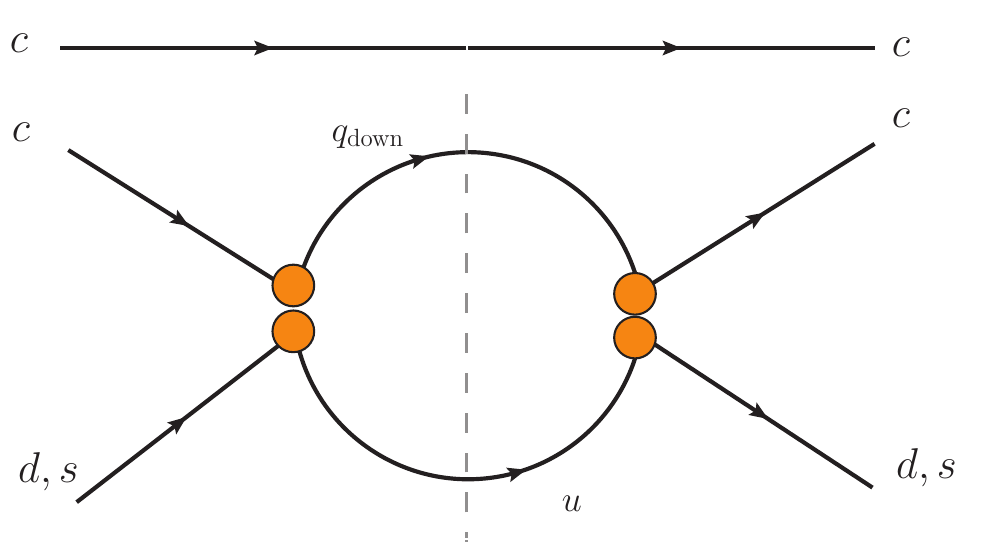}
        \end{subfigure}
        \subcaption*{\footnotesize{(a) WE in $\Xi_{cc}^+$$(ccd)$ and $\Omega_{cc}^+$$(ccs)$.}}
    \end{minipage}%
    \hfill
    \begin{minipage}{0.32\textwidth}
    \vspace*{-0.35cm}
        \centering
        \includegraphics[width=\linewidth]{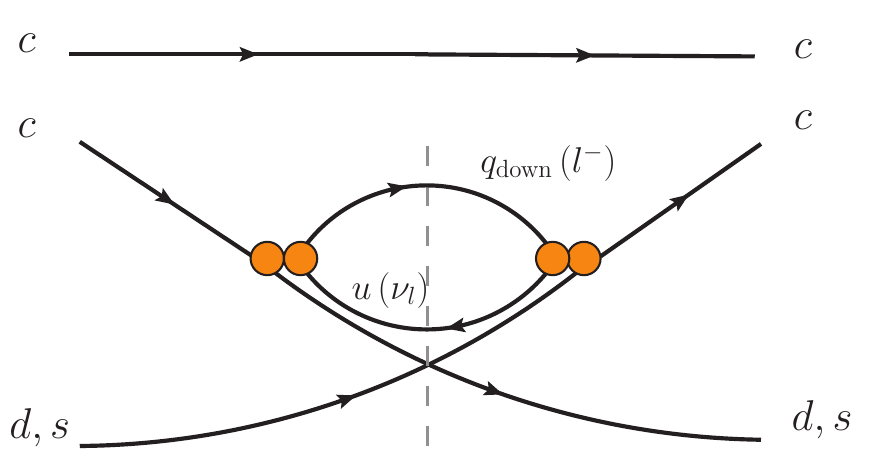}
        \subcaption*{\footnotesize{(b) int$^+$ in $\Xi_{cc}^+$$(ccd)$ and $\Omega_{cc}^+$$(ccs)$.}}
    \end{minipage}
    \hfill
    \begin{minipage}{0.28\textwidth}
    \vspace*{-0.25cm}
        \centering
        \includegraphics[width=\linewidth]{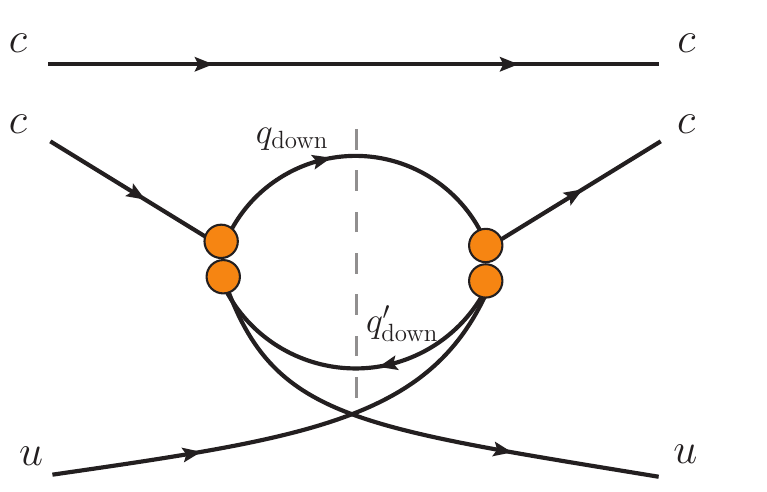}
        \subcaption*{\footnotesize{(c) int$^-$ in $\Xi_{cc}^{++}$$(ccu)$.}}
    \end{minipage}
    \caption{\footnotesize Topologies contributing to the decay of $cc$ baryons via spectator interactions. Internal particles are $q_{\rm down}$, $q_{\rm down}^{\prime} = \{d,s\}$, and for the semileptonic decays $l^- = \{e,\mu\}$ with the corresponding neutrinos $\nu_l$.} 
    \label{fig:cc-topologies}
\end{figure}
\newpage
\begin{figure}[h]
    \centering
    \begin{minipage}{0.32\textwidth}
    \vspace*{0.24cm}
        \centering
     \includegraphics[width=\linewidth]{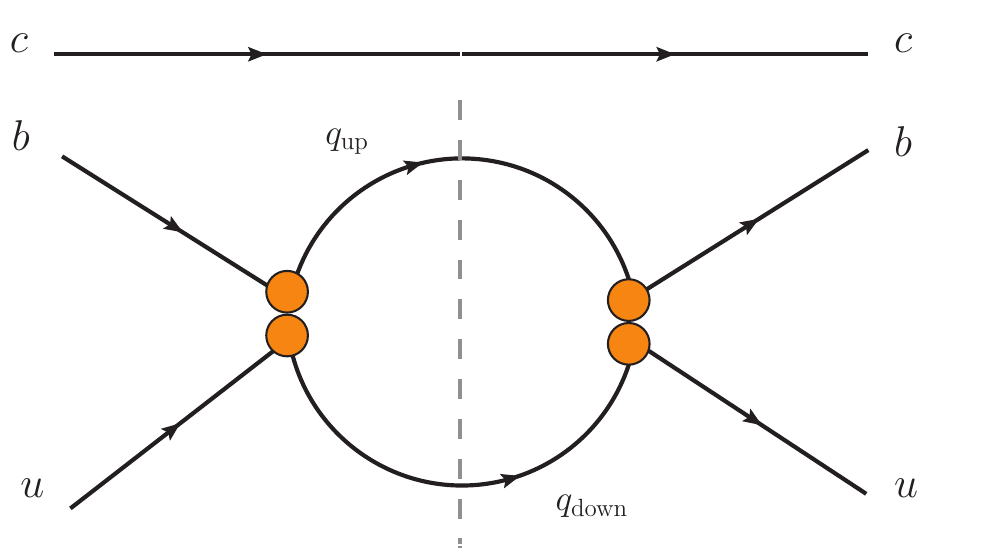}
        \subcaption*{\footnotesize{(a) WE from $b$-quark, with heavy spectator, in $\Xi_{bc}^+$$(bcu)$.}}
    \end{minipage}%
    \hfill
    \begin{minipage}{0.3\textwidth}
       \vspace*{-0.15cm}
        \centering
        \includegraphics[width=\linewidth]{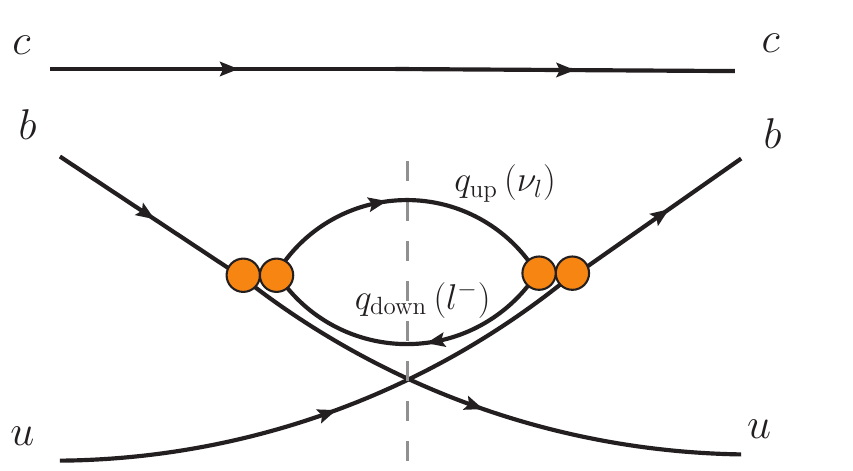}
        \subcaption*{\footnotesize{(b) int$^+$ from $b$-quark, with heavy spectator, in $\Xi_{bc}^+$$(bcu)$.}}
    \end{minipage}
    \hfill
    \begin{minipage}{0.28\textwidth}
        \centering
        \includegraphics[width=\linewidth]{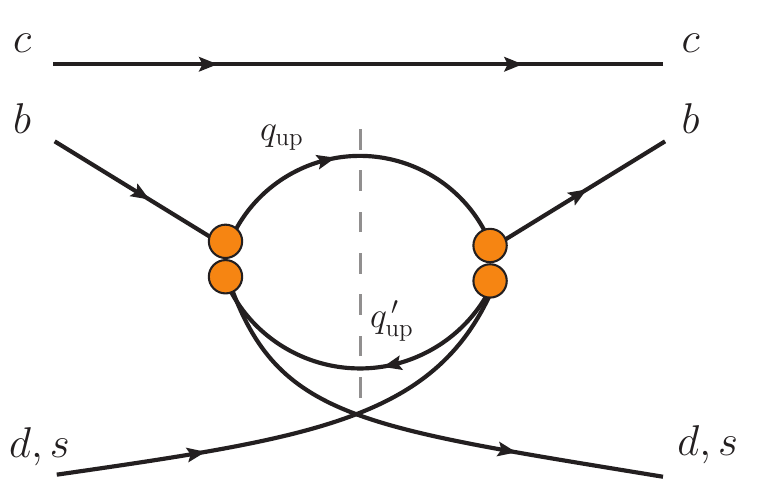}
        \subcaption*{\footnotesize{(h) int$^-$ from $b$-quark in \\$\Xi_{bc}^0$$(bcd)$  and $\Omega_{bc}^0$$(bcs)$.}}
    \end{minipage}
    
    \vspace*{0.3cm}
    
    \begin{minipage}{0.32\textwidth}
     \vspace*{0.33cm}
        \centering
        \includegraphics[width=\linewidth]{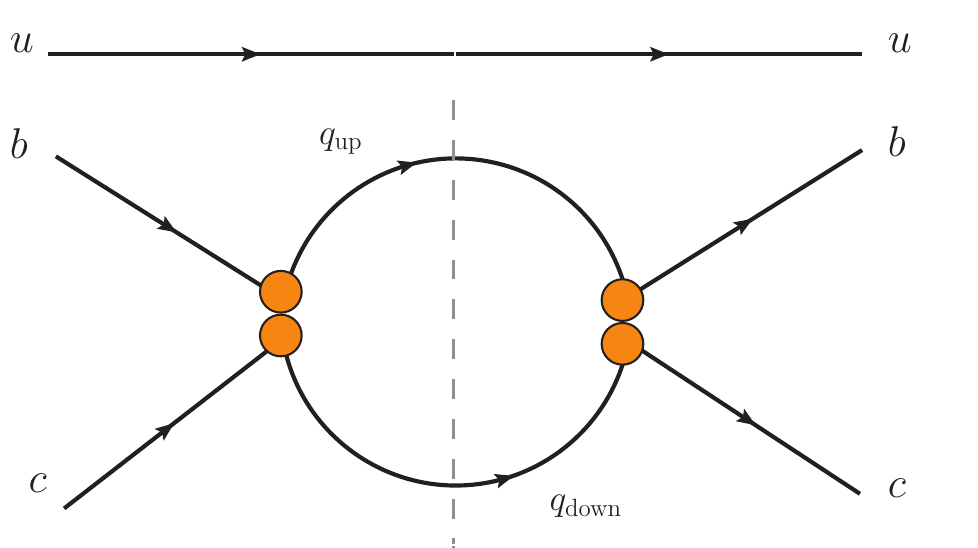}
        \subcaption*{\footnotesize{(d) WE from $b$-quark, with light spectator, in $\Xi_{bc}^+$$(bcu)$.}}
    \end{minipage}%
    \hfill
    \begin{minipage}{0.3\textwidth}
        \centering
        \includegraphics[width=\linewidth]{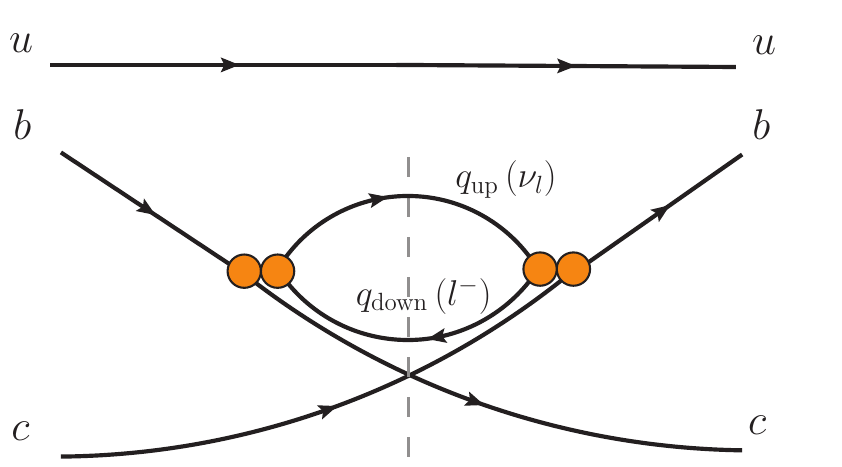}
        \subcaption*{\footnotesize{(e) int$^+$  from $b$-quark, with light spectator, in $\Xi_{bc}^+$$(bcu)$.}}
    \end{minipage}
    \hfill
    \begin{minipage}{0.3\textwidth}
    \hfill
    \end{minipage}

\vspace*{1.5cm}

    \begin{minipage}{0.32\textwidth}
    \vspace*{0.25cm}
        \centering
        \includegraphics[width=\linewidth]{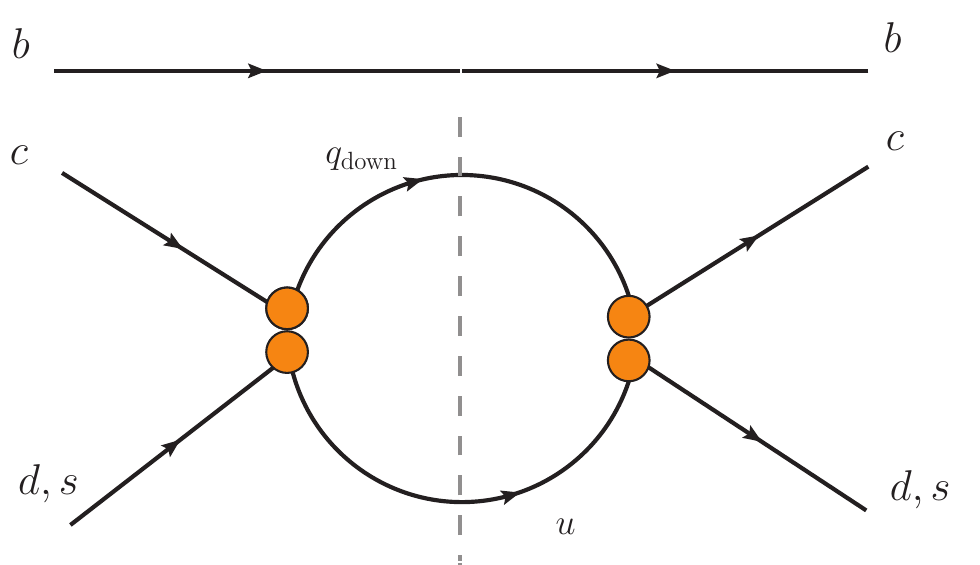}
        \subcaption*{\footnotesize{(f) WE from $c$-quark,  with heavy  spectator,  in $\Xi_{cb}^0$$(bcd)$ and \\$\Omega_{bc}^0$$(bcs)$.}}
    \end{minipage}%
    \hfill
    \begin{minipage}{0.3\textwidth}
         \vspace*{-0.15cm}
        \centering
        \includegraphics[width=\linewidth]{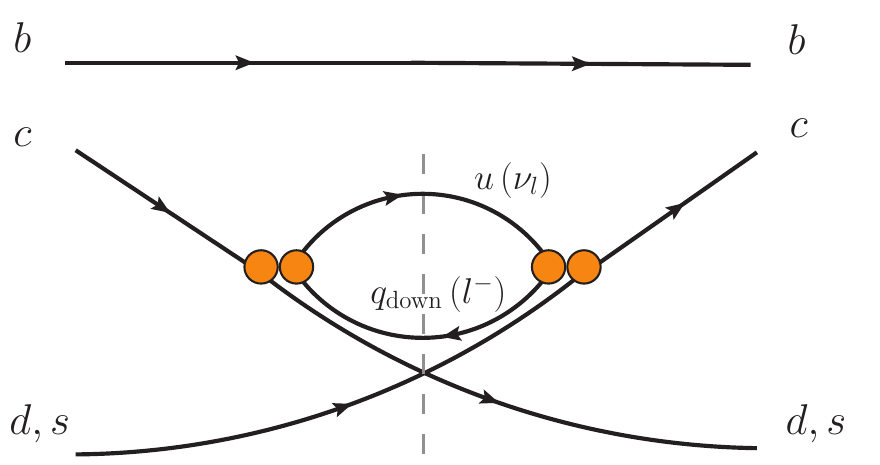}
        \subcaption*{\footnotesize{(g) int$^+$ from $c$-quark, with heavy spectator, in $\Xi_{bc}^0$$(bcd)$  and \\$\Omega_{bc}^0$$(bcs)$.}}
    \end{minipage}
    \hfill
    \begin{minipage}{0.28\textwidth}
       \vspace*{-0.33cm}
        \centering
        \includegraphics[width=\linewidth]{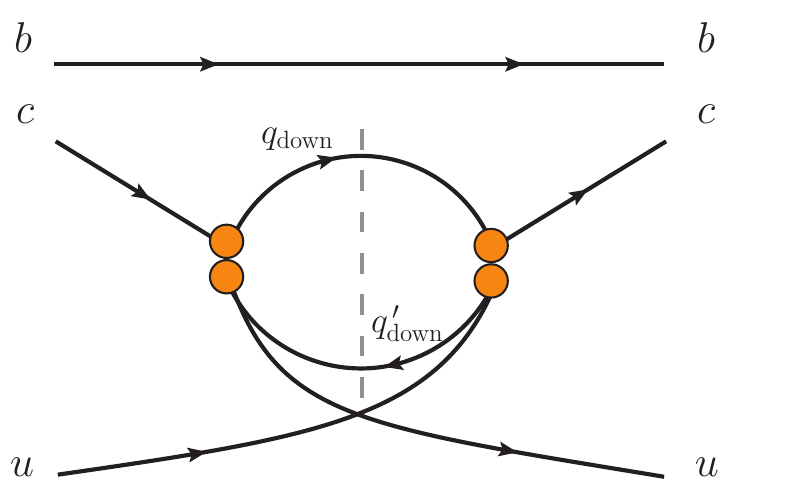}
       \subcaption*{\footnotesize{(c) int$^-$ from $c$-quark in \\$\Xi_{bc}^+$$(bcu)$.}}
    \end{minipage}

   \vspace*{0.3cm}
   
    \begin{minipage}{0.32\textwidth}
       \vspace*{0.33cm}
        \centering
        \includegraphics[width=\linewidth]{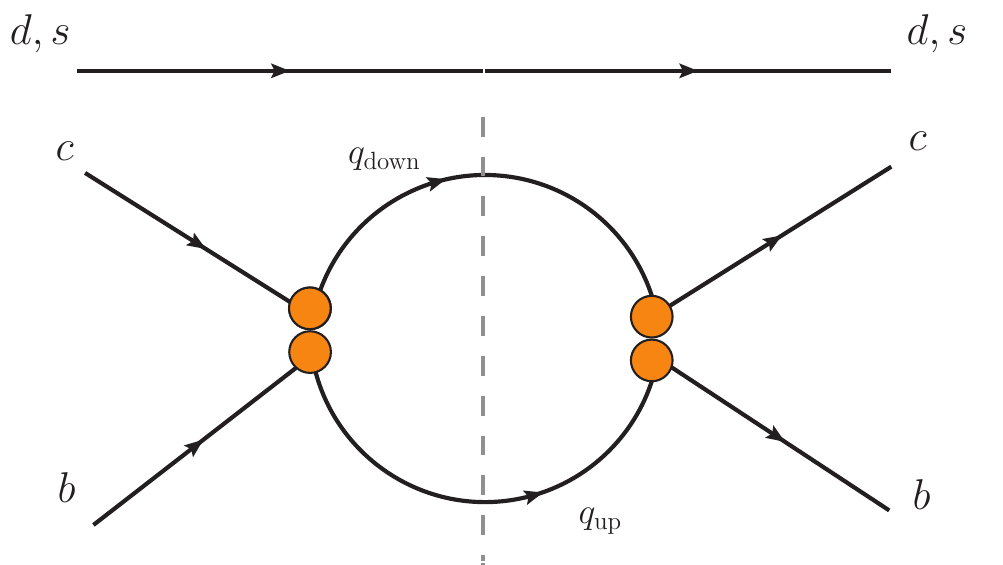}
        \subcaption*{\footnotesize{(i) WE from $c$-quark, with light spectators, in  $\Xi_{bc}^0$$(bcd)$ and \\$\Omega_{bc}^0$$(bcs)$.}}
    \end{minipage}%
    \hfill
    \begin{minipage}{0.3\textwidth}
        \centering
        \includegraphics[width=\linewidth]{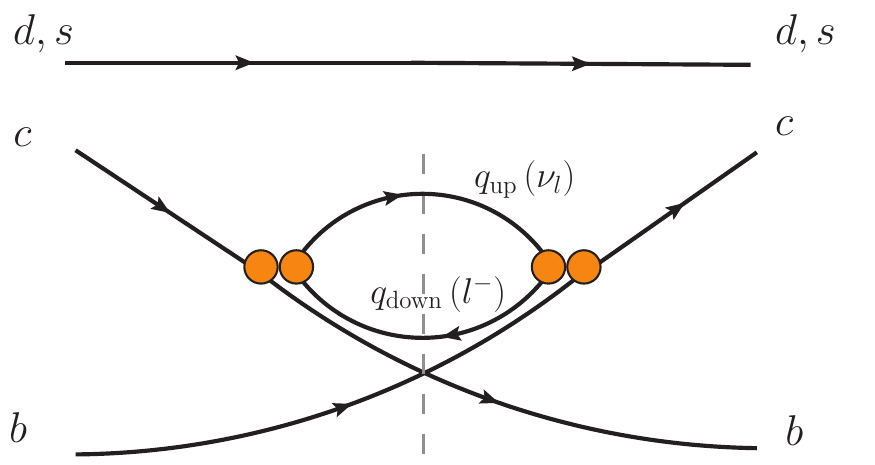}
        \subcaption*{\footnotesize{(j) int$^+$  from $c$-quark, with light spectators, in $\Xi_{bc}^0$$(bcd)$  and \\$\Omega_{bc}^0$$(bcs)$.}}
    \end{minipage}
    \hfill
    \begin{minipage}{0.3\textwidth}
    \hfill
    \end{minipage}
    \caption{\footnotesize Topologies contributing to the decays of $\mathcal{B}_{bc}$ baryons via spectator interactions. Internal lines correspond to $q_{\rm up}$, $q_{\rm up}^{\prime} = \{u,c\}$, and $q_{\rm down}$, $q_{\rm down}^{\prime} = \{d,s\}$.  For the semileptonic decays $\ell^- = \{e,\mu,\tau\}$ for spectator interactions involving $b$-quark and $\ell^- = \{e,\mu\}$ for spectator interactions involving $c$-quark, with the corresponding neutrinos $\nu_\ell$, respectively.} \textbf{}
\label{fig:bc-topologies}
\end{figure}

\newpage
\bibliographystyle{JHEP.bst}
\bibliography{References.bib}
\end{document}